\documentclass[pra,twocolumn,aps,showpacs,floatfix]{revtex4}
\usepackage{graphicx}
\usepackage{amsmath}
\usepackage{amsfonts, amssymb, amstext, amsxtra}
\usepackage{longtable}
\usepackage[dvips]{hyperref}

\def\bfa{{\bf a}}
\def\bfk{{\bf k}}
\def\q{{\bf q}}
\def\bfx{{\bf x}}
\def\Q{{\bf Q}}
\def\bfK{{\bf K}}
\DeclareMathOperator{\Tr}{Tr}

\newcommand{\av}[1]{\langle #1 \rangle}

\begin{document}

\title{Supersolidity of cold-atom Bose-Fermi mixtures in optical lattices}

\author{Peter P. Orth, Doron L. Bergman, and Karyn Le Hur}
\affiliation{Department of Physics, Yale University, New Haven, CT 06520, USA}
 
\date{\today} 

\begin{abstract}
We investigate a cold atomic mixture of spinless bosons and fermions in two-dimensional optical lattices. In the presence of a nested Fermi surface, the bosons may develop a fascinating supersolid behavior characterized by a finite superfluid density as well as a spatial density wave order. Focusing on the triangular lattice geometry and combining a general Landau-Ginzburg-Wilson  approach with microscopically derived mean-field theory, we find an exotic supersolid phase at a fermionic band-filling of $n_f = 3/4$ with a Kagome-type crystalline order. We also address the case of anisotropic hopping amplitudes, and show that striped supersolid phases emerge on the square and triangular lattices. For weak interactions, the supersolid competes with phase separation.
For strong  intra- and inter-species interactions, with the total number of fermions and bosons 
corresponding to one particle per site, the bosons form an alternating Mott insulator ground state.
Finally, for a mixture of $^{87}\text{Rb}^{40}\text{K}$ and $^{23}\text{Na}^6\text{Li}$, we show
that supersolidity can be observed in the range of accessible temperatures in the square lattice geometry. 
\end{abstract}
\pacs{03.75.Lm; 67.80.kb; 67.85.Pq}
\maketitle
\section{Introduction}
\label{sec:introduction}
One of the most intriguing predictions in the theory of quantum mechanics is the possibility of supersolidity - 
superfluid behavior in a rigid crystal (solid).
A supersolid phase involves two unrelated broken symmetries - global $U(1)$ phase invariance breaking 
(superfluidity) and translational invariance breaking 
(density wave)~\cite{andreev_lifshitz_supersolid, PhysRevLett.25.1543}.
In the context of a lattice system, the discrete translational symmetry will be broken (forming a 
superlattice structure). Over the years, much work has been devoted to its experimental realization as well as its theoretical understanding~\cite{doi:10.1038/nature02220,arXiv:0904.2640v1,prokofev_supersolid_he,PhysRevLett.94.155302}. 
While experimental efforts have so far concentrated on solid $^4\text{He}$, there is a variety of theoretical models that exhibit supersolidity, most notably interacting lattice models such as the Bose-Hubbard~\cite{PhysRevB.40.546,PhysRevLett.81.3108} or the Bose-Fermi Hubbard~\cite{PhysRevA.68.023606, PhysRevLett.93.190405} model. The possibility to simulate these models using ultracold atoms in optical lattices~\cite{greiner_nature_2002,bloch:885,Jaksch200552,PhysRevLett.92.050401,PhysRevLett.87.080403,PhysRevLett.92.140405,zaccanti:041605,ospelkaus:020401,ospelkaus:180403,ospelkaus:120403} offers a fascinating alternative experimental route to supersolidity~\cite{PhysRevLett.91.130404,PhysRevLett.95.033003,keilmann:255304,vengalattore:170403,arXiv:0806.1991v1}. 
In particular, time of flight experiments can probe superfluid condensation and density wave order simultaneously~\cite{ScarolaSS}. 

Single-component Hubbard models require a nearest-neighbor interaction, at least, to stabilize a supersolid phase~\cite{PhysRevLett.94.207202,wessel:127205,heidarian:127206,melko:127207,batrouni:087209,arXiv:0901.4366v1,0295-5075-72-2-162,mazzarella:013625,iskin:053634};  a typical example being the frustrated bosonic $U$-$V$ model - where $U$ denotes the on-site Hubbard interaction and $V$ the interaction
between nearest neighbors - on the triangular lattice. Another concrete example is the dipolar boson lattice model~\cite{Goral:2002,buchler:060404}. On the other hand, it has been shown that in two-component mixtures already pure on-site interactions are sufficient to induce a supersolid phase~\cite{PhysRevLett.91.130404,titvinidze:100401,sinha:115124,mathey:011602}. Mixtures with different species, in general, allow to realize new states of matter in a variety of settings~\cite{orth:051601,PhysRevLett.91.073601,PhysRevLett.91.090402,PhysRevLett.91.150403,rizzi:023621,zollner:013629,roscilde:190402,buonsante:240402,maska:060404,mathey:174516,PhysRevLett.93.120404,ottenstein:203202,arXiv:0906.4378v1,bhaseen:135301,girardeau:245303,arXiv:0903.5226v1}, due to effective interactions
that one species induces on another.
Hereafter, we focus on Bose-Fermi mixtures in which the fermions induce an effective 
interaction between the bosons, and vice versa. 
We study in detail the emergence of supersolidity in the Bose-Fermi mixtures; other aspects of Bose-Fermi mixtures have been addressed in the literature~\cite{PhysRevA.69.021601,PhysRevLett.93.090406,PhysRevA.72.051604,refael:144511,lutchyn:220504,klironomos:100401,mering:023601,Fehrmann200423,imambekov:021602,pollet:023608,sengupta:063625,pollet:190402,marchetti:134517,bergman:184520}.

In Bose-Fermi mixtures, one important possible mechanism to achieve boson supersolidity relies on the existence of a nested Fermi surface. 
With nesting, fermions tend to exhibit a density wave at the nesting wavevector(s), and this generates the same ordering tendency on the bosons, through the boson-fermion interaction. Alternatively, the fermions induce interactions between bosons, and the superfluid-to-supersolid transition can also be understood as a condensation of rotons, occuring 
when the roton gap in the superfluid excitation spectrum vanishes upon increasing the interaction strength~\cite{Goral:2002,zhao:105303,arXiv:0806.1991v1}. We will employ two distinct mean-field calculations which follow both points of view. 

Here we focus on two-dimensional lattices with both spatially isotropic and anisotropic hopping amplitudes; the anisotropic cases will allow us to investigate the (quasi-)one-dimensional limit of supersolidity. In one dimension, where fermions are formally equivalent to hard-core bosons, supersolidity was recently predicted to occur in a strongly interacting two-component bosonic mixture~\cite{mathey:011602}, also as a non-equilibrium state~\cite{keilmann:255304}. In three dimensions, our analysis predicts that the supersolid appears at lower temperatures compared to the two-dimensional case, because the presence of van Hove singularities strongly enhance the tendency for density wave formation only in lower than three dimensions.

We begin our analysis with a mixture of spin-polarized bosons and fermions in a two-dimensional triangular optical lattice~\cite{morsch:179}, which exhibits nesting at a particular fermionic band filling of $n_f = 3/4$, as shown in Fig.~\ref{fig:1}. A supersolid phase has already been predicted for the isotropic square lattice \cite{PhysRevLett.91.130404,titvinidze:100401}. For a sufficiently deep optical lattice only nearest-neighbor hopping survives, and the system is described by the ubiquitous single band Bose-Fermi Hubbard Hamiltonian~\cite{bloch:885,Jaksch200552},
\begin{eqnarray}
  \label{eq:1}
    H &=& - \sum_{\av{i,j}} ( t_{f,ij} f^\dag_i f_j + t_{b,ij} b^\dag_i b_j ) \\ \nonumber
     &-& \sum_i \left(\mu_f m_i +  \mu_b n_i \right) \\ \nonumber
     &+& \frac{U_{bb}}{2} \sum_i n_i (n_i - 1) + U_{bf} \sum_i n_i m_i\,,
\end{eqnarray}
where $f^\dag_i (b^\dag_i)$ is the fermionic (bosonic) creation operator at site $i$, while $m_i = f^\dag_i f_i$ $(n_i = b_i^\dag b_i)$ denotes the fermionic (bosonic) number operator and $\mu_{f(b)}$ is the chemical potential of the fermions (bosons). Hopping is restricted to neighboring sites (denoted by the summation over $\av{i,j}$) with amplitudes $t_{f,ij} (t_{b,ij})$ for fermions (bosons) that in general depend on the direction of hopping. The on-site boson-boson and boson-fermion interaction strengths are given by $U_{bb}$ and $U_{bf}$, respectively. 
We do not include interactions between fermions, which due to the Pauli-principle only occur in the p-wave scattering channel, and are frozen out at ultracold temperatures. 

Before we set out to show how and in which parameter regime supersolidity arises from the above Hamiltonian, we give a brief outline of the article.

The paper is organized as follows:
after this introduction, in Sec.~\ref{sec:instability-analysis}, we focus on the isotropic triangular lattice, and carry out an instability analysis of the system in the weak-interaction limit after the fermionic degrees of freedom were integrated out. For the particular fermionic filling $n_f = 3/4$, the Fermi surface both shows nesting and contains van Hove singularities. This triggers two distinct low-temperature instabilities of the superfluid bosons: one towards phase separation and one towards supersolid formation. The density modulation in the supersolid phase is characterized by the \emph{three} nonequivalent nesting vectors $\Q_{1,2,3}$ of the Fermi surface (see Fig.~{\ref{fig:1}(b)), producing a Kagome-type crystalline order in real-space.

In Sec.~\ref{sec:mean-field-theories}, we calculate the low-temperature mean-field phase diagram of the system, containing a Kagome-supersolid and a phase separated regime. We also determine the amplitude of the density wave modulations inside the supersolid. 

In Sec.~\ref{sec:phase-diagram-from}, we use a general Landau-Ginzburg-Wilson mean-field theory 
to find a number of likely supersolid phases in a general phase diagram, assuming possible condensation into the wavevector modes ${\bf 0}, \Q_{1,2,3}$. We match the phenomenological Landau expansion parameters to a microscopically derived mean-field expression and obtain criteria where the supersolid and phase separated regions emerge in the phase diagram, 
taking us beyond the instability analysis. 

We then embark, in Sec.~\ref{sec:supers-dens-wave}, to calculate the density wave modulation in the supersolid using a different mean-field approach that treats the fermions exactly. We find that the fermionic spectrum acquires a gap in the supersolid phase, allowing the system to lower its energy. The density modulation in the supersolid is found to be rather weak, typically involving only $0.1 \%$ of all the bosons. The transition temperatures $T_{\text{SS}}$ are also small compared to $T_F$, typically we find $T_{\text{SS}}/T_F \simeq 0.01$, where we define $T_F= 6 t_f$ as the Fermi temperature of the optical lattice (the Boltzmann constant $k_B=1$). 

To find larger transition temperatures, we turn to investigate the case of anisotropic hopping amplitudes in Sec.~\ref{sec:anisotropic-hopping}, first for the square (Sec.~\ref{sec:square-lattice-1}) and then for the triangular lattice (Sec.~\ref{sec:triangular-lattice-1}). As a result of reduced symmetry, only one nesting vector occurs and the supersolid phase exhibits a striped pattern in real-space. Since the relevant features of anisotropic hopping are already captured by the square lattice geometry, we mainly focus on this case, where analytical results can be derived. Since nesting is fulfilled for a larger fraction of wavevectors, we find larger supersolid transition temperatures in our mean-field analysis. They are as high as $T_{\text{SS}} \simeq t_f \; (3 t_f/5)  \simeq T_F/4 \;(T_F/5)$ for the isotropic (anisotropic) square lattice. The supersolid density wave now involves up to $20 \%$ of all the bosons. 
\begin{figure}[tb]
  \centering
  \includegraphics[width=.7\linewidth]{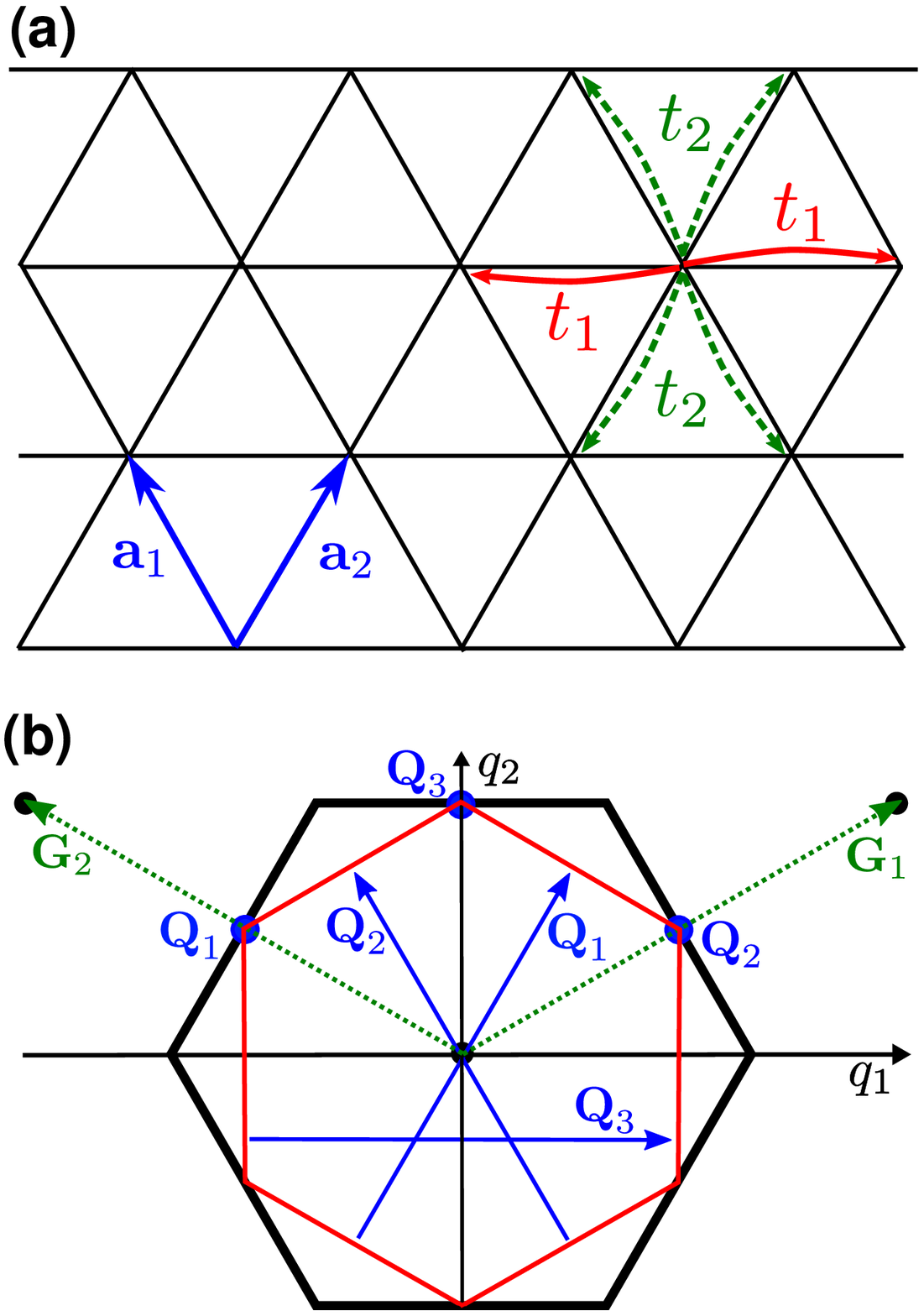}
  \caption{(Color online) (a): Triangular lattice in real space with our conventions of unit length Bravais lattice vectors $\{\bfa_{1}, \bfa_2\}$ and hopping amplitudes $\{t_1, t_2\}$. 
\newline
(b) Reciprocal lattice vectors $\{ {\bf G}_1, {\bf G}_2 \}$ (dotted), first Brillouin zone (thick hexagon), and the Fermi surface at $\mu_f = 2 t_f$ ($n_f = 3/4)$ (thin inner hexagon). The Fermi surface exhibits three nonequivalent nesting vectors ${ \Q_1, \Q_2, \Q_3 }$ (solid arrows), which when folded back into the first Brillouin zone occur at
: $\Q_1 = {\bf G}_2/2$, $\Q_2 = {\bf G}_1/2$, and $\Q_3 = ({\bf G}_1 + {\bf G}_2)/2$. They coincide with the critical points (corners) of the Fermi surface hexagon, which give rise to a van Hove singularity in the density of states at this filling (see Fig. 2(b)).}
\label{fig:1}
\end{figure}

In Sec.~\ref{sec:exper-param-supers}, we predict the supersolid parameter regime for two experimental realizations of Bose-Fermi mixtures, $^{87}\text{Rb}^{40}\text{K}$ and $^{23}\text{Na}^{6}\text{Li}$. We show that while the transition temperatures for the triangular lattice geometry are beyond current cooling limits, the supersolid phase on the square (isotropic and anisotropic) lattice, should be accessible with current technology. We give optimal choices of experimental parameters in order to maximize $T_{\text{SS}}$. Anisotropic hopping offers the possibly crucial advantage to significantly weaken the tendency towards phase separation while retaining supersolid transition temperatures close to current experimental limits. 

We discuss in detail how the supersolid phase can be detected using time-of-flight measurements, and conclude that the detection becomes easier for a smaller ratio of bosonic to fermionic hopping amplitudes $t_b/t_f$, \emph{i.e.} slow bosons.

Finally, in Sec.~\ref{sec:mott-insul-phas}, we address the limit of strong interactions, where the system, at total
unit filling (bosons and fermions combined), can be described by a quantum Heisenberg Hamiltonian with an additional gauge field arising from the celebrated Jordan-Wigner transformation in two dimensions. We show that the bosons
are localized in an (alternating) Mott insulator phase, and the fermions feature a density wave with wavevectors equal to the nesting vectors. 

We summarize our results in Sec.~\ref{sec:conclusions}, and leave the details of a number of our calculations to the appendices.

\section{Low temperature instabilities}
\label{sec:instability-analysis}
In general, one expects a supersolid phase to occur for weak-interspecie interaction $U_{bf}$, since for larger interactions the mixture either phase separates or enters a Mott-insulating state (large $U_{bb}$ and $U_{bf}$)~\cite{titvinidze:100401}. We will separately address the strongly interacting regime in Sec.~\ref{sec:mott-insul-phas}. For now, we focus on sufficiently weak boson-fermion interactions $U_{bf}$ (we will specify the exact condition below).

\subsection{Definitions for the triangular lattice}
\label{sec:defin-triang-latt}
We begin by integrating out the fermionic degrees of freedom in an imaginary time functional integral approach. Beforehand, it is convenient to write the Hamiltonian of Eq.~(\ref{eq:1}) in the Bloch state basis, where the annihilation operators read $f_j = \frac{1}{\sqrt{N_L}} \sum_{\q \in \text{BZ}} f_\q e^{i \q \cdot \bfx_j}$, and $b_j = \frac{1}{\sqrt{N_L}} \sum_{\q \in \text{BZ}} b_\q e^{i \q \cdot \bfx_j}$. Here, the summation is over the first Brillouin zone (BZ) of the triangular lattice, $\bfx_j$ is the real-space vector to lattice site $j$ and $N_L$ is the number of unit cells. Our convention of unit length Bravais lattice vectors is ${\bf a}_1 = \left(\frac12,\frac{\sqrt{3}}{2} \right)$ and ${\bf a}_2 = \left(-\frac12,\frac{\sqrt{3}}{2} \right)$, {\it i.e.}, $|{\bf a}_{1,2}| = 1$; for simplicity, the lattice constant is fixed to one. The reciprocal lattice vectors are then given by ${\bf G}_1 = 2 \pi \left(1, \frac{1}{\sqrt{3}} \right)$ and ${\bf G}_2 = 2 \pi \left( -1, \frac{1}{\sqrt{3}} \right)$. The real space lattice and the first Brillouin zone are shown in Fig.~\ref{fig:1}. 

The Hamiltonian in momentum space then reads $H = H_b + H_f + H_{bf}$ with 
\begin{equation} 
 \label{eq:2}
 \begin{split}
 H_b &= \sum_{\q \in \text{BZ}}\Big[\xi_b(\q) b_\q^{\dag} b_\q  + \sum_{\bfk_1,\bfk_2}  \frac{U_{bb}}{2 N_L} b_{\bfk_1 - \q}^{\dag} b_{\bfk_2 + \q}^{\dag} b_{\bfk_2} b_{\bfk_1}  \Big]  \\ 
 H_f &= \sum_{\q \in \text{BZ}}  \xi_f(\q) f_\q^{\dag} f_\q  \\ 
H_{bf} &= \sum_{\q, \bfk_1, \bfk_2\in \text{BZ}}  \frac{U_{bf}}{N_L} b_{\bfk_2 - \q}^{\dag} b_{\bfk_2} f_{\bfk_1 + \q}^{\dag} f_{\bfk_1}  \,,
\end{split}
\end{equation}
where the dispersion relation for the fermions (bosons) on the triangular lattice reads 
\begin{equation}
  \label{eq:3}
  \xi_{f(b)}(\q) = - \mu_{f(b)} - 2 t_{f(b)1} \cos q_1 - 4 t_{f(b)2} \cos \frac{q_1}{2} \cos \frac{\sqrt{3} q_2}{2} \,.
\end{equation}
The hopping amplitudes $\{t_{f(b)1}, t_{f(b)2}\}$ describe hopping along the direction $\pm ({\bf a}_2 - {\bf a}_1)$ and along $\pm {\bf a}_{1,2}$, respectively. Most generally there are three different hopping parameters for the three directions on the triangular lattice, but we will only consider two of them being different. This already includes the interesting cases of weakly-coupled one-dimensional chains ($t_{f(b)2} \ll t_{f(b)1}$) and the transition to the square lattice ($t_{f(b)2} \gg t_{f(b)1}$). Both scenarios will be discussed in Sec.~\ref{sec:anisotropic-hopping}. Until then, we assume isotropic hopping amplitudes $t_{f(b)1} = t_{f(b)2} \equiv t_{f(b)}$.  

The imaginary time partition function of the system is quadratic in the fermionic degrees of freedom, which can therefore be integrated out exactly. 

\subsection{Effective bosonic theory}
\label{sec:effect-boson-theory}
Integrating out the fermions, yields formally
\begin{equation}
  \label{eq:4}
  \int {\cal D} b_q^* {\cal D} b_q {\cal D}f^*_q {\cal D} f_q  e^{-\left( S_b + S_f + S_{bf} \right)} = \int {\cal D} b_q^* {\cal D} b_q  e^{-S_b^{\text{eff}}}\,.
\end{equation}
The variable $q$ contains a momentum and an imaginary time component $q = (\tau, \q)$, and the bare action derives from the respective parts of the Hamiltonian: 
\begin{equation}
  \label{eq:5}
  \begin{split}
    S_b &= \int_0^\beta d\tau \sum_{\q} \left[ b^*_q \frac{\partial}{\partial \tau} b_q + H_b(b_q^*, b_q) \right]\\
    S_f &= \int_0^\beta d\tau \sum_{\q} \left[ f^*_q \left( \frac{\partial}{\partial \tau} + \xi_f(\q) \right) f_q
      \right] \\
      &= \int_0^\beta d\tau \sum_{\q} f^*_q {\mathcal G}^{-1}(q) f_q
      \\
    S_{bf} &= \int_0^\beta d\tau \frac{U_{bf}}{N_L} \sum_{\q, \bfk_1, \bfk_2}   b_{k_2 - q}^* b_{k_2} f_{k_1 + q}^* f_{k_1}
    \, .
  \end{split}
\end{equation}
Here $\beta = 1/T$ where $T$ is the temperature and implicitly all the fields have the same imaginary time component. Integrating out the fermions, we get a determinant depending on the boson density
\begin{equation}
S_b^{\text{eff}} = S_b - \Tr \ln \Big( {\mathcal G}^{-1}(q) \delta({\bf k-q}) + 
\frac{U_{bf}}{N_L} \sum_{\bf p} b_{p + (k-q)}^* b_{p} \Big)\,.
\end{equation}
Next we expand to second order in $U_{bf}$. To first order in $U_{bf}$, the fermions simply produce a (trivial) shift of the bosonic chemical potential $\mu_b \rightarrow \mu_b - U_{bf} n_f$ that depends on the fermionic filling $n_f = N_f/N_L$, where $N_f$ is the total number of fermions. 
Trading the integration over imaginary time with a summation over the bosonic Matsubara 
frequency domain, defined by
\begin{equation}
  \label{eq:6}
  b_{(\tau, \q)} = \frac{1}{\sqrt{\beta}} \sum_{m = -\infty}^\infty b_{(i\omega_m, \q)} e^{- i \omega_m \tau}
\end{equation}
with $\omega_m = 2 \pi m/\beta$, and simplifying a bit, the effective bosonic action up to second order in $U_{bf}$ takes the form
\begin{equation}
  \label{eq:7}
\begin{split}
  S_b^{\text{eff}} &= \sum_{q=(i\omega_m,\q)} \Big\{ \left[ - i \omega_m + \xi_b(\q) + U_{bf} n_f \right]  b^*_q b_q  \\
&  + \frac{1}{2 N_L \beta} \sum_{k_1,k_2} [U_{bb} + U_{bf}^2 \chi(T, q) ] b^*_{k_1 - q} b^*_{k_2+q} b_{k_2} b_{k_1} \Big\}\,.
\end{split}
\end{equation}
The second order term in $U_{bf}$, involves the fermionic polarization Lindhard function,
\begin{equation}
  \label{eq:8}
  \chi(T,  i \omega_m, \q) = \frac{1}{N_L} \sum_\bfk \frac{ f[\xi_f(\bfk)] - f[\xi_f(\bfk + \q)]}{ i \omega_m + \xi_f(\bfk) - \xi_f(\bfk + \q)}\,,
\end{equation}
which depends on temperature $T$ via the Fermi function $f(\xi_f) = [1+ \exp(\xi_f/T)] ^{-1}$.  The induced interaction is \emph{attractive} in momentum space independently of the sign of $U_{bf}$, as $\chi(\q) < 0$ for all $\q$. In real-space, it is long-range and oscillatory in sign with an interesting (Kagome-lattice-type) structure due to the non-trivial wavevector dependence (see Appendix~\ref{sec:ferm-induc-inter}). 

Higher order terms can be neglected when $M_0 U_{bf} \ll 1$, where $M_0 \sim t_f^{-1}$ is an estimate of the Lindhard function (away from its singularities; see Fig.~\ref{fig:2}(b)). 
\begin{figure}[tb]
  \centering
  \includegraphics[width=\linewidth]{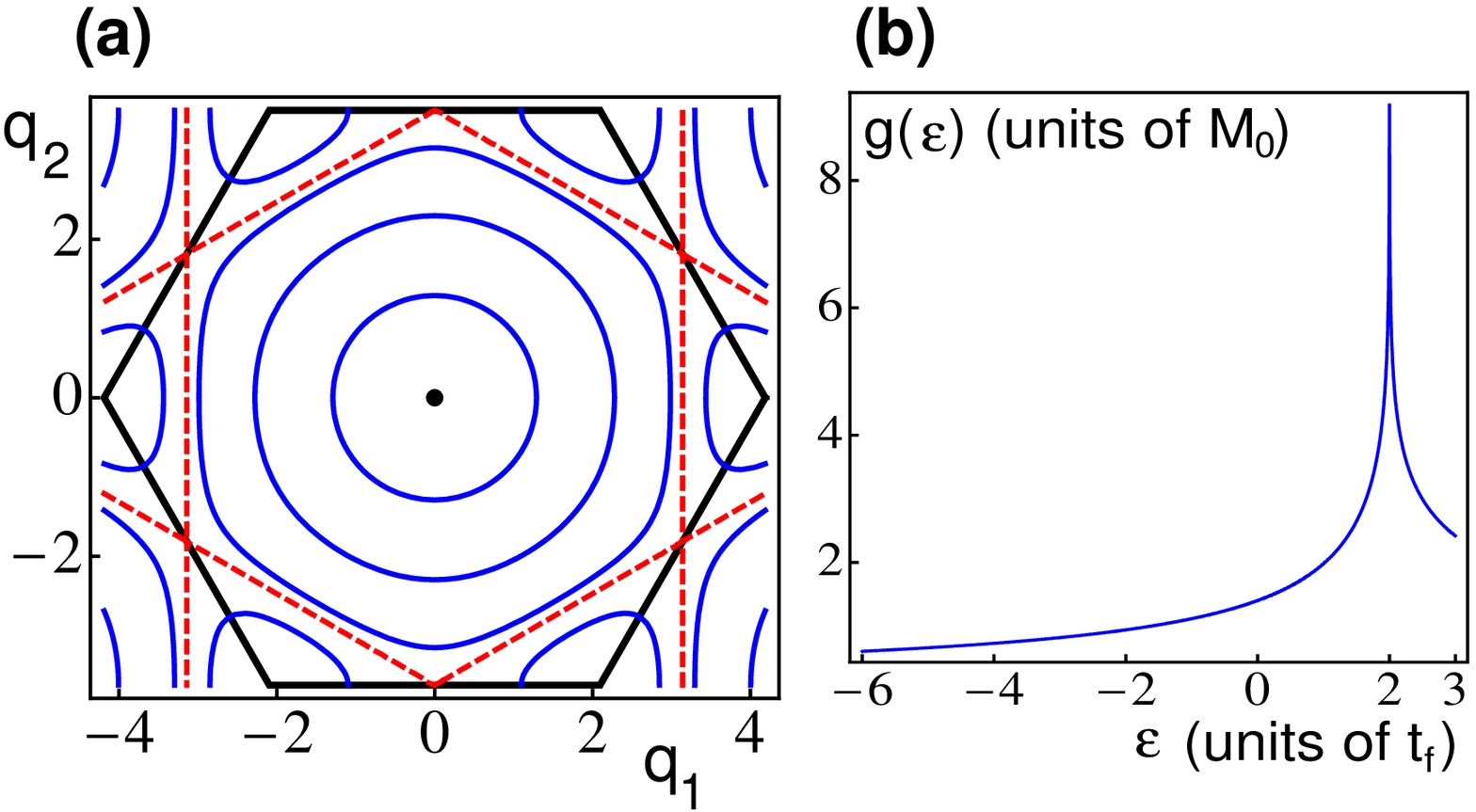}
  \caption{(Color online) (a): Fermi surfaces for $\mu_f/t_f = 2$ (dashed) and $\mu_f/t_f = \{-3.75, -0.375, 0.875, 2.5\}$ (solid). The thick hexagon denotes the first Brillouin zone.
    \newline
    (b) Fermionic density of states $g(\epsilon)$, which diverges logarithmically at energy $\epsilon = 2 t_f$ (van Hove singularity). We take $M_0 = 3/4 \pi^2 t_f$ as a measure of $g(\epsilon)$ away from the divergence.}
  \label{fig:2}
\end{figure}

We will analyze the behavior of this function in some detail in the following and in Appendix~\ref{sec:deta-analys-lindh}, since it provides a basic understanding of the mechanism of supersolid formation. In particular, its static part ($i\omega_m \rightarrow 0$) diverges logarithmically for low temperatures at $\q = 0$ if the Fermi surface contains van Hove singularities. Furthermore, it diverges at special wavevectors $\q = \alpha \Q_i$ if the Fermi surface is nested with nesting vectors $\Q_i$, and $0 \ll \alpha \leq 1$. If both features are present, as it is the case for a fermionic chemical potential of $\mu_f = 2 t_{f}$ or fermionic band filling of $n_f=3/4$ (see Fig.~\ref{fig:1}(b) and Fig.~\ref{fig:2}(a)), the divergence at and close to the nesting vectors gets enhanced to $\chi \sim [\ln (c t_f/T)]^2$, where $c$ is a numerical constant.  
{\it These divergences provide two competing low temperature instabilities in the superfluid (bosonic) phase, one towards supersolid formation and one towards phase separation.}  

First we note, that it is legitimate to only consider the \emph{static limit} ($i \omega_m \rightarrow 0$) of $\chi$, if the fermions are much faster than the bosons ($t_f \gg t_b$). Then, the fermionic response occurs on much faster timescales than the movement of the bosons, and one can safely neglect retardation effects. More formally, the terms with nonzero Matsubara frequencies ($i \omega_m \neq 0$) only contribute subdominantly to the divergences at $\q= {\bf 0}, \Q_{1,2,3}$. The opposite limit of "slow" fermions, where superfluid bosons induce an attractive interaction among the fermions leading to (exotic) superconducting phases, has been discussed in Refs.\cite{PhysRevLett.93.090406,mathey:174516}. 

It was shown in Ref.~\cite{lutchyn:220504} that the static approximation always yields qualitatively correct results, because $(- \chi)$ is positive definite. In general, it improves for smaller fermionic densities $n_f$. 
Note that we take the spatially non-local nature of the induced interaction fully into account. 

In the static limit, one can work with an effective Hamiltonian for the bosons which takes the form:
\begin{equation}
  \label{eq:9}
  \begin{split}
    H_b^{\text{eff}} &= \sum_{\q} \Big[\xi_b(\q) b_\q^{\dag} b_\q  + \sum_{\bfk_1,\bfk_2}  \frac{U(T,\q)}{2 N_L} b_{\bfk_1 - \q}^{\dag} b_{\bfk_2 + \q}^{\dag} b_{\bfk_2} b_{\bfk_1}  \Big],
  \end{split}
\end{equation}
with an interaction $U(T,\q)$ that is given by,
\begin{equation}
U(T,\q) = U_{bb} + U_{bf}^2 \chi(T, \q), 
\end{equation}
where $\chi(T, \q) \equiv \chi(T, i \omega = 0, \q)$. As mentioned above, this perturbative form of the interaction is valid for $M_0 U_{bf} \ll 1 $, where $M_0 = 3/(4 \pi^2 t_f)$ is a measure of the Lindhard function and the density of states away from its singularities (as shown in Fig.~\ref{fig:2}(b)).

At temperatures well below the Kosterlitz-Thouless transition temperature $T_{\text{KT}}$, the bosons form a quasi-condensate, {\it i.e.}, a condensate with a fluctuating phase, described by a wavefunction of the form
$\sqrt{n(\bfx)} e^{i \phi(\bfx)}$. One can diagonalize $H_b^{\text{eff}}$ in the superfluid phase employing the well-established Bogoliubov approximation ${b}_\q = \sqrt{N_L n_0}\delta({\bf q}) + \tilde{b}_\q$ where only terms up to quadratic order in the fluctuation operators $\tilde{b}_{\q \neq 0}$ are kept and $n_0$ represents the finite superfluid density. This yields the spectrum of elementary excitations of the superfluid
\begin{equation}
  \label{eq:10}
  E_b(T, \q) = \sqrt{\xi_b(\q)^2 + 2 n_b \xi_b(\q) \left(U_{bb} + U_{bf}^2 \chi(T, \q) \right)}\,.
\end{equation}
Here, we have assumed that all the bosons are condensed into the zero momentum state by equating $n_0 = n_b$, which is valid to a good approximation for temperatures $T \ll T_{\text{KT}}$. 
In the following, we discuss the two instabilities of the superfluid occurring when $E_b(\q)$ vanishes.

\subsection{Phase separation}
\label{sec:phase-separation}
For small wavevectors $|\q| \ll 1$, the Bogoliubov spectrum is linear $E_b(\q) =  \sqrt{3 n_b t_b [U_{bb} + U_{bf}^2 \chi(T,{\bf 0})] } \, |\q| $, with a sound velocity that vanishes at:
\begin{equation}
  \label{eq:11}
  \chi(T,{\bf 0}) = - U_{bb}/U_{bf}^2\,.
\end{equation}
At this point, the contact interaction $U(T, {\bf 0})$ becomes attractive, which marks the transition to a phase separated regime, since a Bose condensate is thermodynamically unstable for $U(T,{\bf 0}) < 0$~\cite{PhysRevA.69.063603}.  

For a regular density of states $g(\epsilon)$ at the Fermi surface, one finds that $\chi(T, {\bf 0}) = -g(0)$. 
However, due to stationary points ($|\nabla_\q \xi_f(\q)| = 0$) on the Fermi surface for a chemical potential of $\mu_f = 2 t_f$ (see Fig.~\ref{fig:2}(b)), the density of states diverges at this filling like 
\begin{equation}
  \label{eq:12}
  g(\epsilon) \sim M_0 \ln \left| \frac{8 t_f}{\epsilon} \right|\,,
\end{equation}
where $M_0 = 3/(4 \pi^2 t_f)$, resulting in
\begin{equation}
  \label{eq:13}
  \chi(T, {\bf 0}) = - M_0 \ln \left| \frac{8 C_1 t_f}{T} \right|\,,
\end{equation}
with $C_1 = 2 e^C/\pi \approx 1.13$, and $C$ being the Euler-Mascheroni constant. Thus, for any nonzero coupling $U_{bf}$ between the bosons and fermions, there is a temperature $T_{\text{PS}}^{\text{inst.}}$ at which the ${\bf q}=0$
term of the effective interaction becomes attractive $[U(T_{\text{PS}}^{\text{inst.}},{\bf 0}) = 0]$:
\begin{equation}
  \label{eq:14}
   T_{\text{PS}}^{\text{inst.}} = 8 C_1 t_f \exp \left(-\frac{1}{\lambda_{BF}} \right)\,,
\end{equation}
with $\lambda_{BF} = M_0 U_{bf}^2/U_{bb} \ll 1$ describing the ratio of induced attraction to intrinsic repulsion between the bosons. 
\subsection{Supersolid formation}
\label{sec:supersolid-formation}
The other low temperature instability of the superfluid phase occurs only in the presence of a nested Fermi surface. Nesting is defined as the existence of a nesting vector $\Q$ such that for a finite domain of wavevectors $\bfk$, the energy fulfills the prerequisite
\begin{equation}
  \label{eq:15}
  \xi_f(\bfk + \Q) = - \xi_f(\bfk)\,.
\end{equation}
Close to the Fermi surface with $\xi_f(\bfk) \approx 0$, the denominator in the expression of $\chi(T, \Q)$ becomes very small (see Eq.~(\ref{eq:8})). At the same time, the numerator is nonzero, since $\Q$ links an occupied with an unoccupied state, and
\begin{equation}
  \label{eq:16}
  \frac{ f[\xi_f(\bfk)] - f[\xi_f(\bfk + \Q)]}{ \xi_f(\bfk) - \xi_f(\bfk + \Q)} \xrightarrow[\xi_f(\bfk) \approx 0]{} - \frac{1}{4 T}\,.
\end{equation}
Thus, nesting leads to the divergence of $\chi(T\rightarrow 0, \Q)$. 

On the triangular lattice, at the particular band filling of $n_f = 3/4$, as shown in Fig.~\ref{fig:2}(a), 
 the Fermi surface exhibits three nonequivalent nesting vectors which map the different sides of the Fermi surface hexagon onto each other. They read $\Q_1 = (-\pi, \pi/\sqrt{3})$, $\Q_2 = (\pi, \pi/\sqrt{3})$, $\Q_3 = (0, 2 \pi/\sqrt{3})$, and coincide with the location of the van Hove singularities. They fulfill $- \Q_i = \Q_i + {\bf G}_m$, with ${\bf G}_m$ being a reciprocal lattice vector, as well as $\Q_1 + \Q_2 = \Q_3$ (and cyclic permutations). Each of them maps two of the six van Hove points onto another van Hove point, which leads to a significant enhancement of the divergence of $\chi(T,\Q_i)$.
 
We can analytically estimate this divergence of the Lindhard function (Eq.~(\ref{eq:8})) by approximating,
\begin{equation}
  \label{eq:17}
   \chi(T, \Q_i) \approx - \int_0^{\infty} d\epsilon \, \frac{g(\epsilon)}{3} \frac{\tanh(\epsilon/2T)}{2 \epsilon}\,,
\end{equation}
where we have used the nesting relation $\xi_f(\bfk + \Q_i) = - \xi_f(\bfk)$, that strictly holds only for states along a rectangular path ${\cal C}$, that goes in the case of $\Q_3$ along $\{-\Q_1\rightarrow \Q_2 \rightarrow\Q_1 \rightarrow -\Q_2 \rightarrow -\Q_1\}$. 
We have inserted a factor of $1/3$, because the nesting property is only fulfilled for one third of the states on the Fermi surface. In addition, we have ignored the fact that nesting is \emph{not} fulfilled for all $\bfk$-states while replacing the sum over the $\bfk$-states that satisfy the nesting condition by an integral over all states.  Nevertheless, solving the integral gives $ \chi(T, \Q_i) \approx - \frac{M_0}{6} \left[\ln \frac{8 C_1 t_f}{T}\right]^2$, which holds at the three nesting vectors $i=1,2,3$. If we compare this with numerical results using for example Monte-Carlo integration, we observe that the slope $M_0/6$ is in perfect agreement, but the energy scale in the logarithmic function needs to be slightly adjusted. More precisely, we can easily fit our numerical results to the function: 
\begin{equation}
  \label{eq:18}
  \chi(T,\Q_i) \approx - \frac{M_0}{6} \left[\ln \frac{8 \,a C_1 t_f}{T}\right]^2,
\end{equation}
and obtain the fit parameter $a = 2.17$.

If we plug this result into the Bogoliubov dispersion relation (see Eq.~(\ref{eq:10})), we find that $E_b(\Q_i)$ decreases as the temperature is lowered, and finally becomes zero $\left[E_b(T_{\text{SS}}^{\text{inst.}},\Q_i) = 0\right]$ when
\begin{equation}
  \label{eq:19}
  \chi(T_{\text{SS}}^{\text{inst.}}, \Q_i) = - \frac{U_{bb}}{U_{bf}^2} \left[ 1 + \frac{\xi_b(\Q_i)}{ 2 n_b U_{bb}} \right]\,,
\end{equation}
which defines the supersolid transition temperature based on the instability criterion
\begin{equation}
    \label{eq:20}
      T_{\text{SS}}^{\text{inst.}} = 8 a C_1 t_f \exp\left[ - \sqrt{\frac{3}{\lambda_{BF}} \left(2 + \tau_B \right)} \right]\,.
\end{equation}
Here, $\tau_B = 8 t_b/n_b U_{bb}$ is the ratio of kinetic to interaction energy of the pure boson system. 

The transition temperature $T_{\text{SS}}^{\text{inst.}}$ becomes larger for smaller $\tau_B$, favoring slower bosons or larger intrinsic repulsion. However, one has to consider that at strong coupling $\tau_B \ll 1$, there is a competing superfluid to Mott-insulator transition at commensurate densities $n_b$, which occurs on the two-dimensional triangular lattice at the critical ratio $(U_{bb}/t_b)_c = 26.5$~\cite{PhysRevB.59.12184}. For a typical bosonic filling of $n_b = 5/4$, the Mott insulator appears at $\tau_B \approx 1/4$. In addition, weak-coupling requires that $\lambda_{BF} < \tau_B$ which sets an upper limit to the value of $U_{bf}$.

In short, the instability analysis provides an intuitive physical view on why we expect a condensation of rotons, {\it i.e.}, $\av{b_{\Q_{i}}} \neq 0$, in the presence of a nested Fermi surface. 

\subsection{Incommensurate density wave}
\label{sec:incomm-vs.-comm}
It turns out that, at \emph{finite temperatures}, one must be more careful with the analysis of the Lindhard function. We show in detail in Appendix~\ref{sec:diverg-at-nest}, that in an intermediate temperature regime, where the thermal smearing of the Fermi edge is larger than the level spacing ($\sim t_f/N_L$), the minima of the Lindhard function occur at wavevectors slightly different from $\Q_i$. As a result, the roton gap closes (slightly) away from the nesting vectors, at $\bfK_i = \alpha \Q_i$ with $\alpha < 1$ (see Fig.~\ref{fig:3}), which leads to the formation of a density wave that is incommensurate with the lattice structure at intermediate temperatures. 

On the other hand, for a finite lattice composed of $N_L$ unit cells, the level spacing starts to play a role at temperatures $T_L \sim t_f/N_L$, and the minimum of the Lindhard function shifts to $\Q_i$ at temperatures below $T_L$, {\it i.e.}, $\alpha \rightarrow 1$ for $T \ll T_L$, where thermal effects can be ignored. In this sense, the incommensurate regime does not survive for $T<T_L$. 
A more quantitative analysis is given in Appendix~\ref{sec:level-spac-temp}, where we find that,
\begin{equation}
  \label{eq:21}
   T_L \simeq 2 \pi^2 t_f /N_L.
\end{equation}
Choosing an experimentally relevant lattice size of $N_L = 60$~\cite{spielman:120402}, one finds 
$\log_{10} \left(T_{L}/t_f\right) = -2.3$. 
In this paper, we will not address in detail the properties of the (intermediate) incommensurate density wave regime and we will mainly focus on the commensurate supersolid phase that emerges below $T_L$ (see Fig. \ref{fig:3}(b)).

In the next Sec.~\ref{sec:mean-field-theories}, we study the phase diagram using more general bosonic and fermionic mean-field theories.

\begin{figure}[tb]
  \centering
  \includegraphics[width=\linewidth]{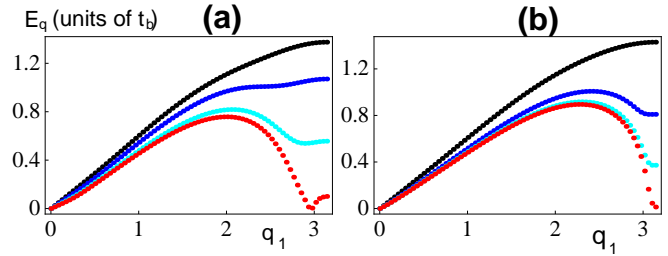}
  \caption{(Color online) Bogoliubov dispersion relation along $\q = (q_1, q_1/\sqrt{3})$ for various temperatures $T$ and fixed parameters $n_b, t_b, U_{bb}, U_{bf}$. (a): Roton gap closes slightly away from the nesting vector $\Q_2 = (\pi, \pi/\sqrt{3})$ for  $n_b = 1.25$, $t_b = 0.1 t_f$, $U_{bb} = 2 t_f$, $U_{bf} = 1.9568 \,t_f$ at the temperature $\log_{10} T/t_f = -1.7$ (lowest curve). Other curves correspond to the temperatures $\log_{10} T/t_f = -1,-1.3,-1.6$ (top to bottom of upper three curves).
(b): Roton gap closes at the nesting vector $\Q_2$ for $n_b = 1.25$, $t_b = 0.1 t_f$, $U_{bb} = 1.15 t_f$, $U_{bf} = 1.117 \,t_f$ at the temperature $\log_{10} T/t_f = -2.9$ (lowest curve). Other curves correspond to $\log_{10} T/t_f = -1,-2.4,-2.8$ (top to bottom of upper three curves).}
  \label{fig:3}
\end{figure}

\section{Phase diagram and properties of Kagome supersolid}
\label{sec:mean-field-theories}
In the following Sec.~\ref{sec:phase-diagram-from}, we derive a low temperature phase diagram of the system using a bosonic mean-field theory that goes beyond the instability analysis. 

We identify a novel, highly symmetric supersolid phase with a Kagome-type density modulation in real-space, and calculate supersolid transition temperatures. 
In Sec.~\ref{sec:supers-dens-wave}, we further study this supersolid and determine the amplitude of the density wave modulation using a different mean-field theory that treats the fermions exactly.

\subsection{Phase diagram from bosonic mean-field theory}
\label{sec:phase-diagram-from}
Here, we employ a Landau-Ginzburg-Wilson mean-field theory to build the low temperature phase diagram of the system. We find a novel Kagome-supersolid phase that competes with phase separation, and derive transition temperatures to both phases generalizing the instability results of Eqs.~(\ref{eq:14}) and~(\ref{eq:20}). 
\subsubsection{Construction of the free energy}
Based on the results of the instability analysis, we expect phase transitions to occur at low temperatures. We therefore construct a general Landau-Ginzburg-Wilson free energy functional for the bosons on the isotropic triangular lattice. The details of this procedure can be found in Appendix~\ref{sec:constr-land-ginzb}. 

We assume that the bosons may have a number of Fourier components condensing at momenta $\q = {\bf 0}, \Q_1, \Q_2, \Q_3$. The complex bosonic order parameters $\{\psi_{0,1,2,3}\}$ are defined as 
\begin{equation}
  \label{eq:22}
  \begin{split}
    \av{b_{\bf 0}} &= \sqrt{N_L} \psi_0 \\
    \av{b_{\Q_{a}}} &= \sqrt{N_L} \psi_\alpha\; (\alpha= 1,2,3) \,,
  \end{split}
\end{equation}
where we have assumed spatially homogeneous order parameters and $\av{\cdot}$ denotes taking the operator's expectation value. We choose $\psi_0$ to be real (and positive). For a fixed number of bosons $N_b$ and at $T = 0$, the fields obey
$\sum_{\alpha=0}^3 |\psi_\alpha|^2 = n_b$, where $n_b = N_b/N_L$ is the bosonic filling factor (density). 

Starting from a bosonic field $\Psi(\bfx)$ which we assume to be slowly varying in real-space continuum, {\it i.e.}, on length scales larger than the lattice spacing $|\bfa_{1,2}| = 1$, (assuming low temperatures compared to $T_{\text{KT}}$) we approximate,
\begin{equation}
  \label{eq:23}
  \Psi(\bfx) \approx \psi_0 + \sum_{\alpha=1}^3 \psi_\alpha e^{i \Q_\alpha \cdot \bfx}\,,
\end{equation}
with a modulation in real-space that is solely due to the wavevectors $\Q_\alpha$. 
We derive the free energy functional ${\cal F}_b$ for the homogeneous system in detail in Appendix~\ref{sec:constr-land-ginzb}. It contains the quadratic and quartic terms, in the $\{\psi_i\}$, that are invariant under all the symmetries of the isotropic triangular lattice. These are one 3-fold rotation, two reflection symmetries and the two translations by ${\bf a}_i$~\cite{PhysRevB.72.134502}. Up to quartic order in the order parameters, it reads
\begin{equation}
  \label{eq:24}
   \frac{{\cal F}_b}{N_L} = m_0 |\psi_0|^2 + m_1 |\boldsymbol{\psi}_Q|^2 + \sum_{i=0}^2 u_i \Theta_i + \sum_{i=1}^4 g_i F_i \,,
\end{equation}
where $\boldsymbol{\psi}_Q = (\psi_1, \psi_2, \psi_3)$ and the different terms read
\begin{equation}
  \label{eq:25}
   \begin{split}
     \Theta_0 &= |\psi_0|^4 \\
    \Theta_1 &= |\boldsymbol{\psi}_Q|^4 = |\psi_1|^4 + |\psi_2|^4 + |\psi_3|^4 \\ 
    &+ 2 \left( |\psi_1|^2 |\psi_2|^2 + |\psi_1|^2 |\psi_3|^2 + |\psi_2|^2 |\psi_3|^2 \right) \\
    \Theta_2 &= |\psi_0|^2 |\boldsymbol{\psi}_Q|^2 = |\psi_0|^2 \left(  |\psi_1|^2 +  |\psi_2|^2 +  |\psi_3|^2 \right)\\
    F_1 &= |\psi_1|^4 + |\psi_2|^4 + |\psi_3|^4 \\
    F_2 &= \left( \psi_1^2 + \psi_2^2 + \psi_3^2 \right)^* \left(\psi_1^2 + \psi_2^2 + \psi_3^2 \right) \\
    F_3 &= \psi_0 \left( \psi_1 \psi_2^* \psi_3^* + \text{cyclic permutations} \right) + c.c. \\
    F_4 &= \psi_0^2 \left( \psi_1^2 + \psi_2^2 + \psi_3^2 \right)^* + c.c. \,,
 \end{split}
\end{equation}
with $\boldsymbol{\psi} = (\psi_0, \psi_1, \psi_2, \psi_3)$. 
The free energy ${\cal F}_b$ contains nine coefficients: two masses $\{m_0, m_1\}$, and seven interaction parameters $\{u_0, u_1, u_2, g_1, g_2, g_3, g_4\}$. 

Instead of giving an exhaustive phase diagram of $\mathcal{F}_b$, we match these coefficients with a microscopically derived bosonic mean-field Hamiltonian $H_b^{\text{eff}}$, in the spirit of Weiss mean-field theory~\cite{goldenfeld_pt}. We obtain $H_b^{\text{eff}}$ from integrating out the fermionic degrees of freedom in the full Hamiltonian of Eq.~(\ref{eq:2}), as before, leading to the effective bosonic Hamiltonian of Eq.~(\ref{eq:9}). 
We then perform a (more general) Bogoliubov approximation of the bosonic operators $ b_\q =\sum_{\alpha=0}^{3}\av{b_{\Q_\alpha}} \delta(\q - \Q_\alpha) + \tilde{b}_{\q\neq \Q_\alpha}$, where $\tilde{b}_\q$ describe the fluctuations around the mean-field values, and we have defined $\Q_0 = {\bf 0}$. Neglecting fluctuations, one arrives at
\begin{multline}
  \label{eq:26}
  \frac{H_b^{\text{eff}}}{N_L} = \sum_{\alpha=0}^{3} \xi_b(\Q_\alpha) |\psi_{\alpha}|^2 \\ + \sideset{}{'}{\sum}^3_{\alpha,\beta,\gamma,\delta=0} \frac{U(\Q_\beta-\Q_\gamma)}{2} \psi^*_\alpha \psi^*_\beta \psi_\gamma \psi_\delta \,,
\end{multline}
where the second sum is restricted to $\Q_\alpha+\Q_\beta=\Q_\gamma+\Q_\delta$.
Since the Lindhard function possesses all the symmetries of the lattice, it is identical at the three nesting vectors $\q = \Q_{1,2,3}$, and we can define the interaction coefficients
\begin{equation}
  \label{eq:27}
  \begin{split}
    u &= U(T, {\bf 0}) =  U_{bb} + U_{bf}^2 \chi(T, {\bf 0}) \\
    g &= U(T,\Q_{1,2,3}) = U_{bb} + U_{bf}^2 \chi(T, \Q_{1,2,3}) \,.
  \end{split}
\end{equation}
This simplifies the Hamiltonian to the general form
 \begin{equation}
  \label{eq:28}
  \frac{H^{\text{eff}}_b}{N_L} = \sum_\alpha \xi_b (\Q_\alpha) |\psi_\alpha|^2 + \frac{1}{2}\left( u + g {\cal W} \right) |\boldsymbol{\psi}|^4\,,
\end{equation}
with
\begin{equation}
  \label{eq:29}
  {\cal W} = \frac{\Theta_1 + 2 \Theta_2 - 2 F_1 + F_2 + 4 F_3 + F_4 }{ |\boldsymbol{\psi}|^4 }\,
\end{equation}
being a function of the direction of the vector $\boldsymbol{\psi}/|\boldsymbol{\psi}|$ only.

\begin{figure}[tb]
  \centering
  \includegraphics[width=.6\linewidth]{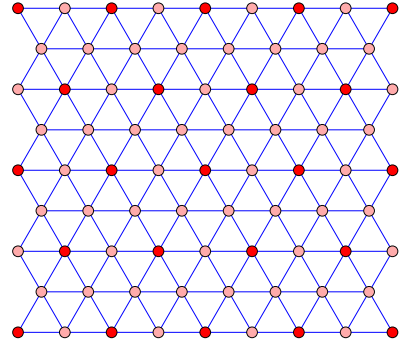}
  \caption{(Color online) Kagome supersolid phase on the triangular lattice. Darker lattice sites exhibit a higher bosonic density. The bosonic density is smaller at the lattice sites that belong to the 
Kagome-sublattice structure. If the boson-fermion interaction is repulsive, the fermionic density is larger where the bosonic density is smaller (lighter lattice sites).} 
  \label{fig:4}
\end{figure}
In this form, one easily derives the \emph{generalized criterion to avoid phase separation}. Stability requires 
that the quartic coefficient is always positive, because the free energy must be bounded from below. This demands 
 $ u + g {\cal W} \geq 0\,,$
and because $3 \geq {\cal W} \geq 0$, the stability conditions are given by
\begin{equation}
  \label{eq:31}
  \begin{split}
    u &\geq 0, \;\text{if} \; g \geq 0\\
    u &\geq 3 |g|, \; \text{if} \;g < 0\,.
  \end{split}
\end{equation}

Matching parameters between Eqs.~(\ref{eq:24}) and (\ref{eq:28}) yields for the mass coefficients the expressions
\begin{equation}
  \label{eq:32}
  \begin{split}
    m_0 &= \xi_b({\bf 0}) = - 6 t_b - \mu_b \\
    m_1 &= \xi_b(\Q_\alpha) = 2 t_b - \mu_b \,.
\end{split}
\end{equation}
We refer to Appendix \ref{sec:deta-analys-land} for the expressions of the interaction coefficients. Since the chemical potential in the superfluid phase is given by $\mu^{(\text{SF)}}_b=-6 t_b + n_b u$, which contains a mean-field energy shift due to interactions, the mass $m_0^{\text{SF}} = - n_b u$ is always negative. This indicates that the system wants to condense into the $\psi_0$-mode independently of the value of $t_b$, because it does not cost any kinetic energy to add a boson to the zero-momentum condensate. In contrast, the mass $m_1^{\text{SF}} = 8 t_b - n_b u$ depends on the ratio of kinetic to interaction energy. It costs a kinetic energy amount of $8 t_b$ to add a boson into one of the nesting modes $\psi_{1,2,3}$. We will later confirm that for smaller hopping amplitudes $t_b$, the supersolid already occurs for a smaller interaction strength $U_{bb}$.
\subsubsection{Mean-field phase diagram}
Minimization of the Hamiltonian $H_b^{\text{eff}}$ of Eq.~(\ref{eq:28}) yields the phase diagram of the system. Here, we present only the main results, that can be obtained by a straightforward numerical minimization. We refer to Appendix~\ref{sec:deta-analys-land} for a detailed analytical study. 

We numerically minimize Eq.~(\ref{eq:28}) using the most general ansatz $\psi_j = r_j e^{i \phi_j}$ with $r_j \geq 0$, ($j=0,1,2,3$), $\phi_0 = 0$, $0 \leq \phi_{1,2,3} < 2 \pi$. We find that while \emph{for $g > 0$, the superfluid has a lower free energy than any supersolid phase}, such a phase occurs \emph{for sufficiently negative $g<0$, where the system tends to order in a symmetric way with respect to the three nesting fields}. 
Furthermore, the phases of the supersolid order parameters are locked to the superfluid phase $\phi_{1,2,3}=0$. In this way, the 3-fold rotational symmetry of the system is preserved. In real space, the resulting density wave order has the \emph{pattern of the Kagome lattice}, as illustrated in Fig.~\ref{fig:4}. Note that this is also consistent with the form of the fermion-induced interaction between the bosons (see Appendix~\ref{sec:ferm-induc-inter}). This Kagome-state can be written as 
\begin{equation}
  \label{eq:33}
  \begin{split}
   \psi_0 = \sqrt{n_b}\cos\theta \\
   \psi_{1,2,3} = \sqrt{\frac{n_b}{3}} \sin \theta\,,
  \end{split}
\end{equation}
with $0<\theta<\pi/2$. 
\begin{figure}[tb]
  \centering
  \includegraphics[width=\linewidth]{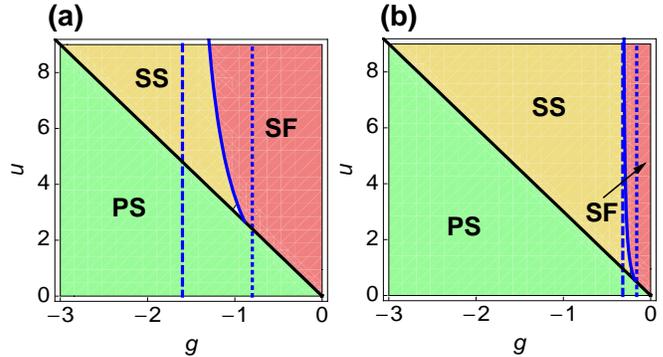}
  \caption{(Color online) Phase diagram from the microscopically matched bosonic mean-field theory in the parameter space $(u,g)$ for $t_b = 0.1\, t_f$ and filling factors $n_b = 1/4$ (left) as well as $n_b = 5/4$ (right). It contains regions of SS, SF and PS. PS occurs below the instability line $u = 3 |g|$ (solid). The two vertical lines denote the critical interaction strengths $g_c^{(1)}=-4t_b/n_b$ (dashed) and $g_c^{(2)}=-2 t_b/n_b$ (dotted), respectively. In between, the phase boundary occurs at $g_c$ (solid), which is given in Eq.~(\ref{eq:34}). The tricritical point occurs at $g_{\text{max}} = - 20 t_b/9 n_b$. }
  \label{fig:5}
\end{figure}
To construct the phase diagram, we first observe that the transition between the superfluid and supersolid phase is between two ordered phases. Starting from the superfluid and replacing $\psi_0 \rightarrow \sqrt{n_b}$ in Eq.~(\ref{eq:28}), we find that the free energy contains a third order term in $\psi_{1,2,3}$. This third-order term leads to a local minimum in the free energy for $g < g_c^{(2)} = - 2 t_b/n_b$. This local minimum, however, becomes the global minimum only at the more negative value
\begin{equation}
  \label{eq:34}
   g_c = - \frac{12 t_b u}{16 t_b + 3 n_b u}\,.
\end{equation}
This equation embodies a \emph{more general condition to enter the supersolid phase} than the instability criterion of Eq.~(\ref{eq:19}). Together with the stability conditions in Eq.~(\ref{eq:31}), it divides the parameter space $(u,g)$ into three phases: superfluid (SF), supersolid (SS) and phase separation (PS). In Fig.~\ref{fig:5}, we show the resulting phase diagram for a fixed value of $t_b=0.1 t_f$ and different bosonic fillings $n_b = 1/4, 5/4$. The maximal value of $g$ to enter the supersolid phase is $g_{\text{max}} = - 20 t_b/9 n_b$, which is the tricritical point of the phase diagram.

 The superfluid-supersolid \emph{phase transition is of first-order}~\cite{zhao:105303, goldenfeld_pt}, because the curvature of the free energy around the superfluid minimum, \emph{i.e.} the effective mass of the fields $\psi_{1,2,3}$, remains positive even below $g_c$ until $g < g_c^{(1)}$. 
The first-order transition region between the two vertical lines $g_c^{(2)} - g_c^{(1)} = 2 t_b/n_b$ shrinks with larger $n_b$, which is fully consistent with numerical results of Ref. \cite{titvinidze:144506}, where a first-order region could only be identified for sufficiently small $n_b$. 
Again, for details of the calculation, we refer the reader to Appendix~\ref{sec:deta-analys-land}. 

\subsubsection{Transition temperatures and low-temperature phase diagram}
Here, we want to derive transition temperatures $T_{\text{PS}}$ and $T_{\text{SS}}$ using the mean-field criteria of Eqs.~(\ref{eq:31}) and Eq.~(\ref{eq:34}), that generalize the instability expressions of Eqs.~(\ref{eq:14},~\ref{eq:19}). This allows us to draw a low-temperature phase diagram of the system.

Starting from the superfluid phase, we can calculate the interaction parameters $u$ and $g$ at temperature $T$ and for fixed $U_{bb},U_{bf}$ using Eqs.~(\ref{eq:27}). Upon lowering the temperature, both $u$ and $g$ decrease and finally reach the superfluid phase boundary (see Fig.~\ref{fig:5}). Phase separation is avoided as long as $u > 3 |g_{\text{max}}|$ or 
\begin{equation}
  \label{eq:35}
  \chi(T_{\text{PS}},{\bf 0}) = - \frac{U_{bb}}{U_{bf}^2} \left[ 1 - \frac{20 t_b}{3n_b U_{bb}} \right]\,.
\end{equation}
Note that the instability analysis only demands the less strict condition $u > 0$.
The supersolid phase appears for $g < g_c$ or
\begin{multline}
  \label{eq:36}
  \chi(T_{\text{SS}},\Q_{1,2,3})= \\ = \frac{- 3 n_b U_{bb} (U_{bb} + c_0 U_{bf}^2) - 4 t_b (7 U_{bb} + 3 c_0 U_{bf}^2  ) }{U_{bf}^2 [16 t_b + 3 n_b (U_{bb} + c_0 U_{bf}^2 )]}\,,
\end{multline}
where we have defined $c_0=\chi(T,\bf{0})$.
Note that the instability analysis required a more negative value of $g < g_c^{(1)} = -4 t_b/n_b$.

We extract the transition temperatures $\{T_{\text{PS}}, T_{\text{SS}}\}$ from Eq.~(\ref{eq:35}) and (\ref{eq:36}), using the expressions of $\chi(T,\Q_{0,1,2,3})$ given in Eqs.~(\ref{eq:13}) and (\ref{eq:18}). One should note that they are only valid in the superfluid phase. Nevertheless, they allow us to divide the $(U_{bb}, U_{bf})$ parameter space into a supersolid and a phase separated region, in the following way.

Assume that $T_{\text{SS}} > T_{\text{PS}}$ and the system is in the superfluid phase at a temperature $T > T_{\text{SS}}$. As the temperature is lowered, it will become supersolid at $T=T_{\text{SS}}$. As we will show in the next Sec.~\ref{sec:supers-dens-wave}, the instability towards phase separation is removed inside the supersolid phase, since the fermionic spectrum acquires a gap at the Fermi energy. The system remains supersolid for all temperatures $T<T_{\text{SS}}$. 
In contrast, if $T_{\text{PS}} > T_{\text{SS}}$ and we again lower the temperature starting in the superfluid phase at some $T> T_{\text{PS}}$, the system will phase separate at $T=T_{\text{PS}}$. Then, the fermionic density deviates locally from $n_f = 3/4$ and the Fermi surface is not nested anymore. Hence, the instability towards supersolid formation is removed.
\begin{figure}[tb]
  \centering
  \includegraphics[width=.7\linewidth]{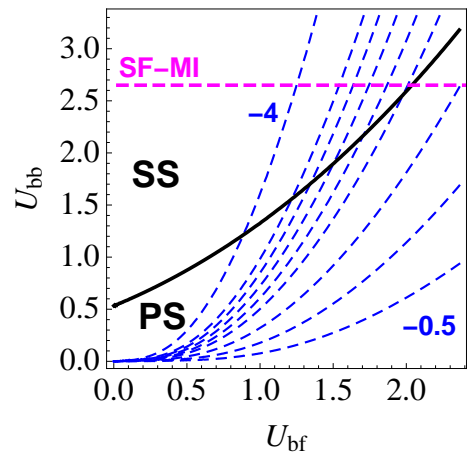}
  \caption{(Color online) Low-temperature phase diagram for parameters $(U_{bf},U_{bb})$ (in units of $t_f$), and fixed values of $t_f = 1$, $t_b = 0.1\, t_f$, $n_b = 5/4$, $n_f = 3/4$, obtained from the bosonic mean-field theory on the triangular lattice. The SS appears where $T_{\text{SS}} > T_{\text{PS}}$, PS appears where $T_{\text{PS}} > T_{\text{SS}}$. Temperatures are from Eqs.~(\ref{eq:35}) and (\ref{eq:36}). The phase boundary (solid line) is defined by $T_{\text{SS}} = T_{\text{PS}}$. 
The horizontal dashed line denotes the critical ratio for a competing SF-MI phase transition, that occurs at $(U_{bb}/t_b)_c = 26.5$~\cite{PhysRevB.59.12184}. 
Other dashed lines are supersolid transition temperature contour lines corresponding (from right to left) to $\log_{10}(T_{\text{SS}}/t_f)$$ = -0.5, -1,$$-1.5,-2,$$-2.25,-2.5,$$-2.75,-3,$$-4$.}
\label{fig:6}
\end{figure}
In Fig.~\ref{fig:6}, we show the resulting phase diagram and contours of constant supersolid transition temperatures $T_{\text{SS}}$ for fixed $n_b=5/4$, $t_b= 0.1 t_f$.
It is important to note that the transition temperatures are invariant under the transformation 
\begin{equation}
  \label{eq:37}
  t_b \rightarrow \alpha t_b,\; U_{bb} \rightarrow \alpha U_{bb},\; U_{bf} \rightarrow \sqrt{\alpha} U_{bf}\,.
\end{equation}
Therefore, the same $T_{\text{SS}}$ can be found for other bosonic hopping amplitudes $t_b$ under proper rescaling of $U_{bf}$ and $U_{bb}$.  

\subsection{Supersolid density wave modulation from fermionic mean-field theory}
\label{sec:supers-dens-wave}
So far, we have located the supersolid parameter regime (see Fig.~\ref{fig:6}) and identified a highly symmetric Kagome-type supersolid phase with equal amplitude modulation in all three nesting wavevector modes (see Eq.~\eqref{eq:33}). In this section, we use a different (fermionic) mean-field approach, that treats the fermions exactly, to calculate the effect of this condensation, {\it i.e.}, $\psi_{1,2,3} \neq 0$, onto the fermions. In general, one finds that the fermionic spectrum acquires a gap at the Fermi surface for nonzero $\psi_{1,2,3}$. This is energetically favorable for the fermionic subsystem, however, notice that this energy gain has to be sufficient to, at least, compensate the kinetic and interaction energy cost of adding a boson with a large wavevector in the system. 

We then determine the amplitude of the density wave modulation in the supersolid phase. We find that the amplitude is generally rather small and increases for decreasing bosonic hopping amplitudes $t_b$. For $t_b = 0.01 t_f$, we find it to be (maximally) about $\Delta n_b/n_b = 0.1$ at zero temperature and involve only about $0.1 \%$ of all the bosons. We like to mention that similar density wave modulations were predicted in Refs.~\cite{titvinidze:100401, titvinidze:144506} using Dynamical Mean-Field theory (DMFT). As we show in Sec.~\ref{sec:detect-supers-phase}, the experimental detection of such a small density wave is not feasible with current technology, however, we note already at this point, that one finds significantly larger density wave amplitudes for the square lattice geometry (see Sec.~\ref{sec:anisotropic-hopping}). 
 
We begin by replacing the bosonic operators $b_{\Q_{0,1,2,3}}$ with $\psi_{0,1,2,3}$ as in Eq.~(\ref{eq:22}). Neglecting any fluctuations, the full Hamiltonian of Eq.~(\ref{eq:2}) then becomes $H_f^{\text{eff}} = H_f^{(1)} + {H}_f^{(2)} + {H}_f^{(3)}$, where \begin{equation}
  \label{eq:38}
  \begin{split}
    \frac{H_f^{(1)}}{N_L} &= \sum_{\alpha=0}^{3} \xi_b(\Q_\alpha) |\psi_{\alpha}|^2 +  \frac{U_{bb}}{2} \sideset{}{'}{\sum}_{\alpha,\beta,\gamma,\delta=0}^3 \psi^*_\alpha \psi^*_\beta \psi_\gamma \psi_\delta \\
    {H}_f^{(2)} &= U_{bf} \sum_{\q\in BZ} \sum_{\alpha\neq \beta=0}^3 \psi_\beta^* \psi_\alpha f_{\q + \Q_\alpha }^\dag f_{\q + \Q_\beta} \\
    {H}_f^{(3)} &= \sum_{\q \in BZ} \xi_f(\q) f_\q^\dag f_\q\,,
\end{split}
\end{equation}
where the primed sum is restricted to $\Q_\alpha+\Q_\beta=\Q_\gamma+\Q_\delta$, and we have incorporated a mean-field energy shift of the fermions due to the presence of the bosons given by $U_{bf} n_b \sum_\q f_\q^\dag f_\q$ into the chemical potential $\mu_f$. 

The first term $H_f^{(1)}$ describes the kinetic and interaction energy of the condensed bosons. The kinetic energy cost to take a boson from the superfluid condensate $\psi_0$ and add it to one of the nesting modes is given by $\xi_b(\Q_{1,2,3}) = 8 t_b$. The (quartic) interaction energy obviously increases with the number of nonzero fields $\psi_i$ and larger $U_{bb}$. To find out whether these energy costs of having nonzero $\psi_{1,2,3}$ can be compensated by the last two (fermionic) terms, we diagonalize them. 

For this, we need to symmetrize the expressions with respect to adding a nesting vector $\Q_\alpha$, such that, in matrix form, we write: $H_f^{(2)} + H_f^{(3)} = \sum'_{\bfk} \sum_{\alpha,\beta} f^\dag_{\bfk + \Q_\alpha} h_{\alpha \beta} f_{\bfk + \Q_\beta}$, where $h_{\alpha\beta}$ is given by
\begin{multline}
  \label{eq:39}
   h_{\alpha\beta} = \delta_{\alpha\beta} \xi_f(\bfk + \Q_\alpha) \\ + (1 - \delta_{\alpha \beta}) U_{bf} \left( \psi_\alpha^* \psi_\beta + \psi_\gamma^* \psi_\delta + c.c. \right)\,,
\end{multline}
where the sum over wavevectors is restricted to $1/4$ of the first Brillouin zone, and $(\gamma, \delta) \neq (\alpha, \beta) \in \{0,..,3\}$ as well as $\gamma \neq \delta$. By diagonalizing $h_{\alpha\beta}$, we obtain the fermionic eigenenergies $\Xi(\bfk, \{\psi_\alpha\})$. 

The free energy at finite temperatures $T > 0$ reads 
\begin{equation}
  \label{eq:40}
\begin{split}
  \frac{{\cal F}}{N_L} &=  \sum_\alpha \xi_b(\Q_\alpha) |\psi_{\alpha}|^2 + \frac{U_{bb}}{2} \sideset{}{'}\sum_{\alpha,\beta,\gamma,\delta} \psi^*_\alpha \psi^*_\beta \psi_\gamma \psi_\delta \\ &- \frac{T}{N_L} \sum_{\bfk} \ln \left(1 + e^{-\Xi(\bfk, \{\psi_\alpha\})/T} \right)\,.
\end{split}
\end{equation}
As the temperature goes to zero ($T \rightarrow 0$), the last term becomes a sum over the lowest $N_L n_f$ eigenstates: $\frac{1}{N_L} \sum_{\text{lowest} j=1}^{N_L n_f} \Xi(\bfk_j, \{\psi_\alpha\})$. Since the supersolid phase appears at temperatures that are much smaller than the fermionic hopping amplitude, $T_{\text{SS}} \ll t_f$, results of finite and zero temperature calculations agree.

We calculate ${\cal F}$ at zero temperature on a finite lattice with $N_L = 500 \times 500$ sites, as a function of the $\{\psi_\alpha\}$, using the Kagome ansatz that we have used before: $\psi_0 = \sqrt{n_b} \cos \theta$, $\psi_{1,2,3} = \sqrt{\frac{n_b}{3} }\,  e^{i \phi}\, \sin \theta$, with angles $0 \leq \theta \leq \pi/2$ and $ 0 \leq \phi < 2 \pi$. We then determine its minimum as a function of $\{\theta, \phi\}$, for a certain choice of parameters $\{U_{bb}, U_{bf}, t_b, n_b\}$. 
\begin{figure}[tb]
  \centering
  \includegraphics[width=0.75\linewidth]{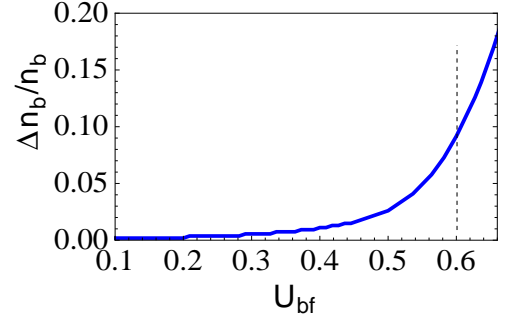}
  \caption{(Color online) Density wave modulation $\Delta n_b/n_b$ as a function of $U_{bf}$ (in units of $t_f$) for fixed $U_{bb} = 0.25 \,t_f$, $n_b = 5/4$, $t_b = 0.01 \,t_f$. The vertical dashed line denotes the phase boundary to the phase separated regime for this value of $U_{bb}$. 
}
  \label{fig:7}
\end{figure}

We find that the minimum always occurs for $\phi = 0$. The location of the minimum as a function of $\theta$ determines whether the system is superfluid or supersolid. If the minimum occurs at $\theta = 0$, condensation into the nesting modes is energetically not favorable and the system remains superfluid. In contrast, if it occurs at $\theta > 0$ the system can lower its energy by establishing $\av{b_{\Q_\alpha}} \neq 0$ and becomes supersolid. The value of $\theta$ determines the amplitude of the density wave in the supersolid phase. 

At zero temperature, the minimum occurs at $\theta > 0$ for any parameter pair $(U_{bb}, U_{bf})$ in the weak-coupling region (see Fig.~\ref{fig:6}). However, the \emph{amplitude of the corresponding density wave modulation} in the supersolid phase, which we define as 
\begin{equation}
  \label{eq:41}
  \Delta n_b = \text{max}_i(b^\dag_i b_i) - \text{min}_i(b^\dag_i b_i)\,,
\end{equation}
with the bosonic operators approximately given by $b_i \simeq \psi_0 + \sum_{\alpha=1}^3 \psi_\alpha e^{i \Q_\alpha \cdot \bfx_i}$, varies significantly with the values of $U_{bb}$ and $U_{bf}$. 

More precisely, $\Delta n_b$ is proportional to $T_{\text{SS}}/U_{bf}$, 
a relation that can be understood from Bardeen-Cooper-Schrieffer theory, where the gap at zero temperature is proportional to the superconducting transition temperature. Here, the gap in the fermionic spectrum is determined by the product $ U_{bf} \psi_\alpha^* \psi_\beta$ with $\alpha \neq \beta \in \{0,..,3\}$, as can be seen from Eq.~(\ref{eq:39}).
For the Kagome supersolid we find,
\begin{equation}
  \label{eq:42}
  \frac{\Delta n_b}{n_b} = \frac83 \sin \theta \left( \sqrt{3} \cos \theta + \sin \theta \right)\,,
\end{equation}
which takes values between $0 \leq \Delta n_b/n_b \leq 4$ and is shown in Fig.~\ref{fig:7}. For parameters $n_b = 5/4$ and $t_b = 0.01 t_f$, it takes values up to $\Delta n_b/n_b = 0.1$. We also extract the ratio $\Delta n_b U_{bf}/T_{\text{SS}} \simeq 9 \pm 1$ using $T_{\text{SS}}$ from Eq.~(\ref{eq:36}). 
For comparison, we mention that one can derive an exact analytical relation on the square lattice~\cite{PhysRevLett.91.130404}: $\Delta n_b U_{bf}/T_{\text{SS}} = 4 /C_1 \approx 3.53$ (see Eq.~(\ref{eq:61})), where $T_{\text{SS}}$ is defined in Eq.~(\ref{eq:49}). 

\section{Striped supersolid phases on anisotropic lattices}
\label{sec:anisotropic-hopping}
\begin{figure}[tb]
  \centering
  \includegraphics[width=\linewidth]{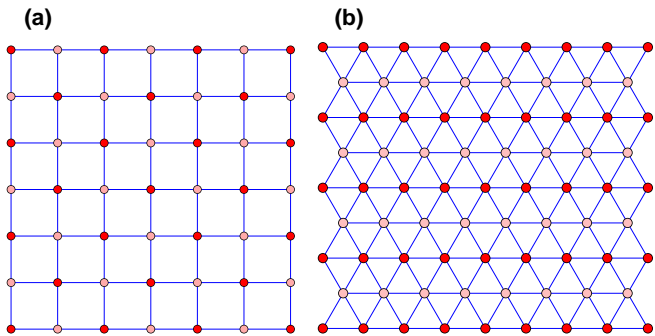}
  \caption{(Color online) Striped supersolid phases on the square (a) and anisotropic triangular (b) lattice. The density wave is characterized by a single nesting vector: (a) $\Q_{\text{sq}} = (\pi,\pi)$, and (b) $\Q_3= (0, 2 \pi/\sqrt{3})$. Like in Fig.~\ref{fig:4}, darker lattice sites exhibit a larger bosonic density.}
  \label{fig:8}
\end{figure}
In this section, we discuss the case of anisotropic hopping amplitudes on the two-dimensional triangular and square lattices, where only one nesting vector remains as a result of the reduced symmetry. The isotropic square lattice also exhibits a single nesting vector. As a result, we find supersolid phases that show a {\it striped pattern in real-space}; see Fig.~\ref{fig:8}. 

On the triangular lattice, there are two ways to introduce anisotropic hopping (see also Fig.~\ref{fig:1}(a)). Either, the hopping amplitude $t_{f1} > t_{f2}$, leading in the limit $t_{f1} \gg t_{f2}$ to an array of weakly coupled chains, or the opposite case of $t_{f1} < t_{f2}$, where the system resembles the isotropic square lattice in the limit $t_{f1} \ll t_{f2}$. We will therefore discuss the square lattice geometry in some detail in Sec.~\ref{sec:square-lattice-1}.

Whereas for the triangular lattice, nesting is always accompanied by the occurrence of van Hove singularities (at the same fermionic filling $n_f$), and thus $\chi(T,\q)$ diverges at zero and the nesting vector, one can separate both phenomena to occur at different $n_f$ on the anisotropic square lattice (compare Figs.~\ref{fig:9} and \ref{fig:13}). As a result, on the anisotropic square lattice, the Lindhard function is regular at $\q = \bf0$ even in the presence of a nested Fermi surface, and the tendency towards phase separation at low-temperatures is suppressed. On the other hand, we will find that the divergence at the nesting vector becomes weaker, which leads to slightly smaller supersolid transition temperatures $T_{\text{SS}}$. We note that this situation is similar to the case of three dimensional lattices. There, critical points on the Fermi surface are integrable and the density of states is regular everywhere. In the case of the 2d anisotropic square lattice, however, the presence of divergencies close by in energy, lead to a significantly increased value of the density of states at the Fermi surface energy (see Fig.~\ref{fig:20}). 

In general, compared to the triangular lattice, the square lattice shows supersolidity at higher temperatures $T_{\text{SS}}$, and with larger density wave modulations $\Delta n_b$, because the nesting relation is fulfilled for more $\bfk$-values. Therefore, the \emph{square lattice geometry is experimentally advantageous over the triangular one}, and will be discussed first. 

\subsection{Square Lattice}
\label{sec:square-lattice-1}
An instability analysis and a fermionic mean-field theory of the supersolid phase on the \emph{isotropic} square lattice has been discussed in Ref.~\cite{PhysRevLett.91.130404}. Here, we will include a bosonic Landau-Ginzburg mean-field treatment and also generalize to the case of spatially anisotropic hopping amplitudes. To obtain a clear presenetation of the main results, the calculational details can be found in Appendix~\ref{sec:deta-calc-anis}.

\subsubsection{Instability analysis}
The fermionic (bosonic) dispersion relation for the square lattice is given by
\begin{equation}
  \label{eq:43}
  \xi_{f(b)}(\q) = - \mu_{f(b)} - 2 [ t_{f(b)1} \cos q_1 + t_{f(b)2} \cos q_2 ]\,.
\end{equation} 
It is useful to introduce the anisotropy parameters $r_{f(b)} = t_{f(b)1}/t_{f(b)2}$ and work with a dimensionless chemical potential $\tilde{\mu}_f = \mu_f/2 t_{f2}$. 

The Fermi surface for different values of $r_f$ and $\tilde{\mu}_f$ is shown in Fig.~\ref{fig:9}. There are two special values of the chemical potential: first, $\tilde{\mu}_f=0$ (dashed  contour), where the system is particle-hole symmetric and shows perfect nesting at wavevector $\Q_{sq} =(\pi,\pi)$ for any value of $r_f$. Second, for $\tilde{\mu}_f = \pm (1 - r_f)$, the density of states has a van Hove singularity due to the critial points $\q=(\pm \pi,0)$ on the Fermi surface (dotted contour). Only in the isotropic system, these values are equal and given by $\tilde{\mu}_f=0$.

The density of states $g(\tilde{\mu}_f, r_f)$ for the isotropic and anisotropic lattice exhibits a logarithmic singularity at $\tilde{\mu}_f = \pm(1-r_f)$. It can be calculated analytically (see Appendix~\ref{sec:square-lattice-2}), which also contains a plot of $g(\tilde{\mu}_f, r_f)$). The anisotropic density of states is regular at $\tilde{\mu}_f=0$, where it is equal to
\begin{equation}
  \label{eq:44}
  g(\tilde{\mu}_f=0, r_f<1) = 2 N_0 K(r_f)\,,
\end{equation}
where $N_0 = 1/2\pi^2t_{f2}$ and $K(x)$ denotes the complete elliptic integral of the first kind. 

Next, we calculate the Lindhard function $\chi(T, \q)$ (Eq.~(\ref{eq:8})). 
On the isotropic square lattice, it was shown in Ref.~\cite{PhysRevLett.91.130404} that the Lindhard function diverges at $\q={\bf 0}, \Q_{sq}$ as
\begin{align}
  \label{eq:45}
    \chi(T, {\bf 0}) &= - N_0 \ln \frac{16 C_1 t_{f}}{T} \\
    \chi(T, \Q_{sq}) &= - \frac{N_0}{2} \left[ \ln \frac{16 C_1 t_f}{T} \right]^2\,,
\end{align}
due to the combination of van Hove singularities and nesting. We note that as opposed to the triangular lattice, here, the minimum of the Lindhard function close to the nesting vector \emph{always} occurs at $\q = \Q_{sq}$ (see Sec.~\ref{sec:incomm-vs.-comm}).

On the anisotropic lattice, the density of states is regular at $\tilde{\mu}_f = 0$, and therefore the Lindhard function is regular at $\q = {\bf 0}$: $\chi(T,{\bf 0}) = - g(0,r_f)$. The tendency towards phase separation at low temperatures is removed for $r_f < 1$.  On the other hand, the divergence of $\chi(T, \Q_{sq})$ in the absence of a van Hove singularity at $\tilde{\mu}_f = 0$, is only linearly logarithmic:
\begin{align}
  \label{eq:46}
    \chi(T,{\bf 0}) &= \int_0^\infty d\epsilon\, g(\epsilon) \frac{\partial f(\epsilon) }{\partial \epsilon} = - 2 N_0 K(r_f) \\
\begin{split}
\label{eq:47}
    \chi(T, \Q_{sq}) &=\int_0^\infty d\epsilon g(\epsilon) \frac{\tanh(\epsilon/2T)}{- 2 \epsilon}  \\ &\simeq  - 2 N_0 K(r_f) \ln \left[\frac{4 e^C (1+r_f) h(r_f) t_{f2}}{\pi T} \right]\,.
\end{split}
\end{align}
The fitting function $h(r_f) = a_0 + a_1 r_f$ with $a_0 = 1.96$, $a_1 = -1.67$ occurs from comparing numerical results to an analytical approximation that replaces $g(\epsilon) \approx g(0)$ in Eq.~\eqref{eq:47}, which neglects the (nearby) divergence in the density of states at $\tilde{\mu}_f= \pm (1-r_f)$. The slope of $\chi(T, \Q_{sq})$ is given by the regular density of states at the Fermi surface.

These divergences result in two instabilities, as we have shown in Sec.~\ref{sec:instability-analysis}. For the isotropic lattice, phase separation occurs at the temperature,
\begin{equation}
  \label{eq:48}
    T^{\text{inst.}}_{\text{PS}} (r_f=1) =  16 C_1 t_f \exp \left[ - \frac{U_{bb}}{N_0 U_{bf}^2} \right],
\end{equation}
whereas the supersolid transition temperature reads
\begin{equation}
  \label{eq:49}
   T_{\text{SS}} (r_f=1) = 16 C_1 t_f \exp\left[ - \sqrt{\frac{2 U_{bb}}{N_0 U_{bf}^2} \left(1 + \frac{4 t_b}{n_b U_{bb}}\right) }\right],
\end{equation}
with $C_1 = 2 \exp{C}/\pi \simeq 1.13$. 

For the anisotropic lattice, phase separation only occurs above a critical interspecie interaction strength, which is given by
\begin{equation}
  \label{eq:50}
  U_{bf}^{\text{PS},\text{inst.}} = \sqrt{\frac{U_{bb}}{2 N_0 K(r_f)}}\,.
\end{equation}
The supersolid transition on the anisotropic lattice occurs at a temperature 
\begin{multline}
  \label{eq:51}
  T_{\text{SS}}(r_f<1) =\\ =A t_{f2} \exp\left[ \frac{-U_{bb}}{g(0,r_f) U_{bf}^2} \left(1 + \frac{2(t_{b1} + t_{b2})}{n_b U_{bb}}\right) \right],
\end{multline}
where $A = 2 C_1 (1+r_f) h(r_f)$. 
\begin{figure}[tb]
  \centering
  \includegraphics[width=.9\linewidth]{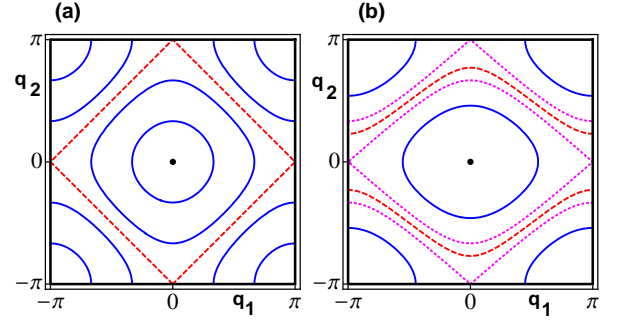}
  \caption{(Color online) Fermi surfaces of the isotropic (a) and anisotropic (b) square lattice with anisotropy parameter $r_f = 0.75$, for different fillings. Nesting occurs at $\tilde{\mu}_f = 0$ (dashed lines) and van Hove singularities at $\tilde{\mu}_f = \pm(1 - r_f)$ (dotted lines). }
  \label{fig:9}
\end{figure}

\subsubsection{Phase diagram from bosonic mean-field analysis}
\label{sec:bosonic-mean-field-2}
The effective bosonic mean-field Hamiltonian on the square lattice (see Sec.~\ref{sec:phase-diagram-from}) reads
\begin{equation}
  \label{eq:52}
  \frac{H_{\text{sq},b}^{\text{eff}}}{N_L} = \sum_{\alpha = 0}^1 \xi_b (\Q_\alpha) |\psi_\alpha|^2 + \frac12 (u + g {\cal W}_{sq}) |\boldsymbol{\psi}|^4\,,
\end{equation}
with $u=U(T, {\bf 0})$, $g=U(T,\Q_{sq})$ and 
\begin{equation}
  \label{eq:53}
  {\cal W}_{\text{sq}} = \frac{(\psi_0 \psi_1^* + c.c.)^2}{(|\psi_0|^2 + |\psi_1|^2)^2} = \sin^2(2\theta) \cos^2(\phi),
\end{equation}
where we have parametrized the bosonic fields as $\boldsymbol{\psi} = (\psi_0, \psi_1) = (\sqrt{n_b} \cos \theta, \sqrt{n_b} e^{i \phi} \sin \theta)$ and $\Q_0 = {\bf 0}$, $\Q_1 = \Q_{\text{sq}}$. For arbitrary $\phi$, we can distinguish between three different phases: the cases $\theta = 0, \pi/2$ correspond to a superfluid and a pure density wave phase, respectively. For $0 < \theta < \pi/2$, the system is supersolid. Furthermore, the \emph{stability requirement}, which is that the fourth order term should be bounded from below, is given by $u \geq 0$ for $g \geq 0$, and $u > - g$ for $g< 0$, because $0 \leq {\cal W}_{sq} \leq 1$.
\begin{figure}[tb]
  \centering
  \includegraphics[width=.6\linewidth]{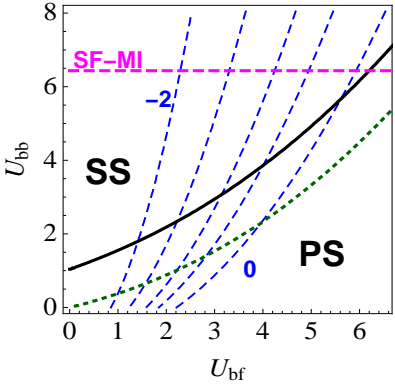}
  \caption{(Color online) Phase diagram for the isotropic square lattice as a function of $(U_{bb}, U_{bf})$ (in units of $t_f$) with fixed $t_b = 0.39 \,t_f$, $n_b = 3/2$, $n_f = 1/2$, so that $\mu_f=0$. Solid and dotted lines indicate the phase boundary obtained from the bosonic mean-field calculation and instability analysis, respectively. They separate a SS from PS regions. Horizontal dashed line denotes the critical SF-MI ratio $(U_{bb}/t_b)_c = 16.5$~\cite{PhysRevA.70.053615}. Other dashed lines indicate constant supersolid transition temperatures: $\log_{10} (T_{\text{SS}}/t_f) = 0, -0.25, -0.5, -1,-2$ (right to left). }
  \label{fig:10}
\end{figure}

To obtain the phase diagram, we minimize the Hamiltonian. This is done analytically in Appendix~\ref{sec:bosonic-mean-field-1}, where we find that \emph{the pure density wave phase always has larger energy than the superfluid}, which is the ground state for $g>0$. Comparing the energy of the superfluid with the supersolid, we find that \emph{the supersolid occurs for sufficiently attractive interactions} at 
 \begin{equation}
  \label{eq:54}
  g \leq g_{\text{sq},c}  = - \frac{2 (t_{b1} + t_{b2})}{n_b}\,.
\end{equation}
In contrast to the triangular lattice, the superfluid-supersolid phase transition on the square lattice is of \emph{second order}. The resulting supersolid transition temperature thus coincides with the transition temperature obtained by the instability analysis $T_{\text{SS}}$ of Eqs.~(\ref{eq:49}) and~(\ref{eq:51}) for both isotropic and anisotropic lattices.

However, the condition to avoid phase separation is modified from $u > 0$ to $u > - g_{sq,c}$, leading to a larger transition temperature towards phase separation on the isotropic lattice
\begin{equation}
  \label{eq:55}
  T_{\text{PS}} = 16 C_1 t_f \exp\left[ - \frac{U_{bb}}{N_0 U_{bf}^2} \left( 1 - \frac{4 t_b}{n_b U_{bb}} \right) \right]\,,
\end{equation}
which obeys $T_{\text{PS}} > T_{\text{PS}}^{\text{inst}}$.  

On the anisotropic lattice, the critical interspecie interaction strength to avoid phase separation is modified as well, and reads
\begin{equation}
  \label{eq:56}
  U_{bf}^{\text{PS}} = \sqrt{\frac{U_{bb} - 2 (t_{b1} + t_{b2})/n_b}{2 N_0 K(r_f)}}\,.
\end{equation}
The resulting phase diagrams for the isotropic and anisotropic square lattice as a function of $\{U_{bb}, U_{bf}\}$, together with contour lines of constant $T_{\text{SS}}$, are shown in Figs.~\ref{fig:10} and~\ref{fig:11}.
\begin{figure}[tb]
  \centering
  \includegraphics[width=.6\linewidth]{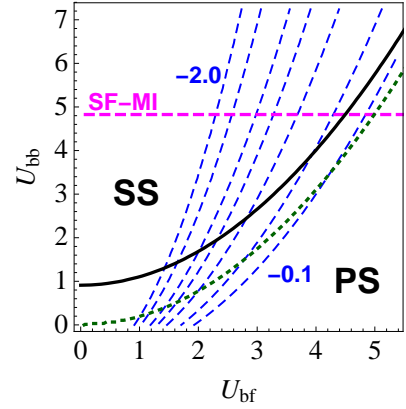}
  \caption{(Color online) Phase diagram for the anisotropic square lattice with $r_f = r_b = 0.75$, as a function of $(U_{bb}, U_{bf})$ (in units of $t_f$) and with fixed $t_{b2} = 0.39 \,t_{f2}$, $n_b = 3/2$, $\mu_f=0$. Solid and dotted lines indicate the phase boundary between SS and PS obtained from the bosonic mean-field calculation and instability analysis, respectively. The horizontal dashed line denotes the critical ratio $(U_{bb}/t_{b1})_c = 16.5$ to the competing MI phase. Other dashed lines indicate constant supersolid transition temperatures $\log_{10}(T_{\text{SS}}/t_f) = -0.1,-0.25,-0.5, -0.75, -1, -1.5,-2.0$ (dashed, from right to left).} 
  \label{fig:11}
\end{figure}
\subsubsection{Density wave modulation from fermionic mean-field theory}
\label{sec:fermionic-mean-field-1}
Following the analysis of Sec.~\ref{sec:supers-dens-wave}, we calculate the amplitude of the supersolid density wave by a fermionic mean-field approach. We replace the bosonic operators $b_{{\bf 0},\Q_1}$ by the complex fields $\psi_{0}=r_0$ and $\psi_1 = r_1 \exp(i \phi)$ ($r_{0,1} \in \mathbb{R}$) and diagonalize the resulting fermionic spectrum exactly. This way, we derive the finite temperature free energy
 \begin{equation}
  \label{eq:57}
\begin{split}
  \frac{{\cal F}}{N_L} &= \frac{(t_{b1} + t_{b2}) \Delta^2}{n_b U_{bf}^2} +  \frac{U_{bb} \Delta^2 \cos^2 \phi}{2 U_{bf}^2} \\ & - T \sum_{\bfk, s=\pm} \ln \left( 1 + e^{- \Xi(\bfk,\Delta)_s/T} \right)\,,
\end{split}
\end{equation}
which contains the gap $\Delta = 2 U_{bf} r_0 r_1$ and the fermionic eigenenergies
\begin{equation}
  \label{eq:58}
  \Xi(\bfk, \Delta)_\pm = \pm \sqrt{\xi_f(\bfk)^2 + \Delta^2 \cos^2 \phi}\,.
\end{equation}
Details of the derivation are given in Appendix~\ref{sec:fermionic-mean-field}.
From $\partial_{\phi} {\cal F} = 0$, one finds $\phi = m \pi$ with integer $m$, and the gap equation arises from $\partial_\Delta {\cal F} = 0$ as
\begin{equation}
  \label{eq:59}
  \frac{2 + \tau_B}{\lambda_{\text{BF}}} = \frac{2}{N_0} \sideset{}{'}{\sum}_\bfk \frac{\tanh [\Xi(\bfk,\Delta)_{+}/2 T]}{\Xi(\bfk,\Delta)_+}\,,
\end{equation}
where $\tau_B = 4 (t_{b1} + t_{b2})/n_b U_{bb}$, $\lambda_{BF} = N_0 U_{bf}^2/U_{bb}$ and the summation is over $1/2$ of the first Brillouin zone. 
Solving for the supersolid transition temperature by setting $\Delta(T_{\text{SS}}) = 0$, reproduces the results from the instability analysis and the bosonic mean-field theory (Eq.~(\ref{eq:49},~\ref{eq:51}). 

The density modulation in the supersolid phase $\Delta n_b$ is proportional to the gap 
\begin{equation}
  \label{eq:60}
  \Delta n_b = \frac{2 \Delta}{U_{bf}}\,,
\end{equation}
because the expectation value of the number operator in the supersolid state reads $\av{b_i^\dag b_i} = |\psi_0|^2 + |\psi_1|^2 + \frac{\Delta}{U_{bf}} \cos (\Q_1 \cdot \bfx_i)$. We can solve for the zero temperature gap $\Delta(T=0)$, using the modified density of states in the gapped system $G(\Xi,\Delta) = g(\sqrt{\Xi^2 - \Delta^2}) |\Xi|/ \sqrt{\Xi^2 - \Delta^2}$. For the isotropic lattice, one finds $\Delta(0) = 2 T_{\text{SS}}(r_f=1)/C_1$ (see Eq.~(\ref{eq:49})), and in the case of the anisotropic lattice, we find $\Delta(0,r_f) = \sqrt{2} T_{\text{SS}}(r_f<1)/C_1 h(r_f)$ (see Eq.~(\ref{eq:51})). The density modulations at $T=0$ thus read for the isotropic lattice 
\begin{equation}
  \label{eq:61}
  \Delta n_b = \frac{4 \;T_{\text{SS}}(r_f=1)}{C_1 U_{bf}}\,,
\end{equation}
and for the anisotropic lattice 
\begin{equation}
  \label{eq:62}
   \Delta n_b = \frac{ 2 \sqrt{2} \;T_{\text{SS}}(r_f<1)}{C_1 h(r_f) U_{bf}}\,,
\end{equation}
where, as defined earlier, $h(r_f) = 1.96 - 1.67 r_f$. In Fig.~\ref{fig:12}, 
we show typical density wave amplitudes $\Delta n_b$ as a function of $U_{bf}$ for fixed $U_{bb}$ and $t_b$.
\begin{figure}[tb]
 \centering
 \includegraphics[width=\linewidth]{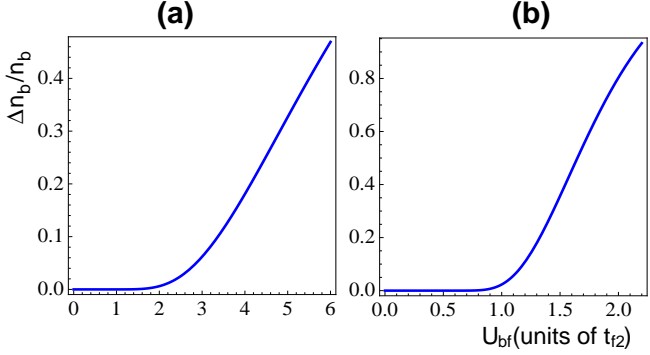}
 \caption{(Color online) Normalized density wave amplitude $\Delta n_b/n_b = (\text{max}\av{b^\dag_i b_i} - \text{min}\av{b^\dag_i b_i})/n_b$ in the supersolid as a function of $U_{bf}$ (in units of $t_{f2}$) for (a): isotropic square lattice with $t_b = 0.39 \, t_f$, $U_{bb} = 5.7 \,t_f $, $n_b=3/2$, and (b): anisotropic square lattice with $r_f = 0.75$, $t_{b1} = 0.06\, t_{f2} $, $t_{b2} = 0.09\, t_{f2}$, $U_{bb} = 0.9 t_{f2}$, $n_b = 3/2$. This choice of parameters corresponds to a particular experimental realization that will be discussed in Sec.~\ref{sec:exper-phase-diagr} (see Tab.~\ref{tab:3},~\ref{tab:4}).}
 \label{fig:12}
 \end{figure}

\subsection{Triangular lattice}
\label{sec:triangular-lattice-1}
For spatially anisotropic hopping on the triangular lattice, the Fermi surface is still nested at the chemical potential $\mu_f = 2 t_{f1}$. As shown in Fig.~\ref{fig:13}, however, only one nesting vector $\Q_3 = (0, 2 \pi /\sqrt{3})$ remains. This is true for both $t_{f1} < t_{f2}$ and $t_{f1} > t_{f2}$. As a result, the Lindhard function only diverges for wavevectors close to $\q = \Q_3$ (and no longer at $\q = \Q_{1,2}$). We show below that the tendency to condense into the modes $\psi_{1,2}$ is then removed. With only $\psi_{0,3}$ being nonzero, the supersolid has a striped pattern in real space; see Fig.~\ref{fig:8}. 
In addition, nesting is \emph{always} accompanied with the occurrence of van Hove singularities at the same chemical potential $\mu_f$. Therefore, also $\chi(T, {\bf 0})$ diverges logarithmically at low temperatures and supersolid formation competes with phase separation.

\begin{figure}[tb]
  \centering
  \includegraphics[width=\linewidth]{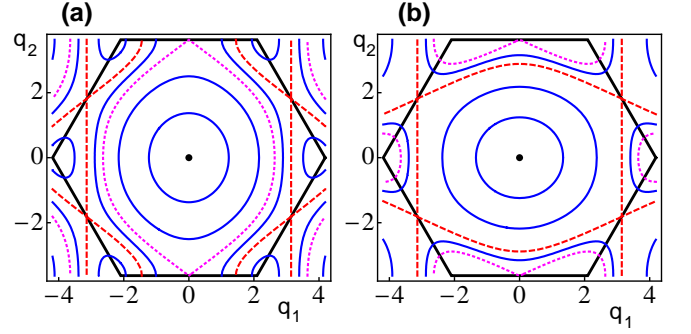}
  \caption{(Color online) Fermi surfaces of the anisotropic triangular lattice for the hopping amplitudes $t_{f1} = 1$, $t_{f2} = 0.75$ (a) and $t_{f1} = 1$, $t_{f2} = 1.25$ (b) and different chemical potentials $\mu_f$. Nesting occurs for $\mu_f = 2 t_{f1}$ (dashed) with a single nesting wavevector $\Q_3 = (0, 2 \pi/\sqrt{3})$. It is accompanied by van Hove singularities in the density of states due to critical points at $\q=\pm \Q_{1,2}$. Critical points also occur for $\mu_f/t_{f2} = 4 - 2 r_f$ (dotted). 
The hexagon (thick line) denotes the first Brillouin zone.}
  \label{fig:13}
\end{figure}
To show that only $\psi_{0,3}$ tend to condense, we derive the effective bosonic mean-field Hamiltonian
\begin{equation}
  \label{eq:63}
\begin{split}
  \frac{H_b^{\text{eff}}}{N_L} &= m_0 |\psi_0|^2 + m_1 \left(|\psi_1|^2 + |\psi_2|^2 \right) + m_3 |\psi_3|^2 \\
  &+ \frac12 (u + v {\cal V} + g {\cal W} ) |\boldsymbol{\psi}|^4\,,
\end{split}
\end{equation}
with the masses $m_i = \xi_b(\Q_i)$ and interaction coefficients $u = U(T,\bf 0)$, $v = U(T,\Q_{1,2})$, $g=U(T,\Q_3)$, and 
\begin{equation}
  \label{eq:64}
  \begin{split}
    {\cal V} |\boldsymbol{\psi}|^4 &= 2 \left[ |\psi_0|^2 (|\psi_1|^2 + |\psi_{2}|^2) + |\psi_3|^2 (|\psi_1|^2 + |\psi_2|^2 ) \right] \\ &+ [ \psi_0 \psi_1 \psi_2^* \psi_3^* + \psi_0^* \psi_1 \psi_2^* \psi_3 + 2 \psi_0 \psi_1^* \psi_2^* \psi_3 + c.c. ] \\ &+ ( \psi_0^2 + \psi_3^2) (\psi_1^2 + \psi_2^2)^* \\
    {\cal W} |\boldsymbol{\psi}|^4 &= (\psi_0 \psi_3^* + \psi_1 \psi_2^* + c.c.)^2\,.
  \end{split}
\end{equation}

We analyze the Hamiltonian in detail in Appendix~\ref{sec:triangular-lattice}, where we find that, again, kinetic energy considerations will select the superfluid state for positive $g$. For negative $g$, the system can possibly lower its energy, compared to the superfluid, by allowing for nonzero $\psi_3$ while still having $\psi_1=\psi_2=0$.
In this subspace of possible field values, the mean-field Hamiltonian is of the same form as on the square lattice and we refer to the discussion in Sec.~\ref{sec:bosonic-mean-field-2} and Appendix~\ref{sec:bosonic-mean-field-1}.

Numerical evaluation of the Lindhard function shows that, for $r_f < 1$ $(r_f > 1)$, the supersolid transition temperatures are above (below) the ones for the isotropic triangular lattice. They are always smaller than $T_{\text{SS}}$ on the isotropic square lattice. 

\subsection{Summary}
\label{sec:summary-1}
On the square lattice, we have found a checkerboard-type supersolid that occurs on the isotropic lattice at temperatures as large as $T_{\text{SS}} \simeq t_f = T_f/4$ (for $n_b=3/2$). For the anisotropic square lattice, we have found only slightly smaller temperatures $T_{\text{SS}} \simeq 0.6 \,t_f \simeq T_F/5$. There, phase separation only occurs above a critical interaction strength $U_{bf}^{\text{PS}}$.

The fermionic energy gap $\Delta$, which is proportional to $T_{\text{SS}}$, is related to the supersolid density wave via $\Delta n_b = 2 \Delta/U_{bf}$. Hence, \emph{larger amplitudes occur for smaller bosonic hopping}, since, according to the transformation of Eq.~\eqref{eq:37}, the same $T_{\text{SS}}$ then occurs at a smaller $U_{bf}$. The amplitudes are similar for isotropic and anisotropic lattices. For $t_b = 0.39 t_f$, we find $\Delta n_b/n_b \approx 0.5$, and for the smaller value of $t_b \approx 0.1 t_f$, we find $\Delta n_b/n_b \approx 0.9$.  

On the anisotropic triangular lattice, we have identified a striped supersolid that competes with phase separation. For $r_f < 1 (r_f > 1)$, it occurs at larger (smaller) temperatures than the Kagome-supersolid on the isotropic lattice. 

In order to compare different lattice geometries, we estimate an \emph{upper bound for the supersolid transition temperatures} as the temperature $T^*_{\text{SS}}$, where 
\begin{equation}
  \label{eq:65}
  \chi(T^*_{\text{SS}},{\bf 0}) = \chi(T^*_{\text{SS}}, \Q_i)\,.,
\end{equation}
where $i = 1 (3)$ for the square (triangular) lattice. 

By numerically computing $\chi(T,\bf{q})$ on the triangular lattice for different anisotropy parameters $r_f= t_{f1}/t_{f2}$, we observe that $T^*_{\text{SS}}$ increases with decreasing values of $r_f < 1$. For the isotropic triangular lattice, one finds $T^*_{\text{SS}}(r_f =1) \approx 0.2 \, t_{f2}$. For the isotropic square lattice, which is the limiting case of $r_f \rightarrow 0$, one calculates $T^*_{\text{SS}}(r_f \rightarrow 0) \approx 1.2 \,t_f$. In between, one has 
\begin{equation}
\label{eq:66}
 0.2 < \frac{T^*_{\text{SS}}(1>r_f > 0)}{t_{f2}} < 1.2  \,.
\end{equation}
In contrast, for $r_f > 1$, the temperature $T^*_{\text{SS}}$ decreases, \emph{i.e.}  $T^*_{\text{SS}} (r_f>1) < T_{\text{SS}}^*(r_f=1)$.

The supersolid transition temperature increases with $n_b$, and with $U_{bf}$ being close to the phase boundary of supersolid to phase separation and $U_{bb}/t_b$ close to the superfluid to Mott-transition ratio $U_{bb}/t_b|_{\text{SF-MI}}$. However, the weak-coupling requirement  $\lambda_{BF}/\tau_B = n_b M_0 U_{bf}^2/8 t_b < 1$ obviously restricts the maximum value of $n_b$.
We find that the transition temperatures consistent with $\lambda_{BF}/\tau_B < 1$ are close to the upper bound $T_{\text{SS}}^*$ for the square lattice, but generally much smaller for the triangular lattice. The difference there, is typically an order of magnitude. 

In conclusion, the isotropic square lattice exhibits the highest supersolid transition temperatures $T_{\text{SS}}$ of all the geometries considered here, with $T_{\text{SS}} \simeq t_f=T_F/4$. 

\section{Experimental predictions for $^{87}\text{Rb}^{40}\text{K}$ and $^{23}\text{Na} ^{6}\text{Li}$}
\label{sec:exper-param-supers}
In this section, we will present results for specific experimental realizations of Bose-Fermi mixtures on the isotropic triangular as well as the isotropic and anisotropic square lattices. We predict the supersolid parameter regime for a mixture of $^{23}\text{Na} ^{6}\text{Li}$~\cite{PhysRevLett.88.160401,PhysRevLett.93.143001} and of $^{87}\text{Rb}^{40}\text{K}$~\cite{best:030408, gunter:180402, 0953-4075-41-20-203001}. We also show that the unambiguous experimental detection of the supersolid phase via time-of-flight imaging (TOF) is feasible for the square lattice geometry. Additional coherence peaks at the nesting vector occur with a size of up to $\approx 0.02$ (measured relative to the main superfluid peak) for both mixtures. Since the weight of these coherent peaks is reduced to about $\lesssim 5 \times 10^{-5}$ in the triangular lattice case, it proves more challenging to detect supersolidity in this case. We therefore propose a combination of usual time-of-flight absorption imaging with noise correlation techniques~\cite{ScarolaSS} to reveal the supersolid phase.     
\subsection{Relating Hamiltonian to experimental quantities}
\label{sec:hamilt-param}
\begin{table}[tb]
   \centering
   \begin{tabular}[t]{cccc}
     & $\quad \lambda_0 [\text{nm}] \quad$ & $\,\gamma_0/2\pi [\text{MHz}]\,$ & $\,I_{\text{sat}} [\text{mW}/\text{cm}^2]\,$ \\
     \hline
     $\quad^{6}\text{Li}\quad$ & $670.96$ & $5.92$ & $2.56$ \\
     \hline
     $^{23}\text{Na}$ & $589.16$ & $10.01$ & $6.40$ \\
     \hline
     $^{40}\text{K}$ & $766.70$ & $6.09 $ & $1.77$ \\
     \hline
     $^{87}\text{Rb}$ & $780.24$ & $5.98$ & $1.64$ \\
     \hline
\end{tabular}
   \caption{Atomic properties of $^{23}\text{Na}$, $^{6}\text{Li}$, $^{87}\text{Rb}$, $^{40}\text{K}$. Transition wavelength $\lambda_0$, natural linewidth $\gamma_0$ and saturation intensity $I_{\text{sat}}$ determines ratio $V_b/V_f$.~\cite{Metcalf-LaserCooling}}
   \label{tab:1}
 \end{table}
The Hamiltonian of Eq.~\eqref{eq:1} contains the parameters 
\begin{equation}
  \label{eq:67}
  \{t_f, t_b, U_{bb}, U_{bf}, \mu_f, \mu_b\}\,,
\end{equation}
which can be expressed by the microscopic and experimentally tunable parameters
\begin{equation}
  \label{eq:68}
  \{m_f, m_b,  n_f, n_b, a_{bb}, a_{bf}, \lambda, V_f^x, V_f^y, V_f^z\}\,.
\end{equation}
Here, $m_{f(b)}$ is the fermionic (bosonic) mass, $n_{f(b)}$ is the fermionic (bosonic) density, and $a_{bb(bf)}$ the s-wave scattering length of boson-boson (boson-fermion) interaction, that can be tuned by an externally applied static magnetic field via Feshbach resonances~\cite{arXiv:0812.1496v1}. $\lambda$ is the wavelength of the optical lattice laser, $V_{f/b}^{x,y,z}$ the optical lattice laser intensity in the $x,y,z$-direction, given in units of the fermionic/bosonic recoil energy $E^r_{f/b} = h^2/(2 m_{f/b} \lambda^2)$, respectively. Here, we focus on the rectangular geometry for notational convenience, but the generalization to the triangular geometry is straightforward. 

The two-dimensional setup is realized by strongly quenching inter-plane hopping via $V_{f(b)}^z \gg V_{f(b)}^{x,y}$, and for the isotropic lattice one sets $V_{f/b}^{x} = V_{f/b}^y=V_{f/b}$. The lattice constant is given by $\lambda/2$.

The ratio of lattice depths experienced by bosons and fermions, respectively, is determined by
\begin{equation}
  \label{eq:69}
  \frac{V_b}{V_f} = \zeta \; \frac{[\lambda^{-1} - \lambda_0(f)^{-1}]}{[\lambda^{-1} - \lambda_0(b)^{-1}]} \frac{\gamma_0(b)}{\gamma_0(f)} \frac{I_{\text{sat}}(f)}{I_{\text{sat}}(b)}  \frac{E^r_f}{E^r_b}\,,
\end{equation}
where $\lambda_0(f/b)$ is the wavelength of the relevant fermionic/bosonic transition, $\gamma_0(f/b)$ its natural linewidth and $I_{\text{sat}}(f/b)$ its saturation intensity. The prefactor $\zeta$ is of order unity and determined by a ratio of Clebsch-Gordon coefficients of the relevant transitions. The experimental values for the relevant transitions in $^{23}\text{Na} ^{6}\text{Li}$ and $^{87}\text{Rb}^{40}\text{K} $ are given in Table~\ref{tab:1}. 

The resulting ratios $V_b/V_f$, together with the recoil energies and collisional properties can be found in Table~\ref{tab:2}. Note that one can tune either one of $a_{bb}, a_{bf}$ over a wide range by applying an external magnetic field close to a Feshbach resonance~\cite{arXiv:0812.1496v1}. In the following, we will consider the case where $a_{bf}$ is tuned, leaving $a_{bb}$ fixed to the (off-resonance) value given in Table~\ref{tab:2}.
\begin{table}[tb]
   \centering
   \begin{tabular}[t]{ccccccc}
   Species & $\lambda [\text{nm}]$ & $V_b/V_f$ & \quad $E^r_f$ \quad & \quad $E^r_b$\quad & $a_{bb}$ [$a_0$]& $a_{bf} [a_0]$  \\
   \hline
   $^{23}\text{Na} ^{6}\text{Li}\quad$ & $1064$ &  $\approx 1.9$ & $\;1.41 \,\mu$K  & $\;368$ nK & $62$ & $13$ \\
   \hline
   $^{87}\text{Rb} ^{40}\text{K}\quad$ & $755$ & $\approx 1$ & $420$ nK & $193$ nK & $100$ & $-284$ \\
   & $1064$ & $\approx 2.5$ & $211$ nK & $97$ nK &  &  \\
\hline
  \end{tabular}
   \caption{Ratio of optical lattice potential for bosons and fermions $V_b[E^r_b]/V_f [E^r_f]$, measured in units of the respective recoil energies $E^r_{b,f}$, and collisional properties of $^{23}\text{Na} ^{6}\text{Li}$~\cite{arXiv:0812.1496v1, PhysRevLett.82.2422,PhysRevLett.93.143001,gacesa:010701},  $^{87}\text{Rb}^{40}\text{K}$~\cite{arXiv:0812.1496v1,ospelkaus:020401,best:030408}. $(a_{bb}$,$a_{bf})$ denote the scattering lengths away from any Feshbach resonance (in units of the Bohr radius $a_0$). }
   \label{tab:2}
 \end{table}

The fermionic (bosonic) hopping amplitude in a direction where the optical lattice depth is given by $V_{f(b)}$, can be expressed in closed form, if the Wannier state of the lowest Bloch band is approximated by a Gaussian ~\cite{bloch:885}
\begin{equation}
  \label{eq:70}
  t_{f(b)} = \frac{4}{\sqrt{\pi}} E^r_{f(b)} V_{f(b)}^{3/4} \exp[ - 2 \sqrt{V_{f(b)}}]\,.
\end{equation}
However, since this approximation fails for $V_{f(b)} \lesssim 10 E_{f(b)}^r$, we calculate the hopping amplitudes from the width of the lowest energy band $W(V_{f(b)})$ via
\begin{equation}
  \label{eq:71}
  t_{f(b)} = W(V_{f(b)})/4\,.
\end{equation}
 
The two interaction parameters are given by~\cite{bloch:885}
\begin{equation}
  \label{eq:72}
  \begin{split}
    \frac{U_{bb}}{E^r_f} &= 4 \sqrt{2 \pi} \; \frac{a_{bb}}{\lambda} (V_b^x V_b^y V_b^z)^{1/4}  \\
    \frac{U_{bf}}{E^r_f} &= 8 \sqrt{\pi} \; \frac{1 + m_f/m_b}{(1 + \sqrt{V_f/V_b})^{3/2}} \frac{a_{bf}}{\lambda} (V_f^x V_f^y V_f^z)^{1/4} \,.
  \end{split}
\end{equation}

The fermionic (bosonic) chemical potential $\mu_{f(b)}$ is determined by the number of fermions (bosons) in the system $N_L n_{f(b)}$. If the system is exposed to an overall harmonic confinement as is often the case experimentally, the chemical potential depends on the spatial location, which can be dealt within the local density approximation (LDA). 
For simplicity, we restrict ourselves to the homogeneous case, which can, in principle, be realized experimentally by compensating the overall confinement by a blue detuned optical lattice~\cite{best:030408} or by working with an external box-like potential.  

\begin{figure}[tb]
  \centering
  \includegraphics[width=.65\linewidth]{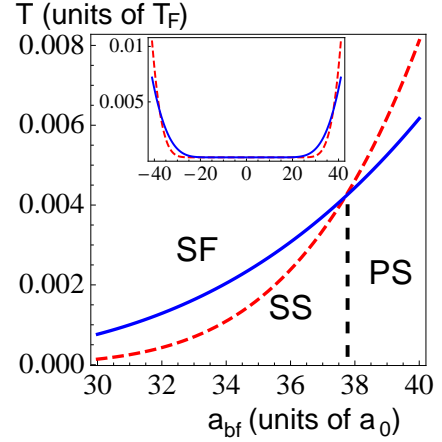}
  \caption{(Color online) Mixture of $^{40}\text{K}^{87}\text{Rb}$ on the triangular lattice: transition temperatures (in units of $T_F = 6 t_f$) towards supersolid $T_{\text{SS}}$ (solid) and phase separation $T_{\text{PS}}$ (dashed) as a function of scattering length $a_{bf}$ (in units of the Bohr radius $a_0$)Vertical dashed line indicates phase boundary between the supersolid and phase separation. For $T > T_{\text{PS}},T_{\text{SS}}$, system is superfluid (SF).
Bosonic filling is $n_b = 3.25$ and other parameters are $\lambda=755 \, \text{nm}$, $V_f = 7.5$, $V_f^z = 20$. Inset shows $a_{bf} \rightarrow - a_{bf}$ symmetry.}
  \label{fig:14}
\end{figure}
\begin{figure}[tb]
  \centering
  \includegraphics[width=\linewidth]{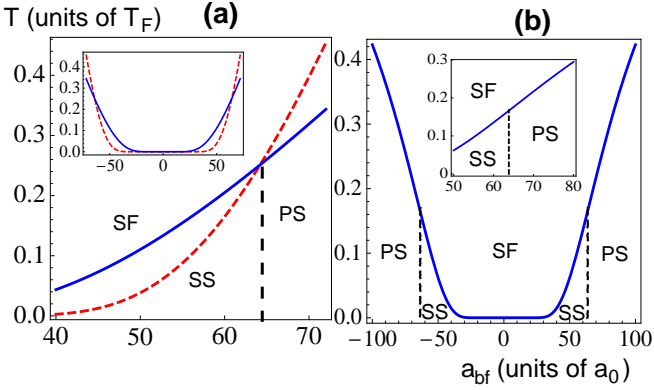}
  \caption{(Color online) Mixture of $^{40}\text{K}^{87}\text{Rb}$ on the isotropic (a) and anisotropic (b) square lattice. Shown are the transition temperatures  towards supersolid $T_{\text{SS}}$ (solid) and phase separation $T_{\text{PS}}$ (dashed, only in (a)) as a function of scattering length $a_{bf}$. Phase boundary between supersolid and phase separation is denoted by the vertical dashed line. (a):  Parameters are $n_b = 3/2$, $\lambda=755 \,\text{nm}$,$V_f = 7.5$, $V_f^z = 20$. Temperature is in units of $T_F = 4 t_f$. Inset shows $a_{bf} \rightarrow - a_{bf}$ symmetry.
(b): Anisotropy parameter $r_f=t_{f1}/t_{f2} = 0.75$ realized by lattice strengths $V_{f1} = 7.5$, $V_{f2} = 6.4$. Other parameters are the same as in (a). Temperature is in units of $T_F = 2( t_{f1} + t_{f2})$. Phase boundary occurs at $U_{bf}^{\text{PS}} = \sqrt{[U_{bb}- 2 (t_{b1} + t_{b2})/n_b]/[2 N_0 K(r_f)]}$ [see Eq.~\eqref{eq:56}]. }
  \label{fig:15}
\end{figure}
\subsection{Experimental phase diagrams}
\label{sec:exper-phase-diagr}
If we choose a particular mixture, the wavelength of the optical lattice $\lambda$ and an external magnetic field value that is far away from any Feshbach resonance of $a_{bb}$, there remain only $(V_f^{x,y}, n_b, a_{bf})$ as free parameters. From the phase diagrams in Figs.~\ref{fig:6}, \ref{fig:10}, \ref{fig:11}, we know that the maximal values of $T_{\text{SS}}$ are to be found where $U_{bf}$ is close to the supersolid-phase separation phase boundary and the ratio $U_{bb}/t_b \approx U_{bb}/t_b|_{\text{SF-MI}}$. Therefore, we determine $V_f^{x,y}$ (for a certain choice of $n_b$) by maximizing the ratio $U_{bb}/t_b$ under the constraints $U_{bb}/t_b < (U_{bb}/t_b)|_{\text{SF-MI}}$ and $\lambda_{BF}/\tau_B < 1$ (weak-coupling).

The finite temperature phase diagram for a $^{40}\text{K}^{87}\text{Rb}$-mixture as a function of the remaining free parameter $a_{bf}$ is shown in Fig.~\ref{fig:14} for the triangular lattice and in Fig.~\ref{fig:15} for the isotropic and anisotropic square lattices. 

We have normalized the temperature scale by the Fermi temperature of the lattice $T_F \sim t_f$, and find a maximal value of $T_{\text{SS}}/T_F \simeq 0.004$ for the triangular and $T_{\text{SS}}/T_F \simeq 0.2$ for the square lattices. The large difference in transition temperatures reflects the fact that the low-temperature divergence of the Lindhard function at the nesting vectors $\chi(T,\Q_i)$ is weaker for the triangular than for the square lattice. 
We note that an estimate of an upper bound of $T_{\text{SS}}$ is given in Sec.~\ref{sec:summary-1}.

In Table~\ref{tab:3} and \ref{tab:4}, we summarize the optimal choice of experimental parameters $V_f, a_{bf}$, which corresponds to the highest supersolid transition temperature $T_{\text{SS}}$ for different mixtures, optical lattice wavelenghts $\lambda$ and bosonic fillings $n_b$, for the cases of isotropic triangular and square (Tab.~\ref{tab:3}) and anisotropic square lattice (Tab.~\ref{tab:4}). 

\subsection{Detection of supersolid phase}
\label{sec:detect-supers-phase}
 \begin{table*}[tb]
   \centering
   \begin{tabular}[t]{c c c c c c c c c c c c c}
\multicolumn{13}{l}{}\\
\multicolumn{13}{l}{ \bf{Triangular Lattice}} \\
\hline
\rule[-8pt]{0pt}{22pt}
Species & $\;\lambda [\text{nm}]\;$ & $\;V_f\;$ &$\;a_{bf}\; $ & $\;n_b\;$ &   $\max \left(\frac{T_{\text{SS}}}{T_F}\right)$ &  $\Delta n_b/n_b$ & $\bar{\rho}_\alpha/\bar{\rho}_0$&   $\;t_b/t_f\;$ & $\; U_{bb}/t_f\;$ & $\;U_{bb}/t_b\;$ & $\; U_{bf}/t_f\;$ & $\; \lambda_{BF}/\tau_B\;$ \\
\hline\hline
   $^{87}\text{Rb} ^{40}\text{K}\quad$ & $755$ &$\;9.0\;$ &  $26$ &   $\;1.25\;$ &  $1 \times 10^{-3}$ &$0.01$ &$1\times 10^{-6}$ &     $0.38$ & $9.0$ & $23.6$ &   $3.6$ & $0.40$ \\
   & &$7.5$&  $38$ &   $3.25$ &  $4 \times 10^{-3}$ &$0.02$ & $4\times 10^{-6} $& $0.39$ &$5.7$ &$14.6$& $3.3$ & $0.88$ \\
\cline{2-13}
   & $ 1064$ & $4.2$&  $72$ &  $1.25$ & $1\times 10^{-3}$ &$0.03$ & $9\times 10^{-6}$&  $0.10$ & $2.4$ & $23.8$ &  $1.9$ & $0.42$ \\
   & & $\;3.5\;$ &  $95$ & $3.25$ &  $4\times 10^{-3}$ &$0.04$ & $1.6\times 10^{-5}$&  $0.12$ & $1.8$ & $14.5$ &$1.9$ & $ 0.87$ \\
   \hline
   $^{23}\text{Na} ^{6}\text{Li}\quad$ & $1064$ & $6.5$ & $32$ & $1.25$ &  $1 \times 10^{-3}$ &$0.03$&$1.0 \times 10^{-5}$&     $0.07$ & $1.6$ & $23.1$ &   $1.5$ & $0.40$ \\
   && $5.8$ &  $42$ & $3.25$ &  $5\times 10^{-3}$ &$0.07$ &$5.1\times 10^{-5}$& $0.08$ & $1.2$ & $16.3$ &  $ 1.6$ & $ 0.99$ \\
\hline
\multicolumn{13}{l}{} \\
\multicolumn{13}{l}{\bf{Isotropic Square Lattice}} \\
\hline
\rule[-8pt]{0pt}{22pt}
Species & $\;\lambda [\text{nm}]\;$ & $\;V_f\;$ &$\;a_{bf}\; $ &  $\;n_b\;$ &   $\max \left(\frac{T_{\text{SS}}}{T_F}\right)$ & $\Delta n_b/n_b$ & $\bar{\rho}_\alpha/\bar{\rho}_0$& $\;t_b/t_f\;$ & $\; U_{bb}/t_f\;$ & $\;U_{bb}/t_b\;$ & $\; U_{bf}/t_f\;$ & $\; \lambda_{BF}/\tau_B\;$ \\
\hline\hline
$^{87}\text{Rb} ^{40}\text{K}\quad$ & $755$ &$7.5$ & $64$ & $1.5$ &  $0.25$ &$0.42$ &$5 \times 10^{-3} $ &      $0.39$ & $5.7$ & $14.6$ &   $5.6$ & $0.77$ \\
& $1064$ & $3.5$ & $161$ & $1.5$ & $0.25$ & $0.74$ &$0.02$ &  $0.12$ & $1.8$ & $14.5$ &   $3.2$ & $ 0.77$ \\
\hline
$^{23}\text{Na} ^{6}\text{Li}\quad$ & $1064$ & $5.6$ & $73$ & $1.5$ &  $ 0.25$ & $0.93$ & $0.03$&      $0.08$ & $1.2$ & $14.7$ &   $2.5$ & $0.78$ \\
\hline
\end{tabular}
   \caption{Choice of optical lattice depth $V_f =V_f^{x,y}$ (in units of $E_f^r$) and scattering length $a_{bf}$ (in units of the Bohr radius $a_0$) which correspond to maximal values of the supersolid transition temperature $T_{\text{SS}}$, the amplitude of the supersolid density wave $\Delta n_b$ (see Eqs.~(\ref{eq:41},~\ref{eq:61})), and the height of the supersolid-superfluid time-of-flight peak ratio $\bar{\rho}_\alpha/\bar{\rho}_0$ (see Eq.~(\ref{eq:78})) on the isotropic triangular and square lattices. We consider different mixtures, optical lattice wavelengths $\lambda$ and bosonic fillings $n_b$. Parameters $\{t_b, U_{bb},U_{bf}\}$ (in units of $t_f$) follow from choice of $V_f,a_{bf}$ ($V_f^z = 20$) via Eqs.~(\ref{eq:71}, \ref{eq:72}). The critical superfluid to Mott-insulator ratio is given by $U_{bb}/t_b|_{\text{SF-MI}} = 26.5\; (16.5)$ for the triangular (square) lattice. Weak-coupling analysis requires $\lambda_{BF}/\tau_B = n_b M_0 U_{bf}^2/8 t_b < 1$ for the triangular lattice. For the square lattice $M_0$ is replaced by $N_0$. }
   \label{tab:3}
 \end{table*}

\begin{table*}[tb]
   \centering
   \begin{tabular}[t]{c c c c c c c c c c c c c c c c }
\multicolumn{11}{l}{\bf{Anisotropic Square Lattice}: $r_f = t_{f1}/t_{f2} = 0.75$, $n_b = 3/2$ } \\
\hline
\rule[-8pt]{0pt}{22pt}
Species &  $\;\lambda [\text{nm}]\;$ & $\;V_{f}^x\;$ & $\;V_{f}^y\;$ &$\;a_{bf}\; $ &  $\max\left(\frac{T_{\text{SS}}}{T_F}\right)$ &$\Delta n_b/n_b$& $\bar{\rho}_\alpha/\bar{\rho}_0$&       $\;t_{b2}/t_{f2}\;$ & $\;t_{b1}/t_{b2}\;$ & $\;U_{bb}/t_{f2}\;$ & $\; U_{bb}/t_{b1}\;$ &  $\; U_{bf}/t_{f2}\;$ & $\lambda_{BF}/\tau_B$ \\ 
\hline \hline
$^{87}\text{Rb} ^{40}\text{K}\quad$& $755$ & $7.5$ & $6.4$ & $64$ &  $0.17$ & $0.34$ & $3 \times 10^{-3} $&     $0.40$ &$0.73$ & $4.1$ & $14.0$ &  $4.1$ &$0.46$  \\
& $1064$ & $3.5$ & $2.4$ & $158$ &  $0.13$ & $0.49$ & $0.01$ & $0.18$ & $0.52$ & $1.2$ & $13.3$ & $2.1$ &$0.31$ \\
\hline
$^{23}\text{Na} ^{6}\text{Li}\quad$ & $1064$ & $5.8$ & $4.7$ &$71$ &  $0.17$ & $0.73$ & $0.02 $& $0.09$ &$0.62$ & $0.9$ & $15.5$ &  $1.9$ & $0.45$ \\
\hline
\end{tabular}
   \caption{Choice of optical lattice depths $V_{f}^x,V_{f}^y$ and scattering length $a_{bf}$ (in units of the Bohr radius $a_0$) which correspond to maximal values of the supersolid transition temperature $T_{\text{SS}}$, the amplitude of the supersolid density wave $\Delta n_b$ (see Eq.~(\ref{eq:62})), and the height of the supersolid-superfluid time-of-flight peak ratio $\bar{\rho}_\alpha/\bar{\rho}_0$ (see Eq.~(\ref{eq:78})) on the anisotropic square lattice. Other parameters follow from generalizations of Eqs.~(\ref{eq:70}, \ref{eq:72}) with $V_f^z = 20$, and weak-coupling analysis requires  $\lambda_{BF}/\tau_B = n_b N_0 U_{bf}^2/8 t_b < 1$.}
   \label{tab:4}
 \end{table*}
The supersolid phase can be detected unambiguously by time-of-flight absorption imaging (TOF), where atoms are suddenly released from the trap and expand approximately freely. After a certain expansion time $s$, the observed spatial density distribution of bosons, averaged over several images,  $\av{{\cal N}_b^s(\bfx)}$, is proportional to the momentum distribution in the lattice~\cite{bloch:885}: $\av{{\cal N}_b^s(\bfx)} = (m_b/\hbar s)^2 |w(\bfk)|^2 \rho(\bfk)$, 
where
\begin{equation}
  \label{eq:73}
  |w(\bfk)|^2 = \frac{1}{2 \pi} e^{- |\bfk|^2/[\pi^2 \sqrt{V_b}]}
\end{equation}
is the Fourier transform of the Wannier function of the lowest Bloch band, and the (dimensionless) momentum $\bfk$ is related to the spatial position in the cloud by $\bfk = \lambda m_b \bfx/2 \hbar s$. The Fourier transform of the one-particle density matrix 
\begin{equation}
  \label{eq:74}
  \rho(\bfk) = \frac{1}{N_L} \sum_{j,k} e^{i(\bfx_j - \bfx_k)\cdot \bfk} \av{b^\dag_j b_k}
\end{equation}
measures the first-order coherence properties of the system. Here, $\bfx_j$ is the (dimensionless) vector to lattice site $j$, \emph{i.e.} $\bfx_j = j_1 {\bf a}_1 + j_2 {\bf a}_2$ with $|{\bf a}_i|=1$. 

For the supersolid phase, one finds that the bosonic operators can be approximated by 
$b_j \simeq \psi_0 + \sum_{\alpha} \psi_{\alpha} e^{i \Q_\alpha \cdot \bfx_j}$ (see Eq.~(\ref{eq:23}), and the normalized momentum distribution $\bar{\rho}(\bfk) = \rho(\bfk) |w(\bfk)|^2/N_L |w({\bf 0})|^2$ takes the form 
\begin{equation}
  \label{eq:75}
  \bar{\rho}(\bfk) = e^{-|\bfk|^2/[\pi^2 \sqrt{V_b}]} \Big( |\psi_0|^2 \delta_{\bfk, {\bf G}_m} + \sum_{\alpha} |\psi_\alpha|^2 \delta_{\bfk, \Q_\alpha} \Big)   \,.
\end{equation}
We see that a nonzero value of the bosonic density wave field $\psi_\alpha$ gives rise to additional coherence peaks at the nesting vector $\Q_\alpha$, where $\alpha=1,2,3$ for the isotropic triangular lattice and $\alpha=1$ for the anisotropic triangular and the square lattice. 
On the other hand, the superfluid component $\psi_0$ manifests itself by peaks at the reciprocal lattice vectors ${\bf G}_m = m_1 {\bf G}_1 + m_2 {\bf G}_2$, with integer $m = (m_1,m_2)$. Note, that this includes ${\bf G}_m = {\bf 0}$, and the reciprocal basis vectors read ${\bf G}_1 = (2 \pi, 0)$, ${\bf G}_2 =(0, 2 \pi)$ for the square lattice, and ${\bf G}_1 = 2 \pi( 1, 1/\sqrt{3})$, ${\bf G}_2 = 2 \pi (-1, 1/\sqrt{3})$ for the triangular lattice.

The number of atoms in the peaks at $\Q_\alpha$ is proportional to $|\psi_\alpha|^2$, which can be expressed by the amplitude of the density modulations $\Delta n_b$ (see Eqs.~(\ref{eq:42}, \ref{eq:61}, \ref{eq:62})). For the Kagome supersolid on the triangular lattice (see Eq.~\eqref{eq:33}), one can approximate 
\begin{equation}
  \label{eq:76}
  |\psi_{1,2,3}|^2 \approx \frac{n_b}{4} \left( \frac{\Delta n_b}{4 n_b} \right)^2\,,
\end{equation}
which is valid if one can neglect quadratic terms in $\theta$, \emph{i.e.} $|\psi_0|^2 \approx n_b$ in Eq.~(\ref{eq:33}). For the square lattice, where $0 \leq \Delta n_b \leq 2 n_b$, one finds
\begin{equation}
  \label{eq:77}
   |\psi_1|^2 = \frac{n_b}{2} \left[ 1 - \sqrt{1 - \left( \frac{\Delta n_b}{2 n_b} \right)^2} \right]\,.
 \end{equation}

Whether the supersolid peak can be detected experimentally, is determined by its weight relative to the superfluid peak 
\begin{equation}
  \label{eq:78}
  \frac{\bar{\rho}_\alpha}{\bar{\rho}_0} \equiv \frac{\bar{\rho}(\Q_\alpha)}{\bar{\rho}({\bf 0})} = \frac{|\psi_\alpha|^2 \exp [- \frac{|\Q_\alpha|^2}{\pi^2 \sqrt{V_b}}]}{|\psi_0|^2}\,.
\end{equation}
We therefore include this ratio in Tables \ref{tab:3} and \ref{tab:4}. 
For comparison, we give the size of the first higher order superfluid peaks at the reciprocal lattice vectors. For a lattice depth of $V_b = 11.4$, one finds for the square lattice $\bar{\rho}({\bf G}_{1,2})/\bar{\rho}({\bf 0}) = 0.31$ and $\bar{\rho}({\bf G}_1 + {\bf G}_2)/\bar{\rho}({\bf 0}) = 0.09$. For the triangular lattice, their size is given by $\bar{\rho}({\bf G}_{1,2})/\bar{\rho}({\bf 0}) = 0.21$, and $\bar{\rho}({2 \bf G}_{1,2})/\bar{\rho}({\bf 0}) = 2 \cdot 10^{-3}$.

From the values given in Table \ref{tab:3} and ~\ref{tab:4}, we conclude that while it is feasible to detect the supersolid peaks for the square lattice geometry, they are too small to be detected for the triangular lattice. Therefore, another way to detect the density wave correlations should be used to confirm the supersolid nature of the system. 

This can for instance be achieved by the analysis of noise correlations in the absorption spectrum, where a density wave also leads to peaks at its characteristic wavevector(s) $\Q_\alpha$~\cite{PhysRevA.70.013603, foelling_nature_2005}. Combined with the observation of a (superfluid) zero momentum peak in TOF, this also proves the existence of the supersolid phase~\cite{ScarolaSS}, if one can exclude the coexistence of multiple phases in the trap. 

The coexistence of phases arises due to spatial inhomogeneities in the chemical potential $\mu_{f(b)} = \mu_{f(b)}(\bfx)$ introduced by an overall (harmonic) confinement.
For example, a pure density wave, \emph{i.e.} with vanishing superfluid component, surrounded by a superfluid shell shows noise correlations similar to a supersolid, however, it does not show any first-order coherence peaks at $\Q_\alpha$. As noted in Ref.\cite{ScarolaSS}, the differences in time-of-flight imaging between a density wave phase coexisting with a superfluid shell and the supersolid phase are merely quantitative.

\section{Mott insulating phases}
\label{sec:mott-insul-phas}
So far we have concentrated on the case of weak-coupling, where the values of the interaction parameters are limited to $U_{bb}/t_b < (U_{bb}/t_b)_{\text{SF-MI}}$, $U_{bf} M_0 \ll 1$, and $\lambda_{BF}/\tau_B < 1$, where $\lambda_{BF} = M_0U_{bf}^2/U_{bb}$  and $\tau_B = 8 t_b/n_b U_{bb}$ for triangular lattice (for the square lattice $M_0$ is replaced by $N_0$). The first inequality assures that the system is superfluid, and not in a Mott insulating (MI) phase, for $T > T_{\text{SS}}, T_{\text{PS}}$. The second and third inequality defines the regime where the effect of the fermions on the bosons can be described in second order perturbation theory. 

In this section we discuss the opposite region of \emph{strong coupling} $U_{bb}, U_{bf} \gg t_f, t_b$, where the system can be described by an effective $t$-$J$-model. At a filling of one particle per site $n_f + n_b = 1$, it reduces to an anisotropic quantum Heisenberg model. For small bosonic hopping $t_b \ll t_f$, it turns out that the in-plane (XY) coupling is much weaker than the coupling of the $z$-components, and we will argue that the system has a stable, and unfrustrated, antiferromagnetic ground state both on the triangular lattice for filling factors $n_f = 3/4$, $n_b = 1/4$ as well as on the square lattice for $n_f = n_b = 1/2$. The fermions form a density wave that is characterized by the nesting wavevectors $\Q_\alpha$, \emph{i.e.} on the triangular lattice this is an antiferromagnet (AF) with a real-space Kagome-pattern, and on the square lattice it is the usual N{\'e}el state. For repulsive $U_{bf} >0$, the bosons become localized at the sites where no fermion is present.  This phase is exactly the alternating Mott insulator phase (AMI) that was described for the square lattice in Ref.~\cite{titvinidze:100401}. 

We conclude that, at unit filling $n_f + n_b = 1$, a Bose-Fermi mixture becomes supersolid only for sufficiently small interspecie interaction $U_{bf}$. 
It will be addressed elsewhere, whether the system enters a supersolid phase in the strong coupling regime away from unit filling, \emph{i.e.} upon doping the AMI phase by adding or removing bosons. Such a behavior was reported recently in one dimensional Bose-Fermi mixtures using quantum Monte-Carlo simulations~\cite{hebert:184505}.

\subsection{Derivation of quantum Heisenberg Hamiltonian}
\label{sec:deriv-quant-heis-1}
In the limit of large $U_{bb}$, where double occupancies are energetically forbidden, one can replace the boson by spin-1/2 operators via $b_j^\dag \rightarrow s_j^+ = \frac12 (s_j^x + i s_j^y)$ and $n_j \rightarrow \frac12 + s_j^z$, where $s_j^\alpha = \sigma_j^\alpha/2$, $(\alpha = x,y,z)$, and  $\sigma_j^\alpha$ are the usual Pauli-matrices. We then fermionize the 'bosonic' spins $s_j^\alpha$ using the celebrated Jordan-Wigner transformation in two-dimensions~\cite{PhysRevLett.63.322}
\begin{equation}
  \label{eq:79}
  \begin{split}
  s_j^+ &= c_j^\dag \exp\left[ -i \sum_{p \neq j} \theta_{pj} N_p\right] \\
  s_j^- &= c_j \exp\left[ i \sum_{p\neq j} \theta_{pj} N_p \right]\\
  s_j^z &= N_j - \frac12\,,
\end{split}
\end{equation}
where $N_j := c_j^\dag c_j = b_j^\dag b_j = n_j$, and $-\pi < \theta_{pj} \leq \pi$ is the argument of the vector from site $j$ to site $p$. It has the important property that $\exp[i \theta_{pj}] \exp[- i \theta_{jp}] = -1$. The purely fermionic Hamiltonian now reads
\begin{equation}
  \label{eq:80}
  \begin{split}
    H_b &= -t_b \sum_{\av{i,j}} \left[ c_i^\dag c_j e^{i A_{ij}} + \text{h.c.} \right] - \mu_b \sum_i N_i\\
    H_f &= - t_f  \sum_{\av{i,j}} \left( f_i^\dag f_j + \text{h.c.} \right) - \mu_f \sum_i m_i\\
    H_{bf} &= U_{bf} \sum_i m_i N_i\,,
  \end{split}
\end{equation}
where $A_{ij} =  \sum_{p\neq i,j} \left( \theta_{pj} - \theta_{pi} \right) N_p$. Except for the additional (gauge) field $A_{ij}$, this is the Hamiltonian of the two-dimensional spin-1/2 fermionic Hubbard model, where $f_i^\dag (c_i^\dag)$ creates a spin-up (down) fermion at site $i$, and the boson-fermion interaction $U_{bf}$ marks the on-site interaction. It is worth noting that the gauge field disappears in a one-dimensional system~\cite{sachdev_qpt_book}. 

In the limit of large $U_{bf}/t_{f,b}$, it is well-known~\cite{auerbach_quantum_magnetism} that one can derive a $t$-$J$-model Hamiltonian that describes the low-energy (spin and charge) excitations of the system. At unit-filling $n_b + n_f = 1$, it reduces to the antiferromagnetic spin-1/2 quantum Heisenberg model. 

If we follow the standard derivation (see Appendix~\ref{sec:deriv-quant-heis} for details) and focus on the unit-filling case, we arrive at the familiar form of the quantum Heisenberg Hamiltonian, except that the XY-coupling terms contain the (Jordan-Wigner) gauge field $A_{ij}$:
\begin{equation}
  \label{eq:81}
  \begin{split}
    H^{\text{eff}} &= \frac12 \sum_{\av{i,j}} \Big\{ \frac{2 t_b t_f}{U_{bf}} \left[S_i^+ S_j^- e^{i A_{ji}} + S_i^- S_j^+ e^{i A_{ij}} \right] \\
        & + \frac{2(t_b^2 + t_f^2)}{U_{bf}} \Big[S_i^z S_j^z - \frac14 \Big] \Big\}  - (\mu_f - \mu_b) \sum_i S^z_i\,.
  \end{split}
\end{equation}
Here, $S_i^+ = f_i^\dag c_i$, $S_i^z = (m_i - N_i)/2$, are proper spin operators, which obey $[S_i^+,S_j^-] = 2 S_i^z \delta_{ij}$. Spin-up corresponds to occupation by a fermion and spin-down to occupation by a boson. Note that the presence of the gauge field reflects the different symmetry of fermions and hard-core bosons under exchange of two particles.

\subsection{Triangular lattice}
For the particular filling of $n_f = 3/4$, $n_b = 1/4$, or a total magnetization of $\av{S^z} = 1/4$, the ground state phase of the system is an alternating Mott-insulator phase for the bosons  
and a density wave phase for the fermions. The real-space configuration is of the Kagome-type that was discussed previously (see Fig.~\ref{fig:4}), only that bosons are now localized. Formulated in the spin-language, the system favors the classical \emph{unfrustrated} Ising ground state for all values of $\{t_b,\, t_f\}$, because of the externally applied magnetic field in the $z$-direction, which is proportional to the number difference of bosons and fermions in the system. 

If the system is doped away from unit filling, \emph{e.g.} by adding or removing bosons, it can be described by an effective $t$-$J$-model with the additional gauge field $A_{ij}$. It remains an open question, whether the system then becomes supersolid, as is the case for a one-dimensional Bose-Fermi mixture~\cite{hebert:184505}. 

\subsection{Square Lattice}
Here, a stable alternating Mott-insulator phase occurs for double half-filling $n_b = n_f = 1/2$. Even without externally applied magnetic field ($\mu_b = \mu_f$), the system enters the (classical) N{\'e}el antiferromagnetically ordered ground state, because of the finite anisotropy in the spin coupling that occurs for $t_f \gg t_b$~\cite{FrohlichLieb-CommMath-1978, PhysRevB.67.104414, PhysRevB.49.15139}. 
The XY-coupling is renormalized to zero, and the spins point along the $z$-axis, \emph{i.e.} the system is a Mott insulator with a site occupation that alternates between bosons and fermions. This agrees with recent DMFT calculations in Ref.~\cite{titvinidze:100401}. 


\section{Conclusions}
\label{sec:conclusions}
We have studied mixtures of spinless bosons and fermions in different two-dimensional optical lattice geometries at fermionic fillings $n_f$ that give rise to a nested Fermi surface. We have shown how nesting can lead to supersolid formation via a density wave instability of the fermions. The resulting density order in the supersolid is characterized by the nesting vectors $\Q_i$. 

On the triangular lattice, we have thereby identified a novel supersolid phase with three ordering wavevectors $\Q_{1,2} = (\pm \pi, \pi/\sqrt{3})$, $\Q_3 = (0, 2 \pi/\sqrt{3})$ that give rise to a Kagome-pattern in real-space. We predict this novel phase to appear at rather low temperatures $T_{\text{SS}} \lesssim t_f/40 = T_F/240$, and the density modulation $\Delta n_b$ in the supersolid, which is proportional to $T_{\text{SS}}/U_{bf}$ to be weak. Typically, we find that the density wave only involves $0.1\%$ of all the bosons, which leads to $\Delta n_b/n_b \simeq 0.05$ . 

Furthermore, for temperatures such that the thermal energy exceeds the characteristic energy level spacing in the system. we have pointed out the possibility of an incommensurate density wave modulation in the supersolid phase. If thermal effects can be ignored, however, only the instability towards the commensurate supersolid remains. 

Higher transition temperatures and larger density wave modulations can be found for spatially anisotropic hopping amplitudes, or if one considers the square lattice geometry. In these cases, the nesting relation is fulfilled for a larger fraction of wavevectors in the first Brillouin zone. We have derived transition temperatures of $T_{\text{SS}} \simeq T_F/4 \;(T_F/5)$ for the isotropic (anisotropic) square lattice. The density wave now involves up to $20 \%$ of all the bosons. We have identified the square lattice as the optimal choice, since it shows the highest values of $T_{\text{SS}}$ and $\Delta n_b$.   

We have pointed out that introducing anisotropic hopping on the square lattice has the advantage that the tendency towards phase separation is weakened, which then only occurs above a certain inter-species interaction threshold $U_{bf}^{\text{PS}}$. The values of $T_{\text{SS}}$ and $\Delta n_b$ on the isotropic and anisotropic ($r_f = 0.75$) square lattice are about the same.

We have also predicted how to experimentally realize and detect the supersolid phase in two commonly used Bose-Fermi mixtures, $^{87}\text{Rb}^{40}\text{K}$ and $^{23}\text{Na}^6\text{Li}$. The square lattice geometry allows for supersolid transition temperatures close to current cooling limits $T_{\text{SS}} \simeq T_F/4$ for both mixtures. However, since the amplitude of the density wave modulations grows for a smaller ratio of bosonic to fermionic hopping amplitudes $t_b/t_f$, we find that the detection of the supersolid phase via additional coherence peaks in time-of-flight absorption images becomes easier for smaller $t_b/t_f$ (slower bosons). Both $^{87}\text{Rb}^{40}\text{K}$, trapped in a $\lambda=1064$ nm optical lattice, as well as $^{23}\text{Na}^6\text{Li}$ are therefore good candidates to observe supersolidity.

Finally, we have considered the strong-coupling regime of $U_{bb}, U_{bf} \gg t_{f},t_b$ and derived a quantum Heisenberg Hamiltonian that includes an additional gauge field due to the Jordan-Wigner transformation in two-dimensions. For filling factors of $n_f=3/4$, $n_b = 1/4$ on the triangular and $n_f = n_b = 1/2$ on the square lattice, the ground state of the strong-coupling Hamiltonian is an alternating Mott-insulating state (AMI) for the bosons. The fermions exhibit a density wave and, for repulsive interaction $U_{bf}>0$, occupy all the other lattice sites. The order is again characterized by the nesting vectors $\Q_i$, which leads on the triangular lattice to the same Kagome-pattern we found previously for the supersolid, however, the bosons are now spatially localized.

\section*{Acknowledgments} We thank I. Bloch, W. Hofstetter, M. Gustavsson, and G. Refael for valuable discussions. This work is supported by NSF through the contract DMR-0803200 and through the Center for Quantum Information Physics at Yale. 

\appendix
\section{Detailed analysis of the Lindhard function on the triangular lattice}
\label{sec:deta-analys-lindh}
This appendix includes a detailed discussion of the \emph{finite temperature behavior} of the Lindhard function $\chi(T, \q)$ on the triangular lattice. In particular, we point out that for temperatures larger than the energy level spacing in the system, the minimum of the function in $\bfk$-space occurs slightly away from the nesting vectors $\Q_i$. This gives rise to an instability towards a supersolid with an incommensurate density wave modulation. 
\subsection{Fermion-mediated interaction}
The instability criteria are based upon analysis of the low temperature divergences of the fermionic Lindhard function $\chi(T, \q) = \sum_\bfk  F(T, \q, \bfk)$ with
\begin{equation}
  \label{eq:82}
F(T, \q, \bfk) = \frac{f[\xi_f(\bfk),T] - f[\xi_f(\bfk+\q),T]}{\xi_f(\bfk) - \xi_f(\bfk+\q) + i \eta} \,,
\end{equation}
and $f(\xi_f,T) = [1 + \exp(\xi_f/T)]^{-1}$ is the Fermi function. 
The Lindhard function describes the part of the effective boson-boson interaction that is induced by the fermions
\begin{equation}
  \label{eq:83}
  U(T,\q) = U_{bb} + U_{bf}^2 \chi(T,\q) \,.
\end{equation}
This effective interaction is obtained after an exact integration of the fermionic degrees of freedom using a functional integral approach, followed by a perturbative expansion to second order in $M_0 U_{bf}$. Therefore, this form of interaction is restricted to the regime of weak boson-fermion coupling $M_0 U_{bf} \ll 1$. Here, $M_0=3/(4\pi^2 t_f)$ is an estimate of the regular part of the fermionic density of states.

At the particular filling of $n_f = 3/4$, the Fermi surface of the system both shows nesting and contains critical points at $\Q_i$, \emph{i.e.} $\nabla_{\q} \xi_f(\q)|_{\Q_i}=0$, that lead to a van Hove singularity in the density of states at that filling (see Fig.~\ref{fig:2}). As a result, the Lindhard function diverges as $T\rightarrow 0$ at the wavevectors that lie on straight lines between $\q=0$ and the three nesting vectors $\Q_{1,2,3}$ (see Fig.~\ref{fig:1}). One can write this set of $\q$-vectors as ${\cal N}=\{\q; \; \exists \alpha \in [0,1]: \q = \alpha \Q_i\}$. 

At $\q={\bf 0}$, the Lindhard function diverges logarithmically
\begin{equation}
  \label{eq:84}
  \chi (T, {\bf 0}) \sim - M_0 \ln \frac{T_0}{T}\,,
\end{equation}
with $T_0 = 8 C_1 t_f$ and $C_1 = 2 e^C/\pi \simeq 1.13$, where $C$ is the Euler-Mascheroni constant.  

At the nesting vectors $\q = \Q_i$, one finds that the divergence is enhanced to 
\begin{equation}
  \label{eq:85}
 \chi(T,\Q_i) \sim - \frac{M_0}{6} \left[\ln \frac{T_1}{T} \right]^2 \,,
\end{equation}
where $T_{1} = a \,T_0$ with $a \simeq 2.17$ being determined from a fit to a numerical calculation of $\chi(T, \Q_i)$ using importance sampling Monte-Carlo integration. 
 
In between, for $\q = \alpha \Q_i$ with $0 < \alpha < 1$, the behavior is different for $\alpha \ll 1$ and for $\alpha \simeq 1$,  \emph{i.e.} for $\q$ being close to the nesting vector.
We investigate the two cases separately in the following sections, and find that at finite temperatures, the thermal width of the Fermi functions in $F(T, \q, \bfk)$ comes into play. For a finite system, however, one can neglect thermal effects for temperatures below the system's characteristic energy level spacing. 

\begin{figure}[tb]
  \centering
  \includegraphics[width=.8\linewidth]{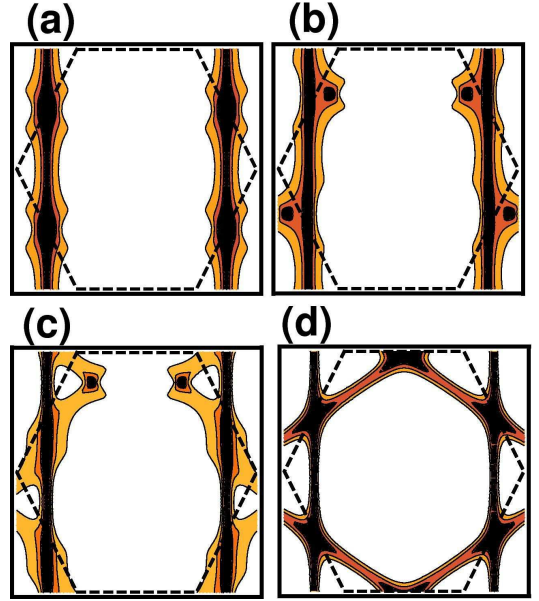}
 \caption{(Color online) Contour plot of the integrand $F(T, \q, \bfk) \leq 0$ of the Lindhard function $\chi(T, \q=(1-\delta) \Q_3 ) = \int d^2 \bfk \, F(T,\q, \bfk)$ for fixed temperature $T = t_f/10$. Parts (a-d) correspond to different deviations of $\q=(1-\delta) \Q_3$ from the nesting vector: $\delta = 0, 0.25,0.5,0.9$ (a-d). Hexagon (dashed) denotes the first Brillouin zone. Bright regions indicate small absolute values of $F$, dark regions indicate large absolute values. In particular, the integrand vanishes in white regions. For small $\delta$ the integrand is peaked along $k_1 = \pm \pi$. An additional peak occurs at $\q_c$ (Eq.~(\ref{eq:90})) for nonzero $\delta$, which moves from $\Q_{1,2}$ towards $\Q_3$ along the Fermi surface. Close to $\q \approx 0$ (d), the integrand is peaked at the critical points  $\pm \Q_{1,2,3}$ of the Fermi surface.   }
  \label{fig:16}
\end{figure}
\begin{figure}[tb]
  \centering
  \includegraphics[width=.55\linewidth]{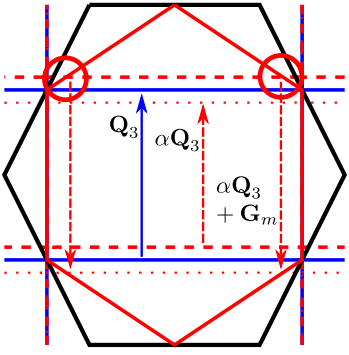}
  \caption{(Color online) Shift of the contour where the denominator of $F(T, \q=(1-\delta) \Q_3, \bfk)$ vanishes in $\bfk$-space up to first order in $\delta \ll 1$.  
For $\delta = 0$, the denominator vanishes along $k_1 = \pm \pi$ and along the horizontal solid lines connected by $\Q_3$ (solid arrow). For nonzero $\delta$, there is no shift of the vertical parts of the contour (to linear order in $\delta$). However, the horizontal parts get shifted to larger values of $k_2$ (horizontal dashed lines). Whereas the numerator of $F(T,\q, \bfk)$ vanishes along the lower (horizontal) part of the contour, the function $F$ becomes peaked at the location $\q_c$ indicated by the circles, since the shifted contour crosses the Fermi surface (connected by $\alpha \Q_3 + {\bf G}_m$ (dashed arrow)).}
  \label{fig:17}
\end{figure}

\subsection{Long-wavelength divergence}
\label{sec:long-wavel-diverg}
In this section, we investigate the regime $0<\alpha \ll 1$. By numerical integration, we find that the temperature dependence of $\chi(T, \alpha \Q_i)$ is always logarithmically, but with a slope that depends on temperature. For temperatures above some $\alpha$-dependent temperature $T'_\alpha$ it is equal to $M_0$. At $T_\alpha'$ the slope decreases abruptly to a slightly smaller value, $\approx 2 M_0/3  $, which holds then for $T < T'_\alpha$. The transition occurs at smaller temperatures for smaller values of $\alpha$. 
\begin{figure}[tb]
  \centering
  \includegraphics[width=.75\linewidth]{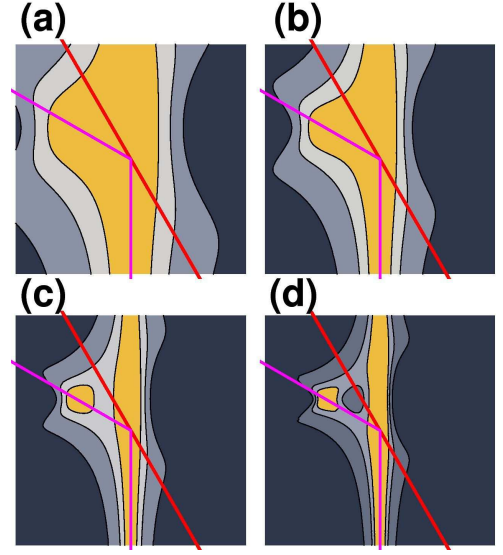}
  \caption{(Color online) Detailed contour plot of $F(T, \q, \bfk) \leq 0$ around $\bfk = \Q_2$ for fixed $\q = 0.9 \, \Q_3$ and various temperatures $\log_{10} T/t_f = -0.8, -1.1, -1.4, -1.7$ (a-d). At $\Q_2$ the Fermi surface (light solid) has a kink and touches the boundary of the first Brillouin zone (darker diagonal line). Brighter colors of the contour indicate larger absolute values of $F$, in particular, the integrand vanishes in black region (inverted color scheme compared to Fig.~\ref{fig:16}). The thermal smearing becomes smaller for decreasing temperature, and the additional peak of $F(T, \q, \bfk)$ is separated from the thermally broadened peak at the critical point $\Q_2$ for $T < T_L$.} 
  \label{fig:18}
\end{figure}

This behavior can easily be understood when the width of the Fermi function in the numerator of $F(T, \q, \bfk)$ (Eq.~(\ref{eq:82})) is taken into account. 
For $\alpha \ll 1$, we find that the function $F(T, \alpha \Q_i, \bfk)$, if considered as a function of the integration variable $\bfk$, is strongly peaked for $\bfk \approx \Q_i$, because these are the critical points, which lead to the van Hove singularity in the density of states (see Fig.~\ref{fig:16}D). 

Let us consider the case of $\q = \alpha \Q_3$ for definiteness. In the limit $\alpha \rightarrow 0$, the function $F(T, \alpha \Q_3, \bfk)$ is equally strongly peaked at all six van Hove singularities. More mathematically speaking, if we write $\bfk = \Q_i + \tilde{\bfk}$, where $\tilde{\bfk}$ is considered small, the energy denominator vanishes only quadratically in the small quantities $(|\tilde{\bfk}|, \alpha)$. One finds $\xi_f(\Q_{1,2} + \tilde{\bfk}) - \xi_f(\Q_{1,2} +\tilde{\bfk} + \alpha \Q_3) = t_f {\cal O}(\alpha |\tilde{\bfk}|)$, and 
\begin{equation}
  \label{eq:86}
  \xi_f(\Q_3 + \tilde{\bfk}) - \xi_f(\Q_3 +\tilde{\bfk} + \alpha \Q_3) = t_f {\cal O}(\alpha |\tilde{\bfk}|, \alpha^2)
\end{equation}
For finite $\alpha$, the transition takes place when the ${\cal O}(\alpha^2)$ term becomes of the order of the width of the Fermi function, which is about $2 T$. At this point, the peak of $F(T, \alpha \Q_3, \bfk)$ at $\pm \Q_3$ is much reduced and the slope thus changes to $2/3$ of its initial value. 

The precise terms read $\xi_f(\Q_3 + \tilde{\bfk}) - \xi_f(\Q_3 +\tilde{\bfk} + \alpha \Q_3) = 2 \pi^2 \alpha^2 t_f + 2 \sqrt{3} \pi \alpha \tilde{k}_2 t_f $, and we can therefore estimate $T_\alpha'$ by
\begin{equation}
  \label{eq:87}
  T_\alpha' = \pi^2 \alpha^2 t_f\,.
\end{equation}

\subsection{Divergence at the nesting vectors}
\label{sec:diverg-at-nest}
In this section, we look at the case of $\alpha \approx 1$. If we write $\alpha = 1 - \delta$, with small $\delta \ll 1$, we find that the divergent behavior of $\chi(T, (1-\delta) \Q_i)$ changes from being of $[\ln T/t_f]^2$-type for temperatures above some $\alpha$-dependent $T_\alpha$ to being proportional to $\ln T/t_f$ for $T < T_\alpha$. 
Generally, one can obtain a $[\ln T]^2$ divergence of $\chi(T, {\bf K})$ at a wavevector ${\bf K}$, if the Fermi surface both shows nesting (with nesting vector ${\bf K}$) and contains critical points where $\nabla_\q \xi_f(\q)|_{\q_i \in \text{FS}}=0$ which lead to a van Hove singularity in the density of states. Additionally it is required though, that the nesting relation is fulfilled for the $\bfk$-states that make up the van Hove peak in the density of states, \emph{i.e.} that are close to the critical points. These $\bfk$-states have the same energy up to linear order in deviations from $\q_i$. 

Thermal smearing of the Fermi edge allows states that only fulfill nesting approximately, \emph{i.e.} $\xi_f(\bfk + \bfK) \approx - \xi_f(\bfk)$, to contribute to the Lindhard integral. However, at $T = 0$, it is required to fulfill the nesting relation exactly which demands ${\bf K} = \Q_{1,2,3}$. 

Let us again specify to the case of ${\bf K} = (1 - \delta) \Q_3= \big(0, \frac{2 \pi (1-\delta)}{\sqrt{3}}\big)$ for definiteness. In Fig.~\ref{fig:16}, we see that, for small $\delta$ (parts (a-b)), the integrand $F(T, {\bf K}, \bfk)$ is strongly peaked along the lines of constant $k_1 = \pm \pi$. The vector ${\bf K}$ only translates along the $k_2$-direction and thus provides a mapping between two points of the Fermi surface (nesting). We also see clearly that the peaks become wider close to the critical points $\Q_{1,2}$, where the energy only varies quadratically with deviations from $\Q_{1,2}$. If we write $\bfk = \Q_i + \tilde{\bfk}$, this reads $\xi_f(\Q_i + \tilde{\bfk}) = \xi_f(\Q_i) + {\cal O}(|\tilde{\bfk}|^2) \approx \xi_f(\Q_i)$.

To obtain a divergent behavior of type $\chi(T,{\bf K}) \sim - [\ln T/t_f]^2$ for \emph{all} $T \rightarrow 0$, it is required that the energy denominator vanishes quadratically in $|\tilde{\bfk}|$:
\begin{equation}
  \label{eq:88}
  \xi_f(\bfk) - \xi_f(\bfk + {\bf K}) = {\cal O}(|\tilde{\bfk}|^2)\,.
\end{equation}
However, this relation is only valid exactly at ${\bf K} = \Q_{1,2,3}$.
Slightly off, at ${\bf K} = (1-\delta) \Q_i$, one finds instead that 
\begin{equation}
  \label{eq:89}
  \xi_f(\bfk) - \xi_f(\bfk + {\bf K}) = \delta \, {\cal O}(\tilde{|\bfk}|)
\end{equation}
The energy denominator now vanishes \emph{linearly} with $|\tilde{\bfk}|$. One can clearly observe that the peak of $F(T, {\bf K}, \bfk)$ close to $\bfk \approx \Q_{1,2}$ narrows as $\delta$ is increased [Fig.~\ref{fig:16}(a-c)].  

The plots also show the emergence of additional peaks close to $\Q_{1,2}$ [see Fig.~\ref{fig:16}(b-c)], which move along the Fermi surface towards $\Q_3$ for increasing $\delta$. They occur because a part of the contour, where the energy denominator vanishes, shifts for nonzero $\delta$, and crosses the Fermi surface. This is shown in Fig.~\ref{fig:17}. The energy denominator vanishes along the rectangle, with vertical sides at $k_1 = \pm \pi$ and horizontal sides at $k_2 = \pi (\pm 1 + \delta)/\sqrt{3}$. For $\delta = 0$, the function $F(T, \Q_3, \bfk)$ vanishes along the horizontal path, since the numerator is zero. This part of the path, however, shifts for nonzero $\delta$ to larger $k_2$ values and crosses the Fermi surface at 
\begin{equation}
  \label{eq:90}
  \q_c = \left(\pi(1 - \delta), (1 + \delta) \pi/\sqrt{3}  \right)\,.
\end{equation}
As a result, a peak in $F(T, {\bf K}, \bfk)$ occurs around $\bfk \approx \q_c$, which is clearly visible in Fig.~\ref{fig:18}. 
\emph{This peak is responsible for the fact that the minimum of the Lindhard function is shifted away from  $\Q_{1,2,3}$ for intermediate temperatures.}

As long as the thermal smearing is larger than the separation $|\q_c - \Q_2|$, the peak of $F(T, (1-\delta) \Q_3, \bfk)$ is actually broader for nonzero $\delta$ than for $\delta=0$ [see Fig.~\ref{fig:18}(a-b)].  However, as the temperature is lowered the separation of the additional peak at $\q_c$ from the critical point $\Q_2$ finally becomes larger than the thermal width [see Fig.~\ref{fig:18}(c-d)]. 
At this point, the nesting relation is no longer fulfilled for all the states close to $\Q_2$ (in the sense defined above), and the divergence of $\chi(T, {\bf K})$ becomes of type $\ln T/t_f$. As a result, eventually one finds $\chi(T, \Q_3) < \chi(T, {\bf K})$, \emph{i.e.} in the limit $T \rightarrow 0$, the minimum of the Lindhard function occurs at $\Q_{1,2,3}$. 

\subsection{Level spacing temperature $T_L$}
\label{sec:level-spac-temp}
In this section we will estimate the temperature $T_\alpha$, where the divergent behavior of $\chi(T, (1-\delta)\Q_i)$ changes from being $[\ln T/t_f]^2$ to being $ln T/t_f$. The arguments we will use are similar to the ones in Sec.~\ref{sec:long-wavel-diverg}. 

First, one can relate the thermal width of the Fermi function to a distance in $\bfk$-space using the dispersion relation. Expanding the fermionic energy around the location of the additional peak $\q_c$ (see above), yields
\begin{equation}
  \label{eq:91}
  \xi_f(\q_c(\delta) + \tilde{\bfk}) \simeq \pi \delta t_f \left(\tilde{k}_1 + \sqrt{3} \tilde{k}_2 \right)\,. 
\end{equation}
The thermal width of the Fermi function $\pm 2 T$ thus relates to $\bfk$-space like
\begin{equation}
  \label{eq:92}
\tilde{k}_1 \simeq \pm \frac{2 T}{t_f \pi \delta}\,, \, \, \tilde{k}_2 \simeq \pm \frac{2 T}{\sqrt{3} t_f\pi \delta }\,.
\end{equation}
The separation of the peak at $\q_c$ from the van Hove singularity is given $|\q_c - \Q_2|= 2 \pi \delta/\sqrt{3}$, we estimate the crossover temperature $T_\alpha$ to occur when $2 \delta \pi/\sqrt{3} = 4 T_\alpha/\pi \delta t_f$, which leads to 
\begin{equation}
  \label{eq:93}
   T_\alpha \simeq \frac{t_f \pi^2 \delta^2}{2}\,,
\end{equation}
where $\delta = 1 - \alpha$. 

For a finite lattice of $N_L=L^2$ unit cells, this defines a (level spacing) temperature $T_L \sim 1/L^2$, below which the roton gap closes at $\Q_{1,2,3}$, by noting that $\min(\delta) = 2/L$:
\begin{equation}
  \label{eq:94}
  T_L = \frac{t_f 2 \pi^2}{L^2}\,.
\end{equation}
This characteristic temperature corresponds to the spacing of energy levels in the finite system, \emph{i.e.} thermal effects can be ignored for temperatures smaller than $T_L$. 

It is worth noting that for the square lattice, the minimum of $\chi(T, \q)$ \emph{always} occurs at the nesting vector $\q=\Q_{\text{sq}} = (\pi,\pi)$.

\section{Fermion induced interaction in real-space}
\label{sec:ferm-induc-inter}
In this section we describe the real-space form of the effective boson-boson interaction $U(T, \q) = U_{bb} + U_{bf}^2 \chi(T, \q)$ on the triangular and square lattice (see Sec.~\ref{sec:phase-diagram-from}). We obtain the Lindhard function for wavevectors $\q$ in the first Brillouin zone by numerical Monte-Carlo integration. The real-space form $U(T, \bfx_i)$ is simply the Fourier transform of $U(T, \q)$, evaluated at lattice sites $\bfx_i$:
\begin{equation}
  \label{eq:95}
  U(T, \bfx_i) = U_{bb} \, \delta_{\bfx_i,{\bf 0}} + U_{bf}^2 \chi(T, \bfx_i)\,,
\end{equation}
where $\delta_{\bfx_i, \bfx_j}$ is the Kronecker delta and the real-space form of the Lindhard function is given by a sum over wavevectors in the first Brillouin zone
\begin{equation}
  \label{eq:96}
  \chi(T, \bfx_i) = \frac{1}{N_L} \sum_{\bfk_j} \chi(T, \bfk_j) e^{i \bfk_j\cdot \bfx_i} \,.
\end{equation}
The intrinsic interaction term $U_{bb} \, \delta_{\bfx_i,{\bf 0}}$ provides a contact interaction in real-space, which is repulsive for $U_{bb} > 0$. On the other hand, the fermion-induced part $U_{bf}^2 \chi(T, \bfx_i)$ provides a long-range interaction between the bosons that is oscillating in sign, as can be seen in Fig.~\ref{fig:19}.
\begin{figure}[t!]
  \centering
  \includegraphics[width=\linewidth]{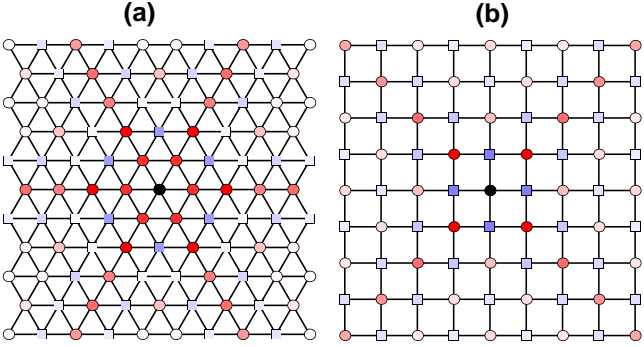}
  \caption{(Color online) Lindhard function in real-space $\chi(T, \bfx_i)$ for triangular (a) and square lattice (b) at a temperature of $\log_{10} T/t_f = -2.9$ (compare with Fig.~\ref{fig:3} (b)). The lattice site $\bfx_i=0$ is located in the center of the lattice. The sign of $\chi(T, \bfx_i)$ is encoded in the shape of the lattice site: circles denote attractive interaction $\chi(T, \bfx_i) < 0$, whereas squares denote repulsive interaction $\chi(T, \bfx_i) > 0$. Darker colors correspond to larger absolute values of $\chi(T, \bfx_i)$.  }
  \label{fig:19}
\end{figure}

On the triangular lattice [see Fig.~\ref{fig:19}(a)], the nearest-neighbor interaction is attractive, but the strongest attractive interaction is at the second nearest-neighbor sites $\pm 2 {\bf a}_{1,2}$ and $\pm 2({\bf a}_1 - {\bf a}_2)$. The interaction is repulsive at all the other second nearest-neighbor sites. Clearly, this form of interaction gives rise to the Kagome-type density wave seen in Fig.~\ref{fig:4}. Note that this form of interaction is quite different from the usual $U$-$V$-model interaction, which is short-ranged. There, the supersolid emerges for large repulsive nearest-neighbor interactions $V$ due to frustration. 

On the square lattice [see Fig.~\ref{fig:19}(b)], the nearest-neighbor interaction in repulsive, whereas the second neighbor interaction is attractive. The emergent bosonic density wave in the supersolid is thus of checkerboard-type [compare with Fig.~\ref{fig:8}(a)]. 

The fact that the sign of the nearest-neighbor interaction is different for triangular and square lattice reflects the fact that the divergence at the nesting vectors is not as pronounced for the triangular lattice as it is for the square lattice. 

\section{Construction of the Landau-Ginzburg-Wilson functional for triangular and square lattice}
\label{sec:constr-land-ginzb}
Motivated by the instability analysis (see Sec.~\ref{sec:instability-analysis}), we assume that the supersolid will cause condensation into a number of Fourier components. In addition to $\av{b_{\bf 0}}$ (superfluid order), we consider the possibility of $\av{b_{\bfK_i}} \neq 0$, where $\bfK_i$ are the wavevectors where the roton gap closes first. For simplicity, we focus on the case $\bfK_i = \Q_i$, \emph{i.e.} on the triangular lattice we assume transition temperatures $T_{\text{SS}} < T_L$. Here, $\Q_i$ are the nesting vectors of the Fermi surface.

\subsection{Isotropic triangular Lattice}
On the isotropic triangular lattice, there are three nesting vectors $\Q_{1,2,3}$, and the real-space wavefunction can be expanded as
\begin{equation}
  \label{eq:97}
  \Psi(\bfx) \approx \psi_0(\bfx) + \sum_{i=1}^3 \psi_i(\bfx) e^{i \Q_i \cdot \bfx}\,.
\end{equation}
The complex fields $\boldsymbol{\psi} = (\psi_0(\bfx), \psi_1(\bfx),\psi_2(\bfx), \psi_3(\bfx))$ are defined as $\psi_i = \sqrt{N_L} \av{b_{\Q_i}}$, where $\Q_0={\bf 0}$. The average is taken over a length scale much larger than the lattice spacing. They are assumed to be slowly varying in space-continuum and below we will eventually focus on the special case of completely homogeneous fields. 

The Landau-Ginzburg-Wilson free energy is constructed from every possible term that is invariant under the symmetry group of the lattice. 
Under operation of a symmetry element of the lattice $\{G, {\bf u}\}$, which acts on a lattice position in real space as $\bfx' = G\cdot \bfx + {\bf u}$, the real-space wavefunction $\Psi(\bfx)$ transforms as 
\begin{equation}
  \label{eq:98}
  \Psi'(\bfx') = \Psi(G \cdot \bfx + {\bf u})\,.
\end{equation}
In momentum space, the wavefunction transforms as 
\begin{equation}
  \label{eq:99}
  \Psi'(\q') = \Psi(G \cdot \q) e^{i \q\cdot G \cdot {\bf u}}\,.
\end{equation}
The symmetry group of the triangular lattice can be spanned by the five generators $\{R, I_1, I_2, T_1, T_2\}$, where 
\begin{equation}
  \label{eq:100}
  R = -\frac12 \begin{pmatrix} 1 & -\sqrt{3} \\ \sqrt{3}& 1 \end{pmatrix}
\end{equation}
is a 3-fold rotation (choosing the origin to be a lattice point), and 
\begin{equation}
  \label{eq:101}
  \begin{split}
  I_{1,2} &= \pm \begin{pmatrix} -1 & 0 \\ 0 & 1 \end{pmatrix}
\end{split}
\end{equation}
are the two inversions, and $T_{1,2}$ are the two translations by a Bravais lattice vector: $\bfx' = \bfx + {\bf a}_{1,2}$. In order to examine how the Fourier components $\psi_i$ transform under these operations, it is convenient to cast the transformation rules into matrix form, acting on the vector space of $\boldsymbol{\psi} = (\psi_0,\psi_1, \psi_2, \psi_3)$. This is in fact a representation of the symmetry group under which these momentum points transform. In this representation, one finds
\begin{equation}
  \label{eq:102}
  R = \begin{pmatrix} 1&0&0&0 \\ 0&0&1&0 \\ 0&0&0&1 \\ 0&1&0&0   \end{pmatrix}
\end{equation}
as well as
\begin{equation}
  \label{eq:103}
  I_1 = I_2 = \begin{pmatrix} 1&0&0&0 \\ 0&0&1&0 \\ 0&1&0&0 \\ 0&0&0&1 \end{pmatrix}\,.
\end{equation}
The translations take the form
\begin{equation}
  \label{eq:104}
  T_1 = \begin{pmatrix} 1&0&0&0 \\ 0&1&0&0 \\ 0&0&-1&0 \\ 0&0&0&-1 \end{pmatrix}
\end{equation}
and
\begin{equation}
  \label{eq:105}
  T_2 = \begin{pmatrix} 1&0&0&0 \\ 0&-1&0&0 \\ 0&0&1&0 \\ 0&0&0&-1 \end{pmatrix}\,.
\end{equation}

We will now shortly describe the general procedure to find all the quadratic and quartic terms that are invariant under the symmetry group. The $N$ different generators can be written as $\{t^{(1)}_{ij}, \ldots, t^{(N)}_{ij}\}$, (here $N=4$), where $t^{(n)}_{ij}$ are $d$-dimensional matrix representations of the group generators, and $d$ is the number of order parameters considered (here $d=4$). The order parameters transform under operation of a group generator like $\psi'_i=t^{(n)}_{ij} \psi_j$ and $(\psi'_i)^* = (t^{(n)}_{ji})^* \psi^*_j$, such that a generic quadratic term $S_2 = \sum_{i,j} A_{ij} \psi_i^* \psi_j$ transforms as 
\begin{equation}
  \label{eq:106}
  S_2' = \sum_{i,j,a,b} A_{ij} (t_{ai}^{(n)})^* t_{jb}^{(n)} \psi^*_a \psi_b \overset{!}{=} S_2\,.
\end{equation}
Invariance under all generators of the symmetry group requires the coefficients to obey 
\begin{equation}
  \label{eq:107}
  \sum_{i,j} A_{ij} t_{ai}^{(n)*} t_{jb}^{(n)} = A_{ab}\,,
\end{equation}
for all $n=1,\ldots,N$, which are $Nd^{2}$ equations constraining the coefficients $A_{ij}$. For the quartic terms $S_4 = \sum_{i,j,k,l} A_{ijkl} \psi_i^* \psi_j^* \psi_k \psi_l$, one finds that invariance requires 
\begin{equation}
  \label{eq:108}
  \sum_{i,j,k,l} A_{ijkl} (t_{ai}^{(n)} t_{bj}^{(n)})^* t_{kc}^{(n)} t_{ld}^{(n)} = A_{abcd}\,,
\end{equation}
for all $n=1,\ldots,N$, which are $Nd^{4}$ equations constraining the coefficients $A_{ijkl}$.

The effective LGW free energy one derives in this way (see Eq.~(\ref{eq:24})), reads
\begin{equation}
  \label{eq:109}
\begin{split}
   {\cal F}_b &= \int d\bfx \Bigl\{ m_0 |\psi_0|^2 + m_1 |\boldsymbol{\psi}_Q|^2 + |\partial \psi_0|^2 + v^2 |\partial \psi_Q|^2 \\ &+ \sum_{i=0}^2 u_i \Theta_i + \sum_{i=1}^4 g_i F_i \Bigr\} \,,
\end{split}
\end{equation}
where $\boldsymbol{\psi}_Q = (\psi_1, \psi_2, \psi_3)$. It contains ten parameters $\{m_{0,1},v,u_{0,1,2},g_{1,2,3,4}\}$ and the different terms read
\begin{equation}
  \label{eq:110}
   \begin{split}
     \Theta_0 &= |\psi_0|^4 \\
    \Theta_1 &= |\boldsymbol{\psi}_Q|^4 = |\psi_1^4| + |\psi_2|^4 + |\psi_3|^4 \\ 
    &+ 2 \left( |\psi_1|^2 |\psi_2|^2 + |\psi_1|^2 |\psi_3|^2 + |\psi_2|^2 |\psi_3|^2 \right) \\
    \Theta_2 &= |\psi_0|^2 |\boldsymbol{\psi}_Q|^2 = |\psi_0|^2 \left(  |\psi_1|^2 +  |\psi_2|^2 +  |\psi_3|^2 \right)\\
    F_1 &= |\psi_1|^4 + |\psi_2|^4 + |\psi_3|^4 \\
    F_2 &= \left( \psi_1^2 + \psi_2^2 + \psi_3^2 \right)^* \left(\psi_1^2 + \psi_2^2 + \psi_3^2 \right) \\
    F_3 &= \psi_0 \left( \psi_1 \psi_2^* \psi_3^* + \text{cyclic} \right) + c.c. \\
    F_4 &= \psi_0^2 \left( \psi_1^2 + \psi_2^2 + \psi_3^2 \right)^* + c.c. \,,
 \end{split}
\end{equation}

\subsection{Anisotropic triangular lattice}
In the case of the triangular lattice with anisotropic hopping, one finds that its symmetry group can be spanned by the same generators, but without the 3-fold rotation $R$. Repeating the analysis with the reduced symmetry will naturally lead to the same invariant terms as in the isotropic case, and some additional terms that are now allowed as a result of lower symmetry. The quadratic terms are given by [see Eq.~\eqref{eq:63}]:
\begin{equation}
  \label{eq:111}
  {\cal F}_b^{(2)} = m_0 |\psi_0|^2 + m_1 (|\psi_1|^2 + |\psi_2|^2) + m_3 |\psi_3|^2\,.
\end{equation}
The quartic term contains, naturally, the mass terms squared, the cross terms between the masses as well as seven additional terms. Altogether, the quartic terms read [see Eq.~\eqref{eq:63}]:
\begin{equation}
  \label{eq:112}
\begin{split}
  {\cal F}_b^{(4)} &= g_0 |\psi_0|^4 + g_1 (|\psi_1|^2 + |\psi_2|^2)^2 + g_2 |\psi_3|^4 \\ & + \lambda_0 |\psi_0|^2 (|\psi_1|^2 + |\psi_2|^2 ) + \lambda_1 |\psi_0|^2 |\psi_3|^2 \\ & + \lambda_2 |\psi_3|^2 (|\psi_1|^2 + |\psi_2|^2 ) +u_1 |\psi_1|^2 |\psi_2|^2 \\ & + u_2 [(\psi_1^2)^* \psi_2^2 + c.c.] + u_3 [(\psi_0^2)^* (\psi_1^2 + \psi_2^2) + c.c.] \\ & + u_4 [\psi_0^* \psi_3^* \psi_1 \psi_2 + c.c.] + u_5[(\psi_0^2)^* \psi_3^2 + c.c.] \\ & + u_6 [(\psi_0^* \psi_1^* \psi_2 \psi_3 + c.c.) + (1 \leftrightarrow 2)] \\ &+ u_7 [(\psi_3^2)^* (\psi_1^2 + \psi_2^2) + c.c.]\,.
\end{split}
\end{equation}
\subsection{Square Lattice}
Only one nesting vector $\Q_{sq} = (\pi,\pi)$ occurs on the square lattice, so the wavefunction is expanded as $\Psi(\bfx) \approx \psi_0 + \psi_1(\bfx) e^{i \Q_{sq} \cdot \bfx}$, \emph{i.e.} we consider condensation into the Fourier components $\q={\bf 0}, \Q_{sq}$. The symmetry group of the square lattice can be generated by the translations along $x,y$, the 4-fold rotation, a reflection with respect to either $x$ or $y$ axis, and inversion. The $\psi_0$ component is invariant under all symmetry operations, and the component $\psi_1$ only changes under the translations, where $T_{x,y} :\psi_1 = - \psi_1$. 

Constructing the most general LGW free energy as outlined above, thus yields for the square lattice (see Eq.~(\ref{eq:52}))
\begin{equation}
  \label{eq:113}
  \begin{split}
   & {\cal F}_{\text{sq},b} = \int d\bfx \Bigl\{ m_0 |\psi_0|^2 + m_1|\psi_1|^2 + |\partial \psi_0|^2 + v^2 |\partial \psi_1|^2 \\ &+ g_0 |\psi_0|^4 + g_1 |\psi_1|^4 + g_2 |\psi_0|^2 |\psi_1|^2 + u [ \psi_0^2 (\psi_1^2)^* + c.c. ] \Bigr\}\,.
  \end{split}
\end{equation}

\section{Details of analysis of the Landau-Ginzburg-Wilson free energy on triangular lattice}
\label{sec:deta-analys-land}
In this appendix, we present the details of the analysis of the LGW free energy on the isotropic triangular lattice. We begin by writing the matched parameters between the general LGW functional of Eq.~\eqref{eq:24} and the microscopically derived mean-field Hamiltonian of Eq.~\eqref{eq:26}. The quadratic coefficients (mass terms) read 
\begin{equation}
  \label{eq:114}
  \begin{split}
    m_0 &= \xi_b({\bf 0}) = - 6 t_b - \mu_b \\
    m_1 &= \xi_b(\Q_\alpha) = 2 t_b - \mu_b\,,
\end{split}
\end{equation}
and the interaction coefficients are given by
\begin{equation}
\begin{split}
    u_0 &= \frac{u}{2} \\
    u_1 &= \frac{u + g}{2} \\
    u_2 &= 2 u_1 = u +g \\
\end{split}
\end{equation}
as well as 
\begin{equation}
\begin{split}
    g_1 &= - g \\
    g_2 &= \frac{g}{2} \\
    g_3 &= 4 g_2 = 2 g \\
    g_4 &= g_2 = \frac{g}{2} \,.
  \end{split}
\end{equation}
The phase diagram is obtained from minimizing the free energy of Eq.~(\ref{eq:28}). Here, we will analytically investigate the free energy expresssion, and begin by identifying the dominant fourth order term. For this, we rewrite the quartic terms of Eq.~(\ref{eq:28}) as 
\begin{equation}
\label{eq:115}  
\begin{split}
&\frac12 (u + g {\cal W}) |\boldsymbol{\psi}|^4 = \frac12 (u - g)  |\boldsymbol{\psi}|^4  \\ &+ \frac{g}{2} \left[  - 2 \Theta_0 + 2 \sum_{i>j=0}^3 |\psi_i|^2 |\psi_j|^2 + 4 F_3 + \left|\sum_{i=0}^3 \psi_i^2\right|^2 \right]\,.
\end{split}
\end{equation}
For a fixed particle number, the first term is just a constant: $\frac12 (u-g) |\boldsymbol{\psi}|^4 = \frac12 (u-g) n_b^2$, and we have to analyze the remaining terms only using the most general form,
\begin{equation}
  \label{eq:116}
  \begin{split}
    \psi_j = r_j e^{i \phi_j}\,,
  \end{split}
\end{equation}
with $r_j \geq 0$, $\phi_0 = 0$, and $0 \leq \phi_j < 2 \pi$ for $j = 1,2,3$. 

Let us begin with $g > 0$. Numerically minimizing the last four terms of Eq.~(\ref{eq:115}), one finds that the global minimum occurs at
\begin{equation}
  \label{eq:117}
  r_0 = \sqrt{n_b}, r_1=r_2=r_3 = 0\,,
\end{equation}
where the last four terms sum up to $ (- g n_b/2)$.
Looking at the individual terms, we see that only the last two terms in the square brackets may favor the supersolid. We assume that the $F_3$ term is dominant due to the prefactor of 
$4$ multiplying it. We will find that this assumption is in accordance with the numerical result,
and leads to the same minimum configuration. The $F_3$ term is minimal when all $\psi_i$ are real and have the relative phases $\phi_0 = 0$, $\phi_{1,2,3} = \pi$, so that  
$2g \psi_0 (\psi_1 \psi_2^* \psi_3^* + \text{cyclic}) + c.c. = - 12 g r_0 r_1 r_2 r_3 $, which is minimized for $r_0=r_1=r_2=r_3= \sqrt{n_b}/4$. Writing more generally  $r_0 = \sqrt{n_b} \cos \theta$ and $r_1=r_2=r_3 = \sqrt{\frac{n_b}{3}} \sin \theta$ $(0\leq \theta\leq \pi/2)$, we find however, that both $ g\sum_{i>j} |\psi_i|^2 |\psi_j|^2 =3 g r_1^2 (r_0^2 + r_1^2) $, which favors $r_1$ to vanish, as well as $-g |\psi_0|^4$, which exclusively favors condensation into $\psi_0$, hinder a supersolid configuration independently of the phases $\{\phi_i\}$. The remaining term is just a constant for $\psi_i\in \mathbb{R}$. 
We conclude that \emph{for $g > 0$, the superfluid has a lower free energy than any supersolid phase}.
This is equivalent to say that a supersolid phase can only occur if $U(T,{\bf Q}_{1,2,3})<0$.

We therefore turn now to the case $g < 0$. Using again the general form of Eq.~(\ref{eq:116}), we numerically find that the global minimum of the last four terms of Eq.~(\ref{eq:115}) occurs for a supersolid configuration
\begin{equation}
  \label{eq:118}
\begin{split}
   r_0 &= \sqrt{n_b} \cos \theta_{min} \\ 
   r_1&=r_2=r_3 = \sqrt{\frac{n_b}{3}} \sin \theta_{min}\,,
\end{split}
\end{equation}
where $\theta_{\text{min}} \approx 1.09 \;\text{rad}$. The system tends to \emph{order in a symmetric way with respect to the three nesting fields} with the phases locked to the superfluid phase $\phi_{1,2,3}=0$, thus preserving the 3-fold rotational symmetry of the system. This is the Kagome-type order shown in Fig.~\ref{fig:4}. Note that $\theta_{\text{min}}$ denotes the lowest energy configuration of the fourth order term only. If we also consider the kinetic (quadratic) terms, the value of the supersolid angle $\theta$ is generally found to be much smaller, since it costs kinetic energy to add a boson with wavevector $\Q_{1,2,3}$ to the system.  

If we look at the free energy term by term, and making the (Kagome) ansatz  $r_0 = \sqrt{n_b} \cos \theta$, $r_{1,2,3}=  \sqrt{n_b/3} \, e^{i \phi} \sin \theta$, we find that now the last three terms of Eq.~(\ref{eq:115}) favor a supersolid configuration. First, $2 g F_3 = 12 g r_0 r_1^3 \cos \phi$ is minimal at $\phi=0$ and $r_0 = r_1$, but also the two terms that were previously, for $g > 0$, opposing the supersolid are now favoring it as well: $g \sum_{i > j} |\psi_i|^2 |\psi_j|^2 = 3 g r_1^2 (r_0^2 + r_1^2)$ wants both components to be nonzero, and $-g |\psi_0|^4$ now opposes condensation into the superfluid mode and hence favors nonzero $\psi_{1,2,3}$. 

We conclude that a \emph{Kagome supersolid phase occurs for a sufficiently negative $g<0$}. The energetically most favorable ordering is \emph{symmetric} with respect to the three nesting vectors with all the phases locked: $\psi_0 = \sqrt{n_b}\cos\theta$, $\psi_{1,2,3} = \sqrt{n_b/3}\;  \sin \theta$ with $0<\theta<\pi/2$. 

The phase boundary between the superfluid and the supersolid phase can be obtained as follows. Assume that the system is in the superfluid phase where the fermionic chemical potential is given by $\mu_b^{(\text{SF})} = - 6 t_b + n_b u$ and $\psi_0 = \sqrt{n_b}$, $\psi_{1,2,3} = 0$. Since the superfluid-supersolid transition is a transition between two ordered phases, we have to consider the influence of the fourth order terms onto the curvature in the $\psi_{1,2,3}$ directions by replacing $\psi_0 \rightarrow \sqrt{n_b}$ and examine the quadratic terms in $\psi_{1,2,3}$. 

We expect the transition to the supersolid phase to take place when the curvature in the $\psi_{1,2,3}$-direction, becomes negative. 
With $\psi_0 \rightarrow \sqrt{n_b}$, and using the knowledge that the minimal energy configuration is given by the field configuration of Eq.~(\ref{eq:118}), \emph{i.e.} setting $\phi = 0$, one finds the quadratic terms $m_{1}^{\text{eff}} (r_1^2 + r_2^2 + r_3^2)$ with 
\begin{equation}
  \label{eq:119}
\begin{split}
  m_{1}^{\text{eff}} =  -\mu_b + 2 t_b + n_b\,[2 g +u]\,.
\end{split}
\end{equation}
The mass $m_1^{\text{eff}}$ becomes negative at the critical chemical potential $\mu_{b}^{\text{crit}} = 2 t_b + n_b [u + 2 g ]$. Since $\mu_b^{(\text{SF})} = - 6 t_b + n_b u$ in the superfluid phase, this critical value is reached for $g_c^{(1)} = - \frac{4 t_b}{n_b}$. The condition of $m_1^{\text{eff}}$ changing sign at the superfluid-supersolid transition, then leads us to conclude that the bosonic system is superfluid for $g > g^{(1)}_c$ whereas it forms a Kagome-type supersolid for $g < g_c^{(1)}$. 

On the other hand, if we use the fact that chemical potentials are equal at the phase boundary: $\mu_b^{(\text{SF})}= \mu_b^{(\text{SS})}$, we find that the phase transition occurs already for larger values of $g$ at  
\begin{equation}
  \label{eq:120}
  g_c^{(2)} = - \frac{ 2 t_b}{n_b} > g_c^{(1)}\,. 
\end{equation}
Indeed, the free energy contains a third order term in $\psi_{1,2,3}$, which reads $12 g r_0 r_1 r_2 r_3 \rightarrow 4 g \sqrt{n_b} r_1^3$, where we have replaced $r_0 \rightarrow \sqrt{n_b}$ and used the Kagome-ansatz with $\phi= 0$ (see Eq.~(\ref{eq:118})).  For values of $g< g_c^{(2)}$, this  term leads to a local minimum at $r_{1,\text{min}} = \frac{- b + \sqrt{b^2 - 4 a c}}{2 c}$ with $a=8 t_b + 2 n_b g$, $b = 6 \sqrt{n_b} g$ and $c = 4 g + 3 u$.
This local minimum eventually becomes the global minimum at (see Eq.~\eqref{eq:34})
\begin{equation}
  \label{eq:121}
   g_c = - \frac{12 t_b u}{16 t_b + 3 n_b u}\,,
\end{equation}
which marks the superfluid-supersolid phase boundary. 

\section{Details of calculations for anisotropic lattices}
\label{sec:deta-calc-anis}
\subsection{Square lattice}
\subsubsection{Density of states}
\label{sec:square-lattice-2}
\begin{figure}[tb]
  \centering
  \includegraphics[width=.9\linewidth]{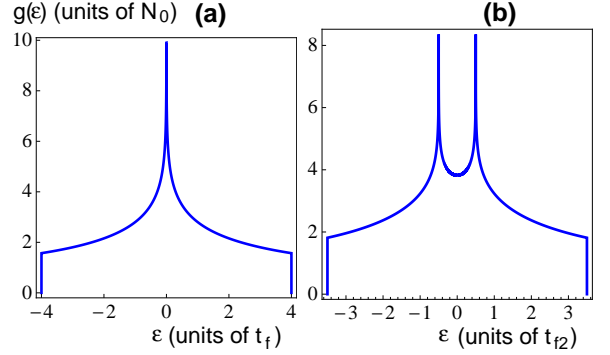}
  \caption{(Color online) Density of states for the isotropic (a) and anisotropic (b) square lattice. The anisotropy parameter in (b) is equal to $r_f = 0.75$.}
  \label{fig:20}
\end{figure}
The density of states for the isotropic and anisotropic cases is shown in Fig.~\ref{fig:20} and can be analytically calculated to be 
\begin{align}
  \label{eq:122}
  g(\tilde{\mu}_f ,r_f=1)  &= N_0 K \left[\sqrt{1 - \tilde{\mu}_f^2/4}\right]  \\
  \label{eq:123}
  g(\tilde{\mu}_f, r_f < 1) &= N_0 \, \frac{4 K\left[k_0\right] + 2 \left( F\left[a,k_2\right] + F\left[b,k_2\right] \right)}{\sqrt{ (r_f-1)^2 - \tilde{\mu}_f^2}}\,.
\end{align}
Here, $F[\phi,k]$ denotes the elliptic integral of the first kind and $K[k]$ denotes the complete elliptic integral of the first kind.  
The expression of Eq.~(\ref{eq:123}) is only valid in the region $-1+r_f < \tilde{\mu}_f < 1 - r_f$, and in particular is not valid in the limit $r_f\rightarrow 1$.

It contains the quantity $N_0 = 1/2 \pi^2 t_{f2}$, which is a measure of the density of states away from the logarithmic singularity at $\tilde{\mu}_f = \pm(1 - r_f)$. Also, for brevity we have defined the expressions $k_0 = \sqrt{4 r_f/[\tilde{\mu}_f^2 - (r_f-1)^2]}$, $k_2 = \sqrt{1 - k_1^2}$, where $k_1 = \sqrt{\frac{\tilde{\mu}_f^2 - (r_f+1)^2}{\tilde{\mu}_f^2 - (r_f-1)^2}}$, as well as $\tan a = \sqrt{\frac{\tilde{\mu}_f + r_f - 1 }{\tilde{\mu}_f - r_f - 1}}$, and $\tan b = \sqrt{\frac{\tilde{\mu}_f - r_f + 1}{\tilde{\mu}_f + r_f + 1}}$. 

Expanding the density of states around the singularity at $\tilde{\mu}_f = -1+r_f$, we find
\begin{align}
  \label{eq:124}
     g\left(\tilde{\mu}_f\simeq 0,r_f=1\right) &\simeq N_0 \ln \frac{8}{\tilde{\mu}_f} = N_0 \ln \frac{16 t_f}{\mu_f} \\
      g\left(\tilde{\mu}_f\simeq (r_f-1), r_f < 1\right) &\simeq N_0 \rho \ln \left( \frac{32 \vartheta r_f}{\tilde{\mu}_f - (r_f-1)}\right)\,,
\end{align}
for the isotropic and anisotropic lattice, respectively. Here, $\rho = 1/2 \sqrt{r_f}$ and $\vartheta = \frac{e^{-2/\sqrt{1-r_f}}}{(-1+\sqrt{2-r_f})^2}$. In contrast to the isotropic case, the anisotropic density of states is regular at $\tilde{\mu}_f = 0$ and equal to 
\begin{equation}
  \label{eq:125}
  g(\tilde{\mu}_f=0, r_f<1) = 2 N_0 K(r_f)\,.
\end{equation}
Note that this value is significantly increased due to the proximity of van Hove singularities at $\tilde{\mu}_f = \pm (1-r_f)$. This is not the case for a three-dimensional optical lattice, since the density of states is there regular everywhere (and divergences only occur in its first derivative). Therefore, supersolidity occurs at higher temperatures in the two-dimensional anisotropic square lattice than in the three-dimensional square lattice.

\subsubsection{Bosonic Mean-field Analysis}
\label{sec:bosonic-mean-field-1}
Here, we analyze the effective bosonic mean-field Hamiltonian on the square lattice (Eq.~(\ref{eq:52}))
\begin{equation}
  \label{eq:126}
  \frac{H_{\text{sq},b}^{\text{eff}}}{N_L} = \sum_{\alpha = 0}^1 \xi_b (\Q_\alpha) |\psi_\alpha|^2 + \frac12 (u + g {\cal W}_{sq}) |\boldsymbol{\psi}|^4\,,
\end{equation}
with
\begin{equation}
  \label{eq:127}
  {\cal W}_{\text{sq}} = \frac{(\psi_0 \psi_1^* + c.c.)^2}{(|\psi_0|^2 + |\psi_1|^2)^2} = \sin^2(2\theta) \cos^2(\phi),
\end{equation}
where $\boldsymbol{\psi} = (\sqrt{n_b} \cos \theta, \sqrt{n_b} e^{i \phi} \sin \theta)$ and $\Q_0 = {\bf 0}$, $\Q_1 = \Q_{\text{sq}}$. 
The coefficients of the second order term read $m_0 = \xi_b({\bf 0}) = - 2(t_{b1} + t_{b2}) - \mu_b$ and $m_1 = \xi_b(\Q_1) = 2(t_{b1} + t_{b2}) - \mu_b$, so from purely kinetic considerations, the system tends to condense solely into the superfluid mode $\psi_0$. However, interactions described by the parameters $(u,g)$ can alter the situation, and we therefore minimize the zero temperature free energy density $f_b(\theta,\phi) = H_{\text{sq},b}^{\text{eff}}/N_L$, as a function of the angles $(\theta,\phi)$. The equation 
\begin{equation}
  \label{eq:128}
  \partial_\phi f_b = - g n_b^2 \sin^2(2 \theta) \sin (2 \phi) = 0\,,
\end{equation}
can be fulfilled for three distinct cases. For $\theta = 0, \pi/2$ and arbitrary $\phi$, which corresponds to a superfluid (SF) or a pure density wave (DW) phase, respectively. The third possibility is that $\phi = n \pi$ with integer $n$ and arbitrary $\theta$, which allows for the supersolid case of $0 < \theta < \pi/2$. At a minimum, it is also required that 
\begin{equation}
  \label{eq:129}
\begin{split}
  \partial_\theta f_b &= n_b \sin 2 \theta \left( 4 (t_{b1} + t_{b2} ) + n_b g \cos 2 \theta \right) \\ & \quad + \frac{g}{2} n_b^2 \sin 4 \theta \cos 2 \phi = 0 \,,
\end{split}
\end{equation}
which can again be fulfilled by $\theta = 0, \pi/2$ (SF, DW) for arbitrary $\phi$, or $\phi = n \pi$ and $\cos^2 \theta_{\text{SS}} = \frac{n_b g - 2 (t_{b1} + t_{b2})}{2n_b g}$, which corresponds to a supersolid, if $0 < \cos^2 \theta < 1$. 

If we compare the energy of the three cases, we immediately find that \emph{the DW always has larger energy than the superfluid} $ f_b (\theta=\pi/2,\phi) -f_b(\theta=0,\phi)  = 4 n_b (t_{b1} + t_{b2}) > 0$. However, if we compare the energy of the superfluid with the supersolid, we find that \emph{the global minimum can occur at $\theta_{\text{SS}}$ only for sufficiently negative $g < 0$}. Specifically,
\begin{equation}
  \label{eq:130}
  f_b(0,\phi)-f_b(\theta_{\text{SS}},n\pi) = - \frac{(n_b g + 2 (t_{b1} + t_{b2}))^2}{2 g}\,,
\end{equation}
is positive for negative $g$. However, in order for the analysis to be self-consistent, it is required that $1 \geq \cos^2 \theta_{\text{SS}} \geq 0$, which for $g<0$ requires that $g \leq - 2(t_{b1} + t_{b2})/n_b$. The superfluid to supersolid phase boundary, which is defined by $f_b(0,\phi) = f_b(\theta_{\text{SS}},n\pi)$, occurs at the critical interaction strength
\begin{equation}
  \label{eq:131}
  g_{\text{sq},c}  = - \frac{2 (t_{b1} + t_{b2})}{n_b}\,,
\end{equation}
with the supersolid occuring for $g \leq g_{\text{sq},c}$.
\subsubsection{Fermionic Mean-field analysis}
\label{sec:fermionic-mean-field}
We start from the Hamiltonian $H = H_f^{(1)} + H_f^{(2)} + H_f^{(3)}$ similar to Eq.~(\ref{eq:38}), where the bosonic operators $b_{{\bf 0}, \Q_1}$ are replaced by the complex fields $\psi_{0,1}$:
\begin{equation}
  \label{eq:132}
  \begin{split}
&    \frac{H_f^{(1)}}{N_L} = 4 (t_{b1} + t_{b2} ) |\psi_1|^2 + \frac{U_{bb}}{2} \left[ n_b^2 + (\psi_0 \psi_1^* + c.c.)^2 \right] \\
  &  H_f^{(2)} + H_f^{(3)} = \sideset{}{'}{\sum}_{\bfk} \sum_{\alpha,\beta} f^\dag_{\bfk + \Q_\alpha} h_{\alpha \beta} f_{\bfk + \Q_\beta} \\
&  \left(h_{\alpha \beta} \right) = \begin{pmatrix} \xi_f(\bfk) & U_{bf} (\psi_0 \psi_1^* + c.c.) \\ U_{bf} (\psi_0 \psi_1^* + c.c.) & - \xi_f(\bfk) \end{pmatrix}\,,
    \end{split}
\end{equation}
where the nesting relation $\xi_f(\bfk + \Q_1) = - \xi_f(\bfk)$ was used, and the sum over wavevectors is restricted to $1/2$ of the first Brillouin zone. With $\psi_0 = r_0$, $\psi_1 = r_1 \exp(i \phi)$ and defining $\Delta = 2 U_{bf} r_0 r_1$, the fermionic eigenenergies read 
\begin{equation}
  \label{eq:133}
  \Xi(\bfk, \Delta)_\pm = \pm \sqrt{\xi_f(\bfk)^2 + \Delta^2 \cos^2 \phi}\,.
\end{equation}
We identify $\Delta$ as the emerging gap in the fermionic spectrum.
\subsection{Triangular lattice}
\label{sec:triangular-lattice}
Here, we analyze the effective Hamiltonian of Eq.~\eqref{eq:63}
\begin{equation}
  \label{eq:134}
\begin{split}
  \frac{H_b^{\text{eff}}}{N_L} &= m_0 |\psi_0|^2 + m_1 \left(|\psi_1|^2 + |\psi_2|^2 \right) + m_3 |\psi_3|^2 \\
  &+ \frac12 (u + v {\cal V} + g {\cal W} ) |\boldsymbol{\psi}|^4\,,
\end{split}
\end{equation}
where $m_i = \xi_b(\Q_i)$, $u = U(\bf 0)$, $v = U(\Q_{1,2})$, $g=U(\Q_3)$ and 
\begin{equation}
  \label{eq:135}
  \begin{split}
    {\cal V} |\boldsymbol{\psi}|^4 &= 2 \left[ |\psi_0|^2 (|\psi_1|^2 + |\psi_{2}|^2) + |\psi_3|^2 (|\psi_1|^2 + |\psi_2|^2 ) \right] \\ &+ [ \psi_0 \psi_1 \psi_2^* \psi_3^* + \psi_0^* \psi_1 \psi_2^* \psi_3 + 2 \psi_0 \psi_1^* \psi_2^* \psi_3 + c.c. ] \\ &+ ( \psi_0^2 + \psi_3^2) (\psi_1^2 + \psi_2^2)^* \\
    {\cal W} |\boldsymbol{\psi}|^4 &= (\psi_0 \psi_3^* + \psi_1 \psi_2^* + c.c.)^2\,.
  \end{split}
\end{equation}
Since $0 \leq {\cal V} \leq 2$ and $0 \leq {\cal W} \leq 1$, \emph{stability requires} $u + 2 v \geq |g|$ for $u,v > 0, g < 0$, where we anticipate that the supersolid only occurs for negative $g$. The Lindhard function is regular at $\Q_{1,2}$, such that $v > u$. In particular, $v$ is positive, since with $\chi(T,\Q_{1,2}) \approx -M_0$, one finds $v = U_{bb} + U_{bf}^2 \chi(T, \Q_{1,2}) > 0$, because $M_0 U_{bf}^2/U_{bb} < 1$. 

The masses read explicitly $m_0 = - \mu_b - 2(t_{b1} + t_{b2})$, $m_1 = -\mu_b + 2 t_{b1}$ and $m_3 = -\mu_b - 2(t_{b1} - 2 t_{b2})$, so from purely kinetic energy considerations, the system preferably condenses into the $\psi_0$-mode only, \emph{i.e.} is superfluid. 

With this is mind, we turn to analyze the interaction terms: as $v>0$ it is energetically favorable to minimize ${\cal V}$, which is achieved by either having the pair of fields $(\psi_0,\psi_3)$ vanish or the pair $(\psi_1, \psi_2 )$. For positive $g>0$, also the ${\cal W}$-term is minimized for either $\psi_0=\psi_3=0$ or $\psi_1=\psi_2 = 0$. As a result, kinetic energy considerations will select the \emph{superfluid state for positive $g$}. 

Let us turn to the case of negative $g$. We observe that in the ${\cal W}$-term, the field $\psi_3$ couples to the superfluid component $\psi_0$. The system can therefore, possibly, lower its energy, compared to the superfluid, by allowing for nonzero $\psi_3$ while still having $\psi_1=\psi_2=0$ such that ${\cal V}=0$. In this subspace of possible field values, the mean-field Hamiltonian is of the same form as on the square lattice (see Sec.~\ref{sec:bosonic-mean-field-2}). 
\section{Details of derivation of quantum Heisenberg Hamiltonian}
\label{sec:deriv-quant-heis}
We start from the purely fermionic Hamiltonian of Eq.~\eqref{eq:80}
\begin{equation}
  \label{eq:136}
  \begin{split}
    H_b &= -t_b \sum_{\av{i,j}} \left[ c_i^\dag c_j e^{i A_{ij}} + \text{h.c.} \right] - \mu_b \sum_i N_i\\
    H_f &= - t_f  \sum_{\av{i,j}} \left( f_i^\dag f_j + \text{h.c.} \right) - \mu_f \sum_i m_i\\
    H_{bf} &= U_{bf} \sum_i m_i N_i\,,
  \end{split}
\end{equation}
where $A_{ij} =  \sum_{p\neq i,j} \left( \theta_{pj} - \theta_{pi} \right) N_p$. Except for the additional (gauge) field $A_{ij}$, this is the Hamiltonian of the two-dimensional spin-1/2 fermionic Hubbard model, where $f_i^\dag (c_i^\dag)$ creates a spin-up (down) fermion at site $i$, and the boson-fermion interaction $U_{bf}$ marks the on-site interaction. 

In the case of unit filling, the ground states of the zeroth order (in $t_{f,b}/U_{bf}$) Hamiltonian ${\cal U} = U_{bf} \sum_i m_i N_i$ are states where each site is occupied by exactly one particle, either a fermion or a boson. Since all other states involve at least one doubly (or multiply) occupied site, one can divide the Fock space into ${\cal F} = {\cal S}  \otimes {\cal D}$, where ${\cal S}$ is the degenerate ground state manifold of ${\cal U}$, and ${\cal D}$ contains all other states. Defining the projection operators onto the two subspaces $P_S$ and $P_D$, one can partition the Hamiltonian $ H = H_M -\sum_i (\mu_b N_i + \mu_f m_i)$ into 
\begin{equation}
\label{eq:137}
H_M = \begin{pmatrix} P_S ({\cal T}+ {\cal U}) P_S & P_S {\cal T} P_D \\ P_D {\cal T} P_S & P_D ({\cal T} + {\cal U}) P_D \end{pmatrix}
\end{equation}
where ${\cal T} = {\cal T}_b + {\cal T}_f = - \sum_{\av{i,j}} ( t_b  c_i^\dag c_j e^{i A_{ij}} + t_f f_i^\dag f_j +  \text{h.c.} )$ contains the hopping terms, that we will treat in perturbation theory, and $H_M$ denotes the Hamiltonian for a fixed particle number difference (or magnetization). The effective Hamiltonian $H_M^{\text{eff}}$, acting in the ground state manifold ${\cal S}$ only, can be obtained by
\begin{equation}
  \label{eq:138}
  P_S (E - H)^{-1} P_S = [E - H_M^{\text{eff}}(E)]^{-1}\,.
\end{equation}
 Using $\left(\begin{smallmatrix} A & B \\ C & D \end{smallmatrix}\right)^{-1} = ( A - B D^{-1} C)^{-1}$, one finds the effective Hamiltonian
 \begin{equation}
   \label{eq:139}
\begin{split}
   &H_M^{\text{eff}} =  P_S {\cal T} P_S + P_S {\cal T} [P_D [ E - ({\cal T} + {\cal U} ) ] P_D]^{-1} {\cal T} P_S \\
   &= -\frac{P_S {\cal T} P_D}{U_{bf}}  \sum_i m_i N_i P_D {\cal T} P_s + {\cal O}(t_{b,f}^3/U^2_{bf}, E/U_{bf}^2)\,,
\end{split}
 \end{equation}
where we have used that $P_S {\cal T} P_S = 0$ for unit filling, and we have expanded to lowest non-trivial order in $t_{f,b}/U_{bf}$. For ${\cal T} = {\cal T}_b + {\cal T}_f$, we obtain four terms $H_M^{\text{eff}} = A + B + C + D$, which read 
\begin{align}
  \label{eq:140}
    A &= - \frac{t_b^2}{U_{bf}} \sum_{\av{\alpha,\beta}} c_\alpha^\dag c_\beta c_\beta^\dag c_\alpha \\
  B &= - \frac{t_b t_f}{U_{bf}} \sum_{\av{\alpha,\beta}} c_\alpha^\dag c_\beta e^{i A_{\alpha\beta}} f_\beta^\dag f_\alpha \\
    C &= - \frac{t_b t_f}{U_{bf}} \sum_{\av{\alpha,\beta}} f_\alpha^\dag f_\beta c_\beta^\dag c_\alpha e^{i A_{\beta\alpha}} \\
    D &= - \frac{t_f^2}{U_{bf}} \sum_{\av{\alpha,\beta}} f_\alpha^\dag f_\beta f_\beta^\dag f_\alpha\,.
\end{align}
Defining proper spin operators via $S_\alpha^+ = f_\alpha^\dag c_\alpha$, $S_\alpha^z = (m_\alpha - N_\alpha)/2$, we arrive at Eq.~\eqref{eq:81}:
\begin{equation}
  \label{eq:141}
  \begin{split}
    H^{\text{eff}} &= \frac12 \sum_{\av{\alpha,\beta}} \Big\{ \frac{2 t_b t_f}{U_{bf}} \left[S_\alpha^+ S_\beta^- e^{i A_{\beta\alpha}} + S_\alpha^- S_\beta^+ e^{i A_{\alpha\beta}} \right] \\
        & + \frac{2(t_b^2 + t_f^2)}{U_{bf}} \Big[S_\alpha^z S_\beta^z - \frac14 \Big] \Big\}  - (\mu_f - \mu_b) \sum_\alpha S^z_\alpha\,.
  \end{split}
\end{equation}
Here, spin-up corresponds to occupation by a fermion and spin-down to occupation by a boson. This Hamiltonian is of the familiar form of the spin-1/2 quantum Heisenberg Hamiltonian, however, it contains the additional Jordan-Wigner gauge field $A_{ij}$ reflecting the different symmetry of hard-core bosons and spinless fermions.


\begin{thebibliography}{97}
\expandafter\ifx\csname natexlab\endcsname\relax\def\natexlab#1{#1}\fi
\expandafter\ifx\csname bibnamefont\endcsname\relax
  \def\bibnamefont#1{#1}\fi
\expandafter\ifx\csname bibfnamefont\endcsname\relax
  \def\bibfnamefont#1{#1}\fi
\expandafter\ifx\csname citenamefont\endcsname\relax
  \def\citenamefont#1{#1}\fi
\expandafter\ifx\csname url\endcsname\relax
  \def\url#1{\texttt{#1}}\fi
\expandafter\ifx\csname urlprefix\endcsname\relax\def\urlprefix{URL }\fi
\providecommand{\bibinfo}[2]{#2}
\providecommand{\eprint}[2][]{\url{#2}}

\bibitem[{\citenamefont{Andreev and
  Lifshitz}(1969)}]{andreev_lifshitz_supersolid}
\bibinfo{author}{\bibfnamefont{A.~F.} \bibnamefont{Andreev}} \bibnamefont{and}
  \bibinfo{author}{\bibfnamefont{I.~M.} \bibnamefont{Lifshitz}},
  \bibinfo{journal}{Sov. Phys. JETP} \textbf{\bibinfo{volume}{29}},
  \bibinfo{pages}{1107} (\bibinfo{year}{1969}).

\bibitem[{\citenamefont{Leggett}(1970)}]{PhysRevLett.25.1543}
\bibinfo{author}{\bibfnamefont{A.~J.} \bibnamefont{Leggett}},
  \bibinfo{journal}{Phys. Rev. Lett.} \textbf{\bibinfo{volume}{25}},
  \bibinfo{pages}{1543} (\bibinfo{year}{1970}).

\bibitem[{\citenamefont{Kim and Chan}(2004)}]{doi:10.1038/nature02220}
\bibinfo{author}{\bibfnamefont{E.}~\bibnamefont{Kim}} \bibnamefont{and}
  \bibinfo{author}{\bibfnamefont{M.}~\bibnamefont{Chan}},
  \bibinfo{journal}{Nature (London)} \textbf{\bibinfo{volume}{427}},
  \bibinfo{pages}{225} (\bibinfo{year}{2004}).

\bibitem[{\citenamefont{Rittner et~al.}(2009)\citenamefont{Rittner, Choi,
  Mueller, and Reppy}}]{arXiv:0904.2640v1}
\bibinfo{author}{\bibfnamefont{A.~S.~C.} \bibnamefont{Rittner}},
  \bibinfo{author}{\bibfnamefont{W.}~\bibnamefont{Choi}},
  \bibinfo{author}{\bibfnamefont{E.~J.} \bibnamefont{Mueller}},
  \bibnamefont{and} \bibinfo{author}{\bibfnamefont{J.~D.} \bibnamefont{Reppy}},
  \bibinfo{journal}{arXiv:0904.2640v1}  (\bibinfo{year}{2009}).

\bibitem[{\citenamefont{Prokofev}(2007)}]{prokofev_supersolid_he}
\bibinfo{author}{\bibfnamefont{N.}~\bibnamefont{Prokofev}},
  \bibinfo{journal}{Adv. Phys.} \textbf{\bibinfo{volume}{56}},
  \bibinfo{pages}{381} (\bibinfo{year}{2007}).

\bibitem[{\citenamefont{Prokofev and Svistunov}(2005)}]{PhysRevLett.94.155302}
\bibinfo{author}{\bibfnamefont{N.}~\bibnamefont{Prokofev}} \bibnamefont{and}
  \bibinfo{author}{\bibfnamefont{B.}~\bibnamefont{Svistunov}},
  \bibinfo{journal}{Phys. Rev. Lett.} \textbf{\bibinfo{volume}{94}},
  \bibinfo{pages}{155302} (\bibinfo{year}{2005}).

\bibitem[{\citenamefont{Fisher et~al.}(1989)\citenamefont{Fisher, Weichman,
  Grinstein, and Fisher}}]{PhysRevB.40.546}
\bibinfo{author}{\bibfnamefont{M.~P.~A.} \bibnamefont{Fisher}},
  \bibinfo{author}{\bibfnamefont{P.~B.} \bibnamefont{Weichman}},
  \bibinfo{author}{\bibfnamefont{G.}~\bibnamefont{Grinstein}},
  \bibnamefont{and} \bibinfo{author}{\bibfnamefont{D.~S.}
  \bibnamefont{Fisher}}, \bibinfo{journal}{Phys. Rev. B}
  \textbf{\bibinfo{volume}{40}}, \bibinfo{pages}{546} (\bibinfo{year}{1989}).

\bibitem[{\citenamefont{Jaksch et~al.}(1998)\citenamefont{Jaksch, Bruder,
  Cirac, Gardiner, and Zoller}}]{PhysRevLett.81.3108}
\bibinfo{author}{\bibfnamefont{D.}~\bibnamefont{Jaksch}},
  \bibinfo{author}{\bibfnamefont{C.}~\bibnamefont{Bruder}},
  \bibinfo{author}{\bibfnamefont{J.~I.} \bibnamefont{Cirac}},
  \bibinfo{author}{\bibfnamefont{C.~W.} \bibnamefont{Gardiner}},
  \bibnamefont{and} \bibinfo{author}{\bibfnamefont{P.}~\bibnamefont{Zoller}},
  \bibinfo{journal}{Phys. Rev. Lett.} \textbf{\bibinfo{volume}{81}},
  \bibinfo{pages}{3108} (\bibinfo{year}{1998}).

\bibitem[{\citenamefont{Albus et~al.}(2003)\citenamefont{Albus, Illuminati, and
  Eisert}}]{PhysRevA.68.023606}
\bibinfo{author}{\bibfnamefont{A.}~\bibnamefont{Albus}},
  \bibinfo{author}{\bibfnamefont{F.}~\bibnamefont{Illuminati}},
  \bibnamefont{and} \bibinfo{author}{\bibfnamefont{J.}~\bibnamefont{Eisert}},
  \bibinfo{journal}{Phys. Rev. A} \textbf{\bibinfo{volume}{68}},
  \bibinfo{pages}{023606} (\bibinfo{year}{2003}).

\bibitem[{\citenamefont{Cramer et~al.}(2004)\citenamefont{Cramer, Eisert, and
  Illuminati}}]{PhysRevLett.93.190405}
\bibinfo{author}{\bibfnamefont{M.}~\bibnamefont{Cramer}},
  \bibinfo{author}{\bibfnamefont{J.}~\bibnamefont{Eisert}}, \bibnamefont{and}
  \bibinfo{author}{\bibfnamefont{F.}~\bibnamefont{Illuminati}},
  \bibinfo{journal}{Phys. Rev. Lett.} \textbf{\bibinfo{volume}{93}},
  \bibinfo{pages}{190405} (\bibinfo{year}{2004}).

\bibitem[{\citenamefont{Greiner et~al.}(2002)\citenamefont{Greiner, Mandel,
  Esslinger, H\"ansch, and Bloch}}]{greiner_nature_2002}
\bibinfo{author}{\bibfnamefont{M.}~\bibnamefont{Greiner}},
  \bibinfo{author}{\bibfnamefont{O.}~\bibnamefont{Mandel}},
  \bibinfo{author}{\bibfnamefont{T.}~\bibnamefont{Esslinger}},
  \bibinfo{author}{\bibfnamefont{T.~W.} \bibnamefont{H\"ansch}},
  \bibnamefont{and} \bibinfo{author}{\bibfnamefont{I.}~\bibnamefont{Bloch}},
  \bibinfo{journal}{Nature} \textbf{\bibinfo{volume}{415}}, \bibinfo{pages}{39}
  (\bibinfo{year}{2002}).

\bibitem[{\citenamefont{Bloch et~al.}(2008)\citenamefont{Bloch, Dalibard, and
  Zwerger}}]{bloch:885}
\bibinfo{author}{\bibfnamefont{I.}~\bibnamefont{Bloch}},
  \bibinfo{author}{\bibfnamefont{J.}~\bibnamefont{Dalibard}}, \bibnamefont{and}
  \bibinfo{author}{\bibfnamefont{W.}~\bibnamefont{Zwerger}},
  \bibinfo{journal}{Rev. Mod. Phys.} \textbf{\bibinfo{volume}{80}},
  \bibinfo{eid}{885} (\bibinfo{year}{2008}).

\bibitem[{\citenamefont{Jaksch and Zoller}(2005)}]{Jaksch200552}
\bibinfo{author}{\bibfnamefont{D.}~\bibnamefont{Jaksch}} \bibnamefont{and}
  \bibinfo{author}{\bibfnamefont{P.}~\bibnamefont{Zoller}},
  \bibinfo{journal}{Ann. Phys.} \textbf{\bibinfo{volume}{315}},
  \bibinfo{pages}{52} (\bibinfo{year}{2005}).

\bibitem[{\citenamefont{Lewenstein et~al.}(2004)\citenamefont{Lewenstein,
  Santos, Baranov, and Fehrmann}}]{PhysRevLett.92.050401}
\bibinfo{author}{\bibfnamefont{M.}~\bibnamefont{Lewenstein}},
  \bibinfo{author}{\bibfnamefont{L.}~\bibnamefont{Santos}},
  \bibinfo{author}{\bibfnamefont{M.~A.} \bibnamefont{Baranov}},
  \bibnamefont{and} \bibinfo{author}{\bibfnamefont{H.}~\bibnamefont{Fehrmann}},
  \bibinfo{journal}{Phys. Rev. Lett.} \textbf{\bibinfo{volume}{92}},
  \bibinfo{pages}{050401} (\bibinfo{year}{2004}).

\bibitem[{\citenamefont{Schreck et~al.}(2001)\citenamefont{Schreck, Khaykovich,
  Corwin, Ferrari, Bourdel, Cubizolles, and Salomon}}]{PhysRevLett.87.080403}
\bibinfo{author}{\bibfnamefont{F.}~\bibnamefont{Schreck}},
  \bibinfo{author}{\bibfnamefont{L.}~\bibnamefont{Khaykovich}},
  \bibinfo{author}{\bibfnamefont{K.~L.} \bibnamefont{Corwin}},
  \bibinfo{author}{\bibfnamefont{G.}~\bibnamefont{Ferrari}},
  \bibinfo{author}{\bibfnamefont{T.}~\bibnamefont{Bourdel}},
  \bibinfo{author}{\bibfnamefont{J.}~\bibnamefont{Cubizolles}},
  \bibnamefont{and} \bibinfo{author}{\bibfnamefont{C.}~\bibnamefont{Salomon}},
  \bibinfo{journal}{Phys. Rev. Lett.} \textbf{\bibinfo{volume}{87}},
  \bibinfo{pages}{080403} (\bibinfo{year}{2001}).

\bibitem[{\citenamefont{Ferlaino et~al.}(2004)\citenamefont{Ferlaino,
  de~Mirandes, Roati, Modugno, and Inguscio}}]{PhysRevLett.92.140405}
\bibinfo{author}{\bibfnamefont{F.}~\bibnamefont{Ferlaino}},
  \bibinfo{author}{\bibfnamefont{E.}~\bibnamefont{de~Mirandes}},
  \bibinfo{author}{\bibfnamefont{G.}~\bibnamefont{Roati}},
  \bibinfo{author}{\bibfnamefont{G.}~\bibnamefont{Modugno}}, \bibnamefont{and}
  \bibinfo{author}{\bibfnamefont{M.}~\bibnamefont{Inguscio}},
  \bibinfo{journal}{Phys. Rev. Lett.} \textbf{\bibinfo{volume}{92}},
  \bibinfo{pages}{140405} (\bibinfo{year}{2004}).

\bibitem[{\citenamefont{Zaccanti et~al.}(2006)\citenamefont{Zaccanti, D'Errico,
  Ferlaino, Roati, Inguscio, and Modugno}}]{zaccanti:041605}
\bibinfo{author}{\bibfnamefont{M.}~\bibnamefont{Zaccanti}},
  \bibinfo{author}{\bibfnamefont{C.}~\bibnamefont{D'Errico}},
  \bibinfo{author}{\bibfnamefont{F.}~\bibnamefont{Ferlaino}},
  \bibinfo{author}{\bibfnamefont{G.}~\bibnamefont{Roati}},
  \bibinfo{author}{\bibfnamefont{M.}~\bibnamefont{Inguscio}}, \bibnamefont{and}
  \bibinfo{author}{\bibfnamefont{G.}~\bibnamefont{Modugno}},
  \bibinfo{journal}{Phys. Rev. A} \textbf{\bibinfo{volume}{74}},
  \bibinfo{eid}{041605(R)} (\bibinfo{year}{2006}).

\bibitem[{\citenamefont{Ospelkaus
  et~al.}(2006{\natexlab{a}})\citenamefont{Ospelkaus, Ospelkaus, Sengstock, and
  Bongs}}]{ospelkaus:020401}
\bibinfo{author}{\bibfnamefont{C.}~\bibnamefont{Ospelkaus}},
  \bibinfo{author}{\bibfnamefont{S.}~\bibnamefont{Ospelkaus}},
  \bibinfo{author}{\bibfnamefont{K.}~\bibnamefont{Sengstock}},
  \bibnamefont{and} \bibinfo{author}{\bibfnamefont{K.}~\bibnamefont{Bongs}},
  \bibinfo{journal}{Phys. Rev. Lett.} \textbf{\bibinfo{volume}{96}},
  \bibinfo{eid}{020401} (\bibinfo{year}{2006}{\natexlab{a}}).

\bibitem[{\citenamefont{Ospelkaus
  et~al.}(2006{\natexlab{b}})\citenamefont{Ospelkaus, Ospelkaus, Wille, Succo,
  Ernst, Sengstock, and Bongs}}]{ospelkaus:180403}
\bibinfo{author}{\bibfnamefont{S.}~\bibnamefont{Ospelkaus}},
  \bibinfo{author}{\bibfnamefont{C.}~\bibnamefont{Ospelkaus}},
  \bibinfo{author}{\bibfnamefont{O.}~\bibnamefont{Wille}},
  \bibinfo{author}{\bibfnamefont{M.}~\bibnamefont{Succo}},
  \bibinfo{author}{\bibfnamefont{P.}~\bibnamefont{Ernst}},
  \bibinfo{author}{\bibfnamefont{K.}~\bibnamefont{Sengstock}},
  \bibnamefont{and} \bibinfo{author}{\bibfnamefont{K.}~\bibnamefont{Bongs}},
  \bibinfo{journal}{Phys. Rev. Lett.} \textbf{\bibinfo{volume}{96}},
  \bibinfo{eid}{180403} (\bibinfo{year}{2006}{\natexlab{b}}).

\bibitem[{\citenamefont{Ospelkaus
  et~al.}(2006{\natexlab{c}})\citenamefont{Ospelkaus, Ospelkaus, Humbert,
  Sengstock, and Bongs}}]{ospelkaus:120403}
\bibinfo{author}{\bibfnamefont{S.}~\bibnamefont{Ospelkaus}},
  \bibinfo{author}{\bibfnamefont{C.}~\bibnamefont{Ospelkaus}},
  \bibinfo{author}{\bibfnamefont{L.}~\bibnamefont{Humbert}},
  \bibinfo{author}{\bibfnamefont{K.}~\bibnamefont{Sengstock}},
  \bibnamefont{and} \bibinfo{author}{\bibfnamefont{K.}~\bibnamefont{Bongs}},
  \bibinfo{journal}{Phys. Rev. Lett.} \textbf{\bibinfo{volume}{97}},
  \bibinfo{eid}{120403} (\bibinfo{year}{2006}{\natexlab{c}}).

\bibitem[{\citenamefont{B\"uchler and Blatter}(2003)}]{PhysRevLett.91.130404}
\bibinfo{author}{\bibfnamefont{H.~P.} \bibnamefont{B\"uchler}}
  \bibnamefont{and} \bibinfo{author}{\bibfnamefont{G.}~\bibnamefont{Blatter}},
  \bibinfo{journal}{Phys. Rev. Lett.} \textbf{\bibinfo{volume}{91}},
  \bibinfo{pages}{130404} (\bibinfo{year}{2003}).

\bibitem[{\citenamefont{Scarola and Das~Sarma}(2005)}]{PhysRevLett.95.033003}
\bibinfo{author}{\bibfnamefont{V.~W.} \bibnamefont{Scarola}} \bibnamefont{and}
  \bibinfo{author}{\bibfnamefont{S.}~\bibnamefont{Das~Sarma}},
  \bibinfo{journal}{Phys. Rev. Lett.} \textbf{\bibinfo{volume}{95}},
  \bibinfo{pages}{033003} (\bibinfo{year}{2005}).

\bibitem[{\citenamefont{Keilmann et~al.}(2009)\citenamefont{Keilmann, Cirac,
  and Roscilde}}]{keilmann:255304}
\bibinfo{author}{\bibfnamefont{T.}~\bibnamefont{Keilmann}},
  \bibinfo{author}{\bibfnamefont{I.}~\bibnamefont{Cirac}}, \bibnamefont{and}
  \bibinfo{author}{\bibfnamefont{T.}~\bibnamefont{Roscilde}},
  \bibinfo{journal}{Phys. Rev. Lett.} \textbf{\bibinfo{volume}{102}},
  \bibinfo{eid}{255304} (\bibinfo{year}{2009}).

\bibitem[{\citenamefont{Vengalattore et~al.}(2008)\citenamefont{Vengalattore,
  Leslie, Guzman, and Stamper-Kurn}}]{vengalattore:170403}
\bibinfo{author}{\bibfnamefont{M.}~\bibnamefont{Vengalattore}},
  \bibinfo{author}{\bibfnamefont{S.~R.} \bibnamefont{Leslie}},
  \bibinfo{author}{\bibfnamefont{J.}~\bibnamefont{Guzman}}, \bibnamefont{and}
  \bibinfo{author}{\bibfnamefont{D.~M.} \bibnamefont{Stamper-Kurn}},
  \bibinfo{journal}{Phys. Rev. Lett.} \textbf{\bibinfo{volume}{100}},
  \bibinfo{eid}{170403} (\bibinfo{year}{2008}).

\bibitem[{\citenamefont{Cherng and Demler}(2008)}]{arXiv:0806.1991v1}
\bibinfo{author}{\bibfnamefont{R.~W.} \bibnamefont{Cherng}} \bibnamefont{and}
  \bibinfo{author}{\bibfnamefont{E.}~\bibnamefont{Demler}},
  \bibinfo{journal}{arXiv:0806.1991v1 [cond-mat.other]}
  (\bibinfo{year}{2008}).

\bibitem[{\citenamefont{Scarola et~al.}(2006)\citenamefont{Scarola, Demler, and
  Das~Sarma}}]{ScarolaSS}
\bibinfo{author}{\bibfnamefont{V.~W.} \bibnamefont{Scarola}},
  \bibinfo{author}{\bibfnamefont{E.}~\bibnamefont{Demler}}, \bibnamefont{and}
  \bibinfo{author}{\bibfnamefont{S.}~\bibnamefont{Das~Sarma}},
  \bibinfo{journal}{Phys. Rev. A} \textbf{\bibinfo{volume}{73}},
  \bibinfo{pages}{051601(R)} (\bibinfo{year}{2006}).

\bibitem[{\citenamefont{Sengupta et~al.}(2005)\citenamefont{Sengupta, Pryadko,
  Alet, Troyer, and Schmid}}]{PhysRevLett.94.207202}
\bibinfo{author}{\bibfnamefont{P.}~\bibnamefont{Sengupta}},
  \bibinfo{author}{\bibfnamefont{L.~P.} \bibnamefont{Pryadko}},
  \bibinfo{author}{\bibfnamefont{F.}~\bibnamefont{Alet}},
  \bibinfo{author}{\bibfnamefont{M.}~\bibnamefont{Troyer}}, \bibnamefont{and}
  \bibinfo{author}{\bibfnamefont{G.}~\bibnamefont{Schmid}},
  \bibinfo{journal}{Phys. Rev. Lett.} \textbf{\bibinfo{volume}{94}},
  \bibinfo{pages}{207202} (\bibinfo{year}{2005}).

\bibitem[{\citenamefont{Wessel and Troyer}(2005)}]{wessel:127205}
\bibinfo{author}{\bibfnamefont{S.}~\bibnamefont{Wessel}} \bibnamefont{and}
  \bibinfo{author}{\bibfnamefont{M.}~\bibnamefont{Troyer}},
  \bibinfo{journal}{Phys. Rev. Lett.} \textbf{\bibinfo{volume}{95}},
  \bibinfo{eid}{127205} (\bibinfo{year}{2005}).

\bibitem[{\citenamefont{Heidarian and Damle}(2005)}]{heidarian:127206}
\bibinfo{author}{\bibfnamefont{D.}~\bibnamefont{Heidarian}} \bibnamefont{and}
  \bibinfo{author}{\bibfnamefont{K.}~\bibnamefont{Damle}},
  \bibinfo{journal}{Phys. Rev. Lett.} \textbf{\bibinfo{volume}{95}},
  \bibinfo{eid}{127206} (\bibinfo{year}{2005}).

\bibitem[{\citenamefont{Melko et~al.}(2005)\citenamefont{Melko, Paramekanti,
  Burkov, Vishwanath, Sheng, and Balents}}]{melko:127207}
\bibinfo{author}{\bibfnamefont{R.~G.} \bibnamefont{Melko}},
  \bibinfo{author}{\bibfnamefont{A.}~\bibnamefont{Paramekanti}},
  \bibinfo{author}{\bibfnamefont{A.~A.} \bibnamefont{Burkov}},
  \bibinfo{author}{\bibfnamefont{A.}~\bibnamefont{Vishwanath}},
  \bibinfo{author}{\bibfnamefont{D.~N.} \bibnamefont{Sheng}}, \bibnamefont{and}
  \bibinfo{author}{\bibfnamefont{L.}~\bibnamefont{Balents}},
  \bibinfo{journal}{Phys. Rev. Lett.} \textbf{\bibinfo{volume}{95}},
  \bibinfo{eid}{127207} (\bibinfo{year}{2005}).

\bibitem[{\citenamefont{Batrouni et~al.}(2006)\citenamefont{Batrouni,
  H\'{e}bert, and Scalettar}}]{batrouni:087209}
\bibinfo{author}{\bibfnamefont{G.~G.} \bibnamefont{Batrouni}},
  \bibinfo{author}{\bibfnamefont{F.}~\bibnamefont{H\'{e}bert}},
  \bibnamefont{and} \bibinfo{author}{\bibfnamefont{R.~T.}
  \bibnamefont{Scalettar}}, \bibinfo{journal}{Phys. Rev. Lett.}
  \textbf{\bibinfo{volume}{97}}, \bibinfo{eid}{087209} (\bibinfo{year}{2006}).

\bibitem[{\citenamefont{Burnell et~al.}(2009)\citenamefont{Burnell, Parish,
  Cooper, and Sondhi}}]{arXiv:0901.4366v1}
\bibinfo{author}{\bibfnamefont{F.~J.}~\bibnamefont{Burnell}},
  \bibinfo{author}{\bibfnamefont{M.~M.} \bibnamefont{Parish}},
  \bibinfo{author}{\bibfnamefont{N.~R.} \bibnamefont{Cooper}},
  \bibnamefont{and} \bibinfo{author}{\bibfnamefont{S.~L.}
  \bibnamefont{Sondhi}}, \bibinfo{journal}{arXiv:0901.4366v1}
  (\bibinfo{year}{2009}).

\bibitem[{\citenamefont{Kovrizhin et~al.}(2005)\citenamefont{Kovrizhin, Pai,
  and Sinha}}]{0295-5075-72-2-162}
\bibinfo{author}{\bibfnamefont{D.~L.} \bibnamefont{Kovrizhin}},
  \bibinfo{author}{\bibfnamefont{G.~V.} \bibnamefont{Pai}}, \bibnamefont{and}
  \bibinfo{author}{\bibfnamefont{S.}~\bibnamefont{Sinha}},
  \bibinfo{journal}{EPL} \textbf{\bibinfo{volume}{72}}, \bibinfo{pages}{162}
  (\bibinfo{year}{2005}).

\bibitem[{\citenamefont{Mazzarella et~al.}(2006)\citenamefont{Mazzarella,
  Giampaolo, and Illuminati}}]{mazzarella:013625}
\bibinfo{author}{\bibfnamefont{G.}~\bibnamefont{Mazzarella}},
  \bibinfo{author}{\bibfnamefont{S.~M.} \bibnamefont{Giampaolo}},
  \bibnamefont{and}
  \bibinfo{author}{\bibfnamefont{F.}~\bibnamefont{Illuminati}},
  \bibinfo{journal}{Phys. Rev. A} \textbf{\bibinfo{volume}{73}},
  \bibinfo{eid}{013625} (\bibinfo{year}{2006}).

\bibitem[{\citenamefont{Iskin and Freericks}(2009)}]{iskin:053634}
\bibinfo{author}{\bibfnamefont{M.}~\bibnamefont{Iskin}} \bibnamefont{and}
  \bibinfo{author}{\bibfnamefont{J.~K.} \bibnamefont{Freericks}},
  \bibinfo{journal}{Phys. Rev. A} \textbf{\bibinfo{volume}{79}},
  \bibinfo{eid}{053634} (\bibinfo{year}{2009}).

\bibitem[{\citenamefont{G\'oral et~al.}(2002)\citenamefont{G\'oral, Santos, and
  Lewenstein}}]{Goral:2002}
\bibinfo{author}{\bibfnamefont{K.}~\bibnamefont{G\'oral}},
  \bibinfo{author}{\bibfnamefont{L.}~\bibnamefont{Santos}}, \bibnamefont{and}
  \bibinfo{author}{\bibfnamefont{M.}~\bibnamefont{Lewenstein}},
  \bibinfo{journal}{Phys. Rev. Lett.} \textbf{\bibinfo{volume}{88}},
  \bibinfo{eid}{170406} (\bibinfo{year}{2002}).

\bibitem[{\citenamefont{B\"uchler et~al.}(2007)\citenamefont{B\"uchler, Demler,
  Lukin, Micheli, Prokof'ev, Pupillo, and Zoller}}]{buchler:060404}
\bibinfo{author}{\bibfnamefont{H.~P.} \bibnamefont{B\"uchler}},
  \bibinfo{author}{\bibfnamefont{E.}~\bibnamefont{Demler}},
  \bibinfo{author}{\bibfnamefont{M.}~\bibnamefont{Lukin}},
  \bibinfo{author}{\bibfnamefont{A.}~\bibnamefont{Micheli}},
  \bibinfo{author}{\bibfnamefont{N.}~\bibnamefont{Prokof'ev}},
  \bibinfo{author}{\bibfnamefont{G.}~\bibnamefont{Pupillo}}, \bibnamefont{and}
  \bibinfo{author}{\bibfnamefont{P.}~\bibnamefont{Zoller}},
  \bibinfo{journal}{Phys. Rev. Lett.} \textbf{\bibinfo{volume}{98}},
  \bibinfo{eid}{060404} (\bibinfo{year}{2007}).

\bibitem[{\citenamefont{Titvinidze et~al.}(2008)\citenamefont{Titvinidze,
  Snoek, and Hofstetter}}]{titvinidze:100401}
\bibinfo{author}{\bibfnamefont{I.}~\bibnamefont{Titvinidze}},
  \bibinfo{author}{\bibfnamefont{M.}~\bibnamefont{Snoek}}, \bibnamefont{and}
  \bibinfo{author}{\bibfnamefont{W.}~\bibnamefont{Hofstetter}},
  \bibinfo{journal}{Phys. Rev. Lett.} \textbf{\bibinfo{volume}{100}},
  \bibinfo{eid}{100401} (\bibinfo{year}{2008}).

\bibitem[{\citenamefont{Sinha and Sengupta}(2009)}]{sinha:115124}
\bibinfo{author}{\bibfnamefont{S.}~\bibnamefont{Sinha}} \bibnamefont{and}
  \bibinfo{author}{\bibfnamefont{K.}~\bibnamefont{Sengupta}},
  \bibinfo{journal}{Phys. Rev. B} \textbf{\bibinfo{volume}{79}},
  \bibinfo{eid}{115124} (\bibinfo{year}{2009}).

\bibitem[{\citenamefont{Mathey et~al.}(2009)\citenamefont{Mathey, Danshita, and
  Clark}}]{mathey:011602}
\bibinfo{author}{\bibfnamefont{L.}~\bibnamefont{Mathey}},
  \bibinfo{author}{\bibfnamefont{I.}~\bibnamefont{Danshita}}, \bibnamefont{and}
  \bibinfo{author}{\bibfnamefont{C.~W.} \bibnamefont{Clark}},
  \bibinfo{journal}{Phys. Rev. A} \textbf{\bibinfo{volume}{79}},
  \bibinfo{eid}{011602(R)} (\bibinfo{year}{2009}).

\bibitem[{\citenamefont{Orth et~al.}(2008)\citenamefont{Orth, Stanic, and
  Le~Hur}}]{orth:051601}
\bibinfo{author}{\bibfnamefont{P.~P.} \bibnamefont{Orth}},
  \bibinfo{author}{\bibfnamefont{I.}~\bibnamefont{Stanic}}, \bibnamefont{and}
  \bibinfo{author}{\bibfnamefont{K.}~\bibnamefont{Le~Hur}},
  \bibinfo{journal}{Phys. Rev. A} \textbf{\bibinfo{volume}{77}},
  \bibinfo{eid}{051601(R)} (\bibinfo{year}{2008}).

\bibitem[{\citenamefont{Dorner et~al.}(2003)\citenamefont{Dorner, Fedichev,
  Jaksch, Lewenstein, and Zoller}}]{PhysRevLett.91.073601}
\bibinfo{author}{\bibfnamefont{U.}~\bibnamefont{Dorner}},
  \bibinfo{author}{\bibfnamefont{P.}~\bibnamefont{Fedichev}},
  \bibinfo{author}{\bibfnamefont{D.}~\bibnamefont{Jaksch}},
  \bibinfo{author}{\bibfnamefont{M.}~\bibnamefont{Lewenstein}},
  \bibnamefont{and} \bibinfo{author}{\bibfnamefont{P.}~\bibnamefont{Zoller}},
  \bibinfo{journal}{Phys. Rev. Lett.} \textbf{\bibinfo{volume}{91}},
  \bibinfo{pages}{073601} (\bibinfo{year}{2003}).

\bibitem[{\citenamefont{Duan et~al.}(2003)\citenamefont{Duan, Demler, and
  Lukin}}]{PhysRevLett.91.090402}
\bibinfo{author}{\bibfnamefont{L.-M.} \bibnamefont{Duan}},
  \bibinfo{author}{\bibfnamefont{E.}~\bibnamefont{Demler}}, \bibnamefont{and}
  \bibinfo{author}{\bibfnamefont{M.~D.} \bibnamefont{Lukin}},
  \bibinfo{journal}{Phys. Rev. Lett.} \textbf{\bibinfo{volume}{91}},
  \bibinfo{pages}{090402} (\bibinfo{year}{2003}).

\bibitem[{\citenamefont{Cazalilla and Ho}(2003)}]{PhysRevLett.91.150403}
\bibinfo{author}{\bibfnamefont{M.~A.} \bibnamefont{Cazalilla}}
  \bibnamefont{and} \bibinfo{author}{\bibfnamefont{A.~F.} \bibnamefont{Ho}},
  \bibinfo{journal}{Phys. Rev. Lett.} \textbf{\bibinfo{volume}{91}},
  \bibinfo{pages}{150403} (\bibinfo{year}{2003}).

\bibitem[{\citenamefont{Rizzi and Imambekov}(2008)}]{rizzi:023621}
\bibinfo{author}{\bibfnamefont{M.}~\bibnamefont{Rizzi}} \bibnamefont{and}
  \bibinfo{author}{\bibfnamefont{A.}~\bibnamefont{Imambekov}},
  \bibinfo{journal}{Phys. Rev. A} \textbf{\bibinfo{volume}{77}},
  \bibinfo{eid}{023621} (\bibinfo{year}{2008}).

\bibitem[{\citenamefont{Z\"ollner et~al.}(2008)\citenamefont{Z\"ollner, Meyer,
  and Schmelcher}}]{zollner:013629}
\bibinfo{author}{\bibfnamefont{S.}~\bibnamefont{Z\"ollner}},
  \bibinfo{author}{\bibfnamefont{H.-D.} \bibnamefont{Meyer}}, \bibnamefont{and}
  \bibinfo{author}{\bibfnamefont{P.}~\bibnamefont{Schmelcher}},
  \bibinfo{journal}{Phys. Rev. A} \textbf{\bibinfo{volume}{78}},
  \bibinfo{eid}{013629} (\bibinfo{year}{2008}).

\bibitem[{\citenamefont{Roscilde and Cirac}(2007)}]{roscilde:190402}
\bibinfo{author}{\bibfnamefont{T.}~\bibnamefont{Roscilde}} \bibnamefont{and}
  \bibinfo{author}{\bibfnamefont{J.~I.} \bibnamefont{Cirac}},
  \bibinfo{journal}{Phys. Rev. Lett.} \textbf{\bibinfo{volume}{98}},
  \bibinfo{eid}{190402} (\bibinfo{year}{2007}).

\bibitem[{\citenamefont{Buonsante et~al.}(2008)\citenamefont{Buonsante,
  Giampaolo, Illuminati, Penna, and Vezzani}}]{buonsante:240402}
\bibinfo{author}{\bibfnamefont{P.}~\bibnamefont{Buonsante}},
  \bibinfo{author}{\bibfnamefont{S.~M.} \bibnamefont{Giampaolo}},
  \bibinfo{author}{\bibfnamefont{F.}~\bibnamefont{Illuminati}},
  \bibinfo{author}{\bibfnamefont{V.}~\bibnamefont{Penna}}, \bibnamefont{and}
  \bibinfo{author}{\bibfnamefont{A.}~\bibnamefont{Vezzani}},
  \bibinfo{journal}{Phys. Rev. Lett.} \textbf{\bibinfo{volume}{100}},
  \bibinfo{eid}{240402} (\bibinfo{year}{2008}).

\bibitem[{\citenamefont{Maska et~al.}(2008)\citenamefont{Maska, Lemanski,
  Freericks, and Williams}}]{maska:060404}
\bibinfo{author}{\bibfnamefont{M.~M.} \bibnamefont{Maska}},
  \bibinfo{author}{\bibfnamefont{R.}~\bibnamefont{Lemanski}},
  \bibinfo{author}{\bibfnamefont{J.~K.} \bibnamefont{Freericks}},
  \bibnamefont{and} \bibinfo{author}{\bibfnamefont{C.~J.}
  \bibnamefont{Williams}}, \bibinfo{journal}{Phys. Rev. Lett.}
  \textbf{\bibinfo{volume}{101}}, \bibinfo{eid}{060404} (\bibinfo{year}{2008}).

\bibitem[{\citenamefont{Mathey et~al.}(2007)\citenamefont{Mathey, Tsai, and
  Castro~Neto}}]{mathey:174516}
\bibinfo{author}{\bibfnamefont{L.}~\bibnamefont{Mathey}},
  \bibinfo{author}{\bibfnamefont{S.-W.} \bibnamefont{Tsai}}, \bibnamefont{and}
  \bibinfo{author}{\bibfnamefont{A.~H.} \bibnamefont{Castro~Neto}},
  \bibinfo{journal}{Phys. Rev. B} \textbf{\bibinfo{volume}{75}},
  \bibinfo{eid}{174516} (\bibinfo{year}{2007}).

\bibitem[{\citenamefont{Mathey et~al.}(2004)\citenamefont{Mathey, Wang,
  Hofstetter, Lukin, and Demler}}]{PhysRevLett.93.120404}
\bibinfo{author}{\bibfnamefont{L.}~\bibnamefont{Mathey}},
  \bibinfo{author}{\bibfnamefont{D.-W.} \bibnamefont{Wang}},
  \bibinfo{author}{\bibfnamefont{W.}~\bibnamefont{Hofstetter}},
  \bibinfo{author}{\bibfnamefont{M.~D.} \bibnamefont{Lukin}}, \bibnamefont{and}
  \bibinfo{author}{\bibfnamefont{E.}~\bibnamefont{Demler}},
  \bibinfo{journal}{Phys. Rev. Lett.} \textbf{\bibinfo{volume}{93}},
  \bibinfo{pages}{120404} (\bibinfo{year}{2004}).

\bibitem[{\citenamefont{Ottenstein et~al.}(2008)\citenamefont{Ottenstein,
  Lompe, Kohnen, Wenz, and Jochim}}]{ottenstein:203202}
\bibinfo{author}{\bibfnamefont{T.~B.} \bibnamefont{Ottenstein}},
  \bibinfo{author}{\bibfnamefont{T.}~\bibnamefont{Lompe}},
  \bibinfo{author}{\bibfnamefont{M.}~\bibnamefont{Kohnen}},
  \bibinfo{author}{\bibfnamefont{A.~N.} \bibnamefont{Wenz}}, \bibnamefont{and}
  \bibinfo{author}{\bibfnamefont{S.}~\bibnamefont{Jochim}},
  \bibinfo{journal}{Phys. Rev. Lett.} \textbf{\bibinfo{volume}{101}},
  \bibinfo{eid}{203202} (\bibinfo{year}{2008}).

\bibitem[{\citenamefont{Wenz et~al.}(2009)\citenamefont{Wenz, Lompe,
  Ottenstein, Serwane, Z\"urn, and Jochim}}]{arXiv:0906.4378v1}
\bibinfo{author}{\bibfnamefont{A.~N.} \bibnamefont{Wenz}},
  \bibinfo{author}{\bibfnamefont{T.}~\bibnamefont{Lompe}},
  \bibinfo{author}{\bibfnamefont{T.~B.} \bibnamefont{Ottenstein}},
  \bibinfo{author}{\bibfnamefont{F.}~\bibnamefont{Serwane}},
  \bibinfo{author}{\bibfnamefont{G.}~\bibnamefont{Z\"urn}}, \bibnamefont{and}
  \bibinfo{author}{\bibfnamefont{S.}~\bibnamefont{Jochim}},
  \bibinfo{journal}{arXiv:0906.4378v1 [cond-mat.quant-gas]}
  (\bibinfo{year}{2009}).

\bibitem[{\citenamefont{Bhaseen et~al.}(2009)\citenamefont{Bhaseen, Hohenadler,
  Silver, and Simons}}]{bhaseen:135301}
\bibinfo{author}{\bibfnamefont{M.~J.} \bibnamefont{Bhaseen}},
  \bibinfo{author}{\bibfnamefont{M.}~\bibnamefont{Hohenadler}},
  \bibinfo{author}{\bibfnamefont{A.~O.} \bibnamefont{Silver}},
  \bibnamefont{and} \bibinfo{author}{\bibfnamefont{B.~D.}
  \bibnamefont{Simons}}, \bibinfo{journal}{Phys. Rev. Lett.}
  \textbf{\bibinfo{volume}{102}}, \bibinfo{eid}{135301} (\bibinfo{year}{2009}).

\bibitem[{\citenamefont{Girardeau}(2009)}]{girardeau:245303}
\bibinfo{author}{\bibfnamefont{M.~D.} \bibnamefont{Girardeau}},
  \bibinfo{journal}{Phys. Rev. Lett.} \textbf{\bibinfo{volume}{102}},
  \bibinfo{eid}{245303} (\bibinfo{year}{2009}).

\bibitem[{\citenamefont{Mering and Fleischhauer}(2009)}]{arXiv:0903.5226v1}
\bibinfo{author}{\bibfnamefont{A.}~\bibnamefont{Mering}} \bibnamefont{and}
  \bibinfo{author}{\bibfnamefont{M.}~\bibnamefont{Fleischhauer}},
  \bibinfo{journal}{arXiv:0903.5226v1 [cond-mat.other]}
  (\bibinfo{year}{2009}).

\bibitem[{\citenamefont{Roth and Burnett}(2004)}]{PhysRevA.69.021601}
\bibinfo{author}{\bibfnamefont{R.}~\bibnamefont{Roth}} \bibnamefont{and}
  \bibinfo{author}{\bibfnamefont{K.}~\bibnamefont{Burnett}},
  \bibinfo{journal}{Phys. Rev. A} \textbf{\bibinfo{volume}{69}},
  \bibinfo{pages}{021601(R)} (\bibinfo{year}{2004}).

\bibitem[{\citenamefont{Illuminati and Albus}(2004)}]{PhysRevLett.93.090406}
\bibinfo{author}{\bibfnamefont{F.}~\bibnamefont{Illuminati}} \bibnamefont{and}
  \bibinfo{author}{\bibfnamefont{A.}~\bibnamefont{Albus}},
  \bibinfo{journal}{Phys. Rev. Lett.} \textbf{\bibinfo{volume}{93}},
  \bibinfo{pages}{090406} (\bibinfo{year}{2004}).

\bibitem[{\citenamefont{Wang et~al.}(2005)\citenamefont{Wang, Lukin, and
  Demler}}]{PhysRevA.72.051604}
\bibinfo{author}{\bibfnamefont{D.-W.} \bibnamefont{Wang}},
  \bibinfo{author}{\bibfnamefont{M.~D.} \bibnamefont{Lukin}}, \bibnamefont{and}
  \bibinfo{author}{\bibfnamefont{E.}~\bibnamefont{Demler}},
  \bibinfo{journal}{Phys. Rev. A} \textbf{\bibinfo{volume}{72}},
  \bibinfo{pages}{051604(R)} (\bibinfo{year}{2005}).

\bibitem[{\citenamefont{Refael and Demler}(2008)}]{refael:144511}
\bibinfo{author}{\bibfnamefont{G.}~\bibnamefont{Refael}} \bibnamefont{and}
  \bibinfo{author}{\bibfnamefont{E.}~\bibnamefont{Demler}},
  \bibinfo{journal}{Phys. Rev. B} \textbf{\bibinfo{volume}{77}},
  \bibinfo{eid}{144511} (\bibinfo{year}{2008}).

\bibitem[{\citenamefont{Lutchyn et~al.}(2008)\citenamefont{Lutchyn, Tewari, and
  Das~Sarma}}]{lutchyn:220504}
\bibinfo{author}{\bibfnamefont{R.~M.} \bibnamefont{Lutchyn}},
  \bibinfo{author}{\bibfnamefont{S.}~\bibnamefont{Tewari}}, \bibnamefont{and}
  \bibinfo{author}{\bibfnamefont{S.}~\bibnamefont{Das~Sarma}},
  \bibinfo{journal}{Phys. Rev. B} \textbf{\bibinfo{volume}{78}},
  \bibinfo{eid}{220504(R)} (\bibinfo{year}{2008}).

\bibitem[{\citenamefont{Klironomos and Tsai}(2007)}]{klironomos:100401}
\bibinfo{author}{\bibfnamefont{F.~D.} \bibnamefont{Klironomos}}
  \bibnamefont{and} \bibinfo{author}{\bibfnamefont{S.-W.} \bibnamefont{Tsai}},
  \bibinfo{journal}{Phys. Rev. Lett.} \textbf{\bibinfo{volume}{99}},
  \bibinfo{eid}{100401} (\bibinfo{year}{2007}).

\bibitem[{\citenamefont{Mering and Fleischhauer}(2008)}]{mering:023601}
\bibinfo{author}{\bibfnamefont{A.}~\bibnamefont{Mering}} \bibnamefont{and}
  \bibinfo{author}{\bibfnamefont{M.}~\bibnamefont{Fleischhauer}},
  \bibinfo{journal}{Phys. Rev. A} \textbf{\bibinfo{volume}{77}},
  \bibinfo{eid}{023601} (\bibinfo{year}{2008}).

\bibitem[{\citenamefont{Fehrmann et~al.}(2004)\citenamefont{Fehrmann, Baranov,
  Damski, Lewenstein, and Santos}}]{Fehrmann200423}
\bibinfo{author}{\bibfnamefont{H.}~\bibnamefont{Fehrmann}},
  \bibinfo{author}{\bibfnamefont{M.}~\bibnamefont{Baranov}},
  \bibinfo{author}{\bibfnamefont{B.}~\bibnamefont{Damski}},
  \bibinfo{author}{\bibfnamefont{M.}~\bibnamefont{Lewenstein}},
  \bibnamefont{and} \bibinfo{author}{\bibfnamefont{L.}~\bibnamefont{Santos}},
  \bibinfo{journal}{Optics Communications} \textbf{\bibinfo{volume}{243}},
  \bibinfo{pages}{23} (\bibinfo{year}{2004}).

\bibitem[{\citenamefont{Imambekov and Demler}(2006)}]{imambekov:021602}
\bibinfo{author}{\bibfnamefont{A.}~\bibnamefont{Imambekov}} \bibnamefont{and}
  \bibinfo{author}{\bibfnamefont{E.}~\bibnamefont{Demler}},
  \bibinfo{journal}{Phys. Rev. A} \textbf{\bibinfo{volume}{73}},
  \bibinfo{eid}{021602(R)} (\bibinfo{year}{2006}).

\bibitem[{\citenamefont{Pollet et~al.}(2008)\citenamefont{Pollet, Kollath,
  Schollwock, and Troyer}}]{pollet:023608}
\bibinfo{author}{\bibfnamefont{L.}~\bibnamefont{Pollet}},
  \bibinfo{author}{\bibfnamefont{C.}~\bibnamefont{Kollath}},
  \bibinfo{author}{\bibfnamefont{U.}~\bibnamefont{Schollwock}},
  \bibnamefont{and} \bibinfo{author}{\bibfnamefont{M.}~\bibnamefont{Troyer}},
  \bibinfo{journal}{Phys. Rev. A} \textbf{\bibinfo{volume}{77}},
  \bibinfo{eid}{023608} (\bibinfo{year}{2008}).

\bibitem[{\citenamefont{Sengupta et~al.}(2007)\citenamefont{Sengupta, Dupuis,
  and Majumdar}}]{sengupta:063625}
\bibinfo{author}{\bibfnamefont{K.}~\bibnamefont{Sengupta}},
  \bibinfo{author}{\bibfnamefont{N.}~\bibnamefont{Dupuis}}, \bibnamefont{and}
  \bibinfo{author}{\bibfnamefont{P.}~\bibnamefont{Majumdar}},
  \bibinfo{journal}{Phys. Rev. A} \textbf{\bibinfo{volume}{75}},
  \bibinfo{eid}{063625} (\bibinfo{year}{2007}).

\bibitem[{\citenamefont{Pollet et~al.}(2006)\citenamefont{Pollet, Troyer,
  Van~Houcke, and Rombouts}}]{pollet:190402}
\bibinfo{author}{\bibfnamefont{L.}~\bibnamefont{Pollet}},
  \bibinfo{author}{\bibfnamefont{M.}~\bibnamefont{Troyer}},
  \bibinfo{author}{\bibfnamefont{K.}~\bibnamefont{Van~Houcke}},
  \bibnamefont{and} \bibinfo{author}{\bibfnamefont{S.~M.~A.}
  \bibnamefont{Rombouts}}, \bibinfo{journal}{Phys. Rev. Lett.}
  \textbf{\bibinfo{volume}{96}}, \bibinfo{eid}{190402} (\bibinfo{year}{2006}).

\bibitem[{\citenamefont{Marchetti et~al.}(2008)\citenamefont{Marchetti, Mathy,
  Huse, and Parish}}]{marchetti:134517}
\bibinfo{author}{\bibfnamefont{F.~M.} \bibnamefont{Marchetti}},
  \bibinfo{author}{\bibfnamefont{C.~J.~M.} \bibnamefont{Mathy}},
  \bibinfo{author}{\bibfnamefont{D.~A.} \bibnamefont{Huse}}, \bibnamefont{and}
  \bibinfo{author}{\bibfnamefont{M.~M.} \bibnamefont{Parish}},
  \bibinfo{journal}{Phys. Rev. B} \textbf{\bibinfo{volume}{78}},
  \bibinfo{eid}{134517} (\bibinfo{year}{2008}).

\bibitem[{\citenamefont{Bergman and Le~Hur}(2009)}]{bergman:184520}
\bibinfo{author}{\bibfnamefont{D.~L.} \bibnamefont{Bergman}} \bibnamefont{and}
  \bibinfo{author}{\bibfnamefont{K.}~\bibnamefont{Le~Hur}},
  \bibinfo{journal}{Phys. Rev. B} \textbf{\bibinfo{volume}{79}},
  \bibinfo{eid}{184520} (\bibinfo{year}{2009}).

\bibitem[{\citenamefont{Zhao and Paramekanti}(2006)}]{zhao:105303}
\bibinfo{author}{\bibfnamefont{E.}~\bibnamefont{Zhao}} \bibnamefont{and}
  \bibinfo{author}{\bibfnamefont{A.}~\bibnamefont{Paramekanti}},
  \bibinfo{journal}{Phys. Rev. Lett.} \textbf{\bibinfo{volume}{96}},
  \bibinfo{eid}{105303} (\bibinfo{year}{2006}).

\bibitem[{\citenamefont{Morsch and Oberthaler}(2006)}]{morsch:179}
\bibinfo{author}{\bibfnamefont{O.}~\bibnamefont{Morsch}} \bibnamefont{and}
  \bibinfo{author}{\bibfnamefont{M.}~\bibnamefont{Oberthaler}},
  \bibinfo{journal}{Rev. Mod. Phys.} \textbf{\bibinfo{volume}{78}},
  \bibinfo{eid}{179} (\bibinfo{year}{2006}).

\bibitem[{\citenamefont{B\"uchler and Blatter}(2004)}]{PhysRevA.69.063603}
\bibinfo{author}{\bibfnamefont{H.~P.} \bibnamefont{B\"uchler}}
  \bibnamefont{and} \bibinfo{author}{\bibfnamefont{G.}~\bibnamefont{Blatter}},
  \bibinfo{journal}{Phys. Rev. A} \textbf{\bibinfo{volume}{69}},
  \bibinfo{pages}{063603} (\bibinfo{year}{2004}).

\bibitem[{\citenamefont{Elstner and Monien}(1999)}]{PhysRevB.59.12184}
\bibinfo{author}{\bibfnamefont{N.}~\bibnamefont{Elstner}} \bibnamefont{and}
  \bibinfo{author}{\bibfnamefont{H.}~\bibnamefont{Monien}},
  \bibinfo{journal}{Phys. Rev. B} \textbf{\bibinfo{volume}{59}},
  \bibinfo{pages}{12184} (\bibinfo{year}{1999}).

\bibitem[{\citenamefont{Spielman et~al.}(2008)\citenamefont{Spielman, Phillips,
  and Porto}}]{spielman:120402}
\bibinfo{author}{\bibfnamefont{I.~B.} \bibnamefont{Spielman}},
  \bibinfo{author}{\bibfnamefont{W.~D.} \bibnamefont{Phillips}},
  \bibnamefont{and} \bibinfo{author}{\bibfnamefont{J.~V.} \bibnamefont{Porto}},
  \bibinfo{journal}{Phys. Rev. Lett.} \textbf{\bibinfo{volume}{100}},
  \bibinfo{eid}{120402} (\bibinfo{year}{2008}).

\bibitem[{\citenamefont{Burkov and Balents}(2005)}]{PhysRevB.72.134502}
\bibinfo{author}{\bibfnamefont{A.~A.} \bibnamefont{Burkov}} \bibnamefont{and}
  \bibinfo{author}{\bibfnamefont{L.}~\bibnamefont{Balents}},
  \bibinfo{journal}{Phys. Rev. B} \textbf{\bibinfo{volume}{72}},
  \bibinfo{pages}{134502} (\bibinfo{year}{2005}).

\bibitem[{\citenamefont{Goldenfeld}(1992)}]{goldenfeld_pt}
\bibinfo{author}{\bibfnamefont{N.}~\bibnamefont{Goldenfeld}},
  \emph{\bibinfo{title}{Lectures on phase transitions and the renormalization
  group}} (\bibinfo{publisher}{Westview Press}, \bibinfo{address}{Boulder, CO},
  \bibinfo{year}{1992}).

\bibitem[{\citenamefont{Titvinidze et~al.}(2009)\citenamefont{Titvinidze,
  Snoek, and Hofstetter}}]{titvinidze:144506}
\bibinfo{author}{\bibfnamefont{I.}~\bibnamefont{Titvinidze}},
  \bibinfo{author}{\bibfnamefont{M.}~\bibnamefont{Snoek}}, \bibnamefont{and}
  \bibinfo{author}{\bibfnamefont{W.}~\bibnamefont{Hofstetter}},
  \bibinfo{journal}{Phys. Rev. B} \textbf{\bibinfo{volume}{79}},
  \bibinfo{eid}{144506} (\bibinfo{year}{2009}).

\bibitem[{\citenamefont{Wessel et~al.}(2004)\citenamefont{Wessel, Alet, Troyer,
  and Batrouni}}]{PhysRevA.70.053615}
\bibinfo{author}{\bibfnamefont{S.}~\bibnamefont{Wessel}},
  \bibinfo{author}{\bibfnamefont{F.}~\bibnamefont{Alet}},
  \bibinfo{author}{\bibfnamefont{M.}~\bibnamefont{Troyer}}, \bibnamefont{and}
  \bibinfo{author}{\bibfnamefont{G.~G.} \bibnamefont{Batrouni}},
  \bibinfo{journal}{Phys. Rev. A} \textbf{\bibinfo{volume}{70}},
  \bibinfo{pages}{053615} (\bibinfo{year}{2004}).

\bibitem[{\citenamefont{Hadzibabic et~al.}(2002)\citenamefont{Hadzibabic, Stan,
  Dieckmann, Gupta, Zwierlein, G\"orlitz, and
  Ketterle}}]{PhysRevLett.88.160401}
\bibinfo{author}{\bibfnamefont{Z.}~\bibnamefont{Hadzibabic}},
  \bibinfo{author}{\bibfnamefont{C.~A.} \bibnamefont{Stan}},
  \bibinfo{author}{\bibfnamefont{K.}~\bibnamefont{Dieckmann}},
  \bibinfo{author}{\bibfnamefont{S.}~\bibnamefont{Gupta}},
  \bibinfo{author}{\bibfnamefont{M.~W.} \bibnamefont{Zwierlein}},
  \bibinfo{author}{\bibfnamefont{A.}~\bibnamefont{G\"orlitz}},
  \bibnamefont{and} \bibinfo{author}{\bibfnamefont{W.}~\bibnamefont{Ketterle}},
  \bibinfo{journal}{Phys. Rev. Lett.} \textbf{\bibinfo{volume}{88}},
  \bibinfo{pages}{160401} (\bibinfo{year}{2002}).

\bibitem[{\citenamefont{Stan et~al.}(2004)\citenamefont{Stan, Zwierlein,
  Schunck, Raupach, and Ketterle}}]{PhysRevLett.93.143001}
\bibinfo{author}{\bibfnamefont{C.~A.} \bibnamefont{Stan}},
  \bibinfo{author}{\bibfnamefont{M.~W.} \bibnamefont{Zwierlein}},
  \bibinfo{author}{\bibfnamefont{C.~H.} \bibnamefont{Schunck}},
  \bibinfo{author}{\bibfnamefont{S.~M.~F.} \bibnamefont{Raupach}},
  \bibnamefont{and} \bibinfo{author}{\bibfnamefont{W.}~\bibnamefont{Ketterle}},
  \bibinfo{journal}{Phys. Rev. Lett.} \textbf{\bibinfo{volume}{93}},
  \bibinfo{pages}{143001} (\bibinfo{year}{2004}).

\bibitem[{\citenamefont{Best et~al.}(2009)\citenamefont{Best, Will, Schneider,
  Hackerm\"{u}ller, van Oosten, Bloch, and L\"{u}hmann}}]{best:030408}
\bibinfo{author}{\bibfnamefont{T.}~\bibnamefont{Best}},
  \bibinfo{author}{\bibfnamefont{S.}~\bibnamefont{Will}},
  \bibinfo{author}{\bibfnamefont{U.}~\bibnamefont{Schneider}},
  \bibinfo{author}{\bibfnamefont{L.}~\bibnamefont{Hackerm\"{u}ller}},
  \bibinfo{author}{\bibfnamefont{D.}~\bibnamefont{van Oosten}},
  \bibinfo{author}{\bibfnamefont{I.}~\bibnamefont{Bloch}}, \bibnamefont{and}
  \bibinfo{author}{\bibfnamefont{D.-S.} \bibnamefont{L\"{u}hmann}},
  \bibinfo{journal}{Phys. Rev. Lett.} \textbf{\bibinfo{volume}{102}},
  \bibinfo{eid}{030408} (\bibinfo{year}{2009}).

\bibitem[{\citenamefont{G{\"u}nter et~al.}(2006)\citenamefont{G{\"u}nter,
  St{\"o}ferle, Moritz, K{\"o}hl, and Esslinger}}]{gunter:180402}
\bibinfo{author}{\bibfnamefont{K.}~\bibnamefont{G{\"u}nter}},
  \bibinfo{author}{\bibfnamefont{T.}~\bibnamefont{St{\"o}ferle}},
  \bibinfo{author}{\bibfnamefont{H.}~\bibnamefont{Moritz}},
  \bibinfo{author}{\bibfnamefont{M.}~\bibnamefont{K{\"o}hl}}, \bibnamefont{and}
  \bibinfo{author}{\bibfnamefont{T.}~\bibnamefont{Esslinger}},
  \bibinfo{journal}{Phys. Rev. Lett.} \textbf{\bibinfo{volume}{96}},
  \bibinfo{eid}{180402} (\bibinfo{year}{2006}).

\bibitem[{\citenamefont{Ospelkaus and
  Ospelkaus}(2008)}]{0953-4075-41-20-203001}
\bibinfo{author}{\bibfnamefont{C.}~\bibnamefont{Ospelkaus}} \bibnamefont{and}
  \bibinfo{author}{\bibfnamefont{S.}~\bibnamefont{Ospelkaus}},
  \bibinfo{journal}{J. Phys. B} \textbf{\bibinfo{volume}{41}},
  \bibinfo{pages}{203001} (\bibinfo{year}{2008}).

\bibitem[{\citenamefont{Metcalf and van~der
  Straten}(1999)}]{Metcalf-LaserCooling}
\bibinfo{author}{\bibfnamefont{H.~J.} \bibnamefont{Metcalf}} \bibnamefont{and}
  \bibinfo{author}{\bibfnamefont{P.}~\bibnamefont{van~der Straten}},
  \emph{\bibinfo{title}{Laser Cooling and Trapping}}
  (\bibinfo{publisher}{Springer}, \bibinfo{year}{1999}).

\bibitem[{\citenamefont{Chin et~al.}(2008)\citenamefont{Chin, Grimm, Julienne,
  and Tiesinga}}]{arXiv:0812.1496v1}
\bibinfo{author}{\bibfnamefont{C.}~\bibnamefont{Chin}},
  \bibinfo{author}{\bibfnamefont{R.}~\bibnamefont{Grimm}},
  \bibinfo{author}{\bibfnamefont{P.}~\bibnamefont{Julienne}}, \bibnamefont{and}
  \bibinfo{author}{\bibfnamefont{E.}~\bibnamefont{Tiesinga}},
  \bibinfo{journal}{eprint arXiv:0812.1496}  (\bibinfo{year}{2008}).

\bibitem[{\citenamefont{Stenger et~al.}(1999)\citenamefont{Stenger, Inouye,
  Andrews, Miesner, Stamper-Kurn, and Ketterle}}]{PhysRevLett.82.2422}
\bibinfo{author}{\bibfnamefont{J.}~\bibnamefont{Stenger}},
  \bibinfo{author}{\bibfnamefont{S.}~\bibnamefont{Inouye}},
  \bibinfo{author}{\bibfnamefont{M.~R.} \bibnamefont{Andrews}},
  \bibinfo{author}{\bibfnamefont{H.-J.} \bibnamefont{Miesner}},
  \bibinfo{author}{\bibfnamefont{D.~M.} \bibnamefont{Stamper-Kurn}},
  \bibnamefont{and} \bibinfo{author}{\bibfnamefont{W.}~\bibnamefont{Ketterle}},
  \bibinfo{journal}{Phys. Rev. Lett.} \textbf{\bibinfo{volume}{82}},
  \bibinfo{pages}{2422} (\bibinfo{year}{1999}).

\bibitem[{\citenamefont{Gacesa et~al.}(2008)\citenamefont{Gacesa, Pellegrini,
  and Cote}}]{gacesa:010701}
\bibinfo{author}{\bibfnamefont{M.}~\bibnamefont{Gacesa}},
  \bibinfo{author}{\bibfnamefont{P.}~\bibnamefont{Pellegrini}},
  \bibnamefont{and} \bibinfo{author}{\bibfnamefont{R.}~\bibnamefont{Cote}},
  \bibinfo{journal}{Phys. Rev. A} \textbf{\bibinfo{volume}{78}},
  \bibinfo{eid}{010701(R)} (\bibinfo{year}{2008}).

\bibitem[{\citenamefont{Altman et~al.}(2004)\citenamefont{Altman, Demler, and
  Lukin}}]{PhysRevA.70.013603}
\bibinfo{author}{\bibfnamefont{E.}~\bibnamefont{Altman}},
  \bibinfo{author}{\bibfnamefont{E.}~\bibnamefont{Demler}}, \bibnamefont{and}
  \bibinfo{author}{\bibfnamefont{M.~D.} \bibnamefont{Lukin}},
  \bibinfo{journal}{Phys. Rev. A} \textbf{\bibinfo{volume}{70}},
  \bibinfo{pages}{013603} (\bibinfo{year}{2004}).

\bibitem[{\citenamefont{F\"olling et~al.}(2005)\citenamefont{F\"olling,
  Gerbier, Widera, Mandel, Gericke, and Bloch}}]{foelling_nature_2005}
\bibinfo{author}{\bibfnamefont{S.}~\bibnamefont{F\"olling}},
  \bibinfo{author}{\bibfnamefont{F.}~\bibnamefont{Gerbier}},
  \bibinfo{author}{\bibfnamefont{A.}~\bibnamefont{Widera}},
  \bibinfo{author}{\bibfnamefont{O.}~\bibnamefont{Mandel}},
  \bibinfo{author}{\bibfnamefont{T.}~\bibnamefont{Gericke}}, \bibnamefont{and}
  \bibinfo{author}{\bibfnamefont{I.}~\bibnamefont{Bloch}},
  \bibinfo{journal}{Nature} \textbf{\bibinfo{volume}{434}},
  \bibinfo{pages}{481} (\bibinfo{year}{2005}).

\bibitem[{\citenamefont{H\'{e}bert et~al.}(2008)\citenamefont{H\'{e}bert,
  Batrouni, Roy, and Rousseau}}]{hebert:184505}
\bibinfo{author}{\bibfnamefont{F.}~\bibnamefont{H\'{e}bert}},
  \bibinfo{author}{\bibfnamefont{G.~G.} \bibnamefont{Batrouni}},
  \bibinfo{author}{\bibfnamefont{X.}~\bibnamefont{Roy}}, \bibnamefont{and}
  \bibinfo{author}{\bibfnamefont{V.~G.} \bibnamefont{Rousseau}},
  \bibinfo{journal}{Phys. Rev. B} \textbf{\bibinfo{volume}{78}},
  \bibinfo{eid}{184505} (\bibinfo{year}{2008}).

\bibitem[{\citenamefont{Fradkin}(1989)}]{PhysRevLett.63.322}
\bibinfo{author}{\bibfnamefont{E.}~\bibnamefont{Fradkin}},
  \bibinfo{journal}{Phys. Rev. Lett.} \textbf{\bibinfo{volume}{63}},
  \bibinfo{pages}{322} (\bibinfo{year}{1989}).

\bibitem[{\citenamefont{Sachdev}(1999)}]{sachdev_qpt_book}
\bibinfo{author}{\bibfnamefont{S.}~\bibnamefont{Sachdev}},
  \emph{\bibinfo{title}{Quantum Phase Transitions}}
  (\bibinfo{publisher}{Cambridge University Press},
  \bibinfo{address}{Cambridge, U.K.}, \bibinfo{year}{1999}).

\bibitem[{\citenamefont{Auerbach}(1994)}]{auerbach_quantum_magnetism}
\bibinfo{author}{\bibfnamefont{A.}~\bibnamefont{Auerbach}},
  \emph{\bibinfo{title}{Interacting Electrons and Quantum Magnetism}}
  (\bibinfo{publisher}{Springer-Verlag}, \bibinfo{address}{New York},
  \bibinfo{year}{1994}).

\bibitem[{\citenamefont{Fr\"ohlich and
  Lieb}(1978)}]{FrohlichLieb-CommMath-1978}
\bibinfo{author}{\bibfnamefont{J.}~\bibnamefont{Fr\"ohlich}} \bibnamefont{and}
  \bibinfo{author}{\bibfnamefont{E.~H.} \bibnamefont{Lieb}},
  \bibinfo{journal}{Comm. Math. Phys.} \textbf{\bibinfo{volume}{60}},
  \bibinfo{pages}{233} (\bibinfo{year}{1978}).

\bibitem[{\citenamefont{Cuccoli et~al.}(2003)\citenamefont{Cuccoli, Roscilde,
  Tognetti, Vaia, and Verrucchi}}]{PhysRevB.67.104414}
\bibinfo{author}{\bibfnamefont{A.}~\bibnamefont{Cuccoli}},
  \bibinfo{author}{\bibfnamefont{T.}~\bibnamefont{Roscilde}},
  \bibinfo{author}{\bibfnamefont{V.}~\bibnamefont{Tognetti}},
  \bibinfo{author}{\bibfnamefont{R.}~\bibnamefont{Vaia}}, \bibnamefont{and}
  \bibinfo{author}{\bibfnamefont{P.}~\bibnamefont{Verrucchi}},
  \bibinfo{journal}{Phys. Rev. B} \textbf{\bibinfo{volume}{67}},
  \bibinfo{pages}{104414} (\bibinfo{year}{2003}).

\bibitem[{\citenamefont{Lopez et~al.}(1994)\citenamefont{Lopez, Rojo, and
  Fradkin}}]{PhysRevB.49.15139}
\bibinfo{author}{\bibfnamefont{A.}~\bibnamefont{Lopez}},
  \bibinfo{author}{\bibfnamefont{A.~G.} \bibnamefont{Rojo}}, \bibnamefont{and}
  \bibinfo{author}{\bibfnamefont{E.}~\bibnamefont{Fradkin}},
  \bibinfo{journal}{Phys. Rev. B} \textbf{\bibinfo{volume}{49}},
  \bibinfo{pages}{15139} (\bibinfo{year}{1994}).

\end{thebibliography}

\end{document}